\documentclass[12pt,a4paper]{article}

\usepackage{epsfig,graphicx}
\usepackage{amsmath,amsfonts,amssymb}
\usepackage{citesort}

\newcommand{\unit}{\leavevmode\hbox{\small1\kern-3.6pt\normalsize1}}

\def\lsim{\raise0.3ex\hbox{$\;<$\kern-0.75em\raise-1.1ex\hbox{$\sim\;$}}}
\def\gsim{\raise0.3ex\hbox{$\;>$\kern-0.75em\raise-1.1ex\hbox{$\sim\;$}}}
\newcommand{\captions}{\sf\caption}

\def\higgsu{m_{H_u}^2}
\def\higgsd{m_{H_d}^2}
\def\higgsuew{m_{H_u}^2}
\def\higgsdew{m_{H_d}^2}

\def\neumass{m_{\tilde\chi_1^0}}

\newcommand{\crosssec}{\sigma_{\tilde\chi^0_1-p}}

\def\neut{\tilde\chi_1^0}
\def\bsg{$b\to s\gamma$}

\def\asusy{a^{\rm SUSY}_\mu}
\def\bmumu{B_s^0\to\mu^+\mu^-}

\def\relic{\Omega_{\tilde{\chi}_1^0}}

\allowdisplaybreaks

\def\higgsu{m_{H_u}^2}
\def\higgsd{m_{H_d}^2}

\def\neumass{m_{\tilde\chi_1^0}}

\def\neut{\tilde\chi_1^0}
\def\neumass{m_{\tilde\chi_1^0}}
\def\bsg{$b\to s\gamma$}

\def\asusy{a^{\rm SUSY}_\mu}

\begin{document}

\thispagestyle{empty}
\begin{flushright}
  FTUAM 07/02\\
  IFT-UAM/CSIC-07-09\\
  KUNS-2091\\
  arXiv:yymm.nnnn\\
  \vspace*{2.5mm}{6 September 2007}
\end{flushright}

\begin{center}
  {\Large \textbf{Prospects for the direct detection of neutralino
  dark matter in orbifold scenarios} }  
  
  \vspace{0.5cm}
  David\,G.~Cerde\~no\,$^{a,b}$, Tatsuo Kobayashi\,$^{c}$,
  Carlos Mu\~noz\,$^{a,b}$\\[0.2cm] 
    
  {$^{a}$\textit{Departamento de F\'{\i}sica Te\'{o}rica C-XI,
      Universidad Aut\'{o}noma de Madrid,\\[0pt] Cantoblanco, E-28049
      Madrid, Spain}}\\[0pt] 
  {$^{b}$\textit{Instituto de F\'{\i}sica Te\'{o}rica C-XVI, Universidad
      Aut\'{o}noma de Madrid,\\[0pt] Cantoblanco, E-28049 Madrid,
      Spain}}\\[0pt] 
  {$^{c}$\textit{Department of Physics, Kyoto University, Kyoto
    606-8502, Japan.}} \\[0pt]

  \vspace*{1cm}
  \begin{abstract}
    We analyse the phenomenology of orbifold scenarios from the
    heterotic superstring, and the resulting theoretical predictions
    for the direct detection of neutralino dark matter. 
    In particular, we study the parameter space of these
    constructions, computing the low-energy spectrum and taking into
    account the most recent experimental and 
    astrophysical constraints, as well as imposing the absence of
    dangerous charge and colour breaking minima. In the remaining
    allowed regions
    the spin-independent part of the neutralino-proton cross section
    is calculated and compared with the sensitivity of dark matter
    detectors. In addition to the
    usual non universalities of the soft terms in orbifold scenarios
    due to the modular weight dependence, we also consider D-term
    contributions to scalar masses. 
    These are generated by the presence of an anomalous
    $U(1)$, providing more flexibility in the resulting soft terms,
    and are crucial in order to avoid charge and colour breaking
    minima. Thanks to the D-term contribution, large neutralino
    detection cross sections can be found, within the reach of
    projected dark matter detectors. 
  \end{abstract}
\end{center}

\vspace*{5mm}\hspace*{3mm}
	
	\vspace*{-2mm}
	\hspace*{3mm}
	\newpage

\section{Introduction}
\label{wmap}

One of the most interesting candidates for the dark matter in the
Universe is a
Weakly Interacting Massive Particle (WIMP), and in fact
many underground 
experiments are being carried out around the world in order to detect 
its flux on the Earth \cite{mireview}.
These try to observe the elastic scattering of
WIMPs on target nuclei through nuclear recoils.
Although one of the experiments, the DAMA collaboration \cite{dama}, 
reported data favouring the existence of a signal with
WIMP-proton cross section $\approx 10^{-6}-10^{-5}$ pb 
for a WIMP mass smaller than $500-900$ GeV \cite{dama,halo}, 
other collaborations such as 
CDMS Soudan \cite{experimento2}, EDELWEISS \cite{edelweiss}, and
ZEPLIN I \cite{zeplin1} 
claim to have excluded important regions of the DAMA 
parameter space\footnote{
  For attempts to show that DAMA and these experiments
  might not be in conflict, see Ref.~\cite{conflict}.}.
Recently, the 
XENON10 experiment at the Gran Sasso National Laboratory
\cite{xenon10} has set the strongest upper limit for the WIMP-proton
cross section, further disfavouring the DAMA result. 
This controversy will be solved in the future since many
experiments are running or in preparation around the world.
For example, LIBRA \cite{libra} and ANAIS \cite{anais} will probe the
region compatible with DAMA result. 
Moreover, CDMS Soudan will be able to explore a WIMP-proton cross
section $\sigma \gsim 2 \times 10^{-8}$ pb, and
planned 1 tonne Ge/Xe detectors are expected to reach cross
sections as low as 
$10^{-10}$ pb \cite{xenon}.

The leading candidate within the class of WIMPs is 
the lightest neutralino, $\neut$, 
a particle predicted by 
supersymmetric (SUSY) extensions of the standard model. 
Given the experimental situation, and assuming that the dark matter 
is a neutralino, it is natural to wonder how big 
the cross section for its direct detection can be.
This analysis is crucial in order to know the
possibility of detecting dark matter 
in the experiments.
In fact, the analysis of the neutralino-proton cross section 
has been carried out by many authors and during many 
years \cite{mireview}.
The most recent studies take into account the present
experimental and astrophysical constraints
on the parameter space. 
Concerning the former, 
the lower bound on the Higgs mass,
the $b\to s\gamma$ and $\bmumu$ branching ratios, and the
muon anomalous magnetic moment have been considered.
The astrophysical bounds on the dark matter density,
$0.1\lsim \Omega h^2\lsim 0.3$ ($0.095\lsim\Omega
h^2\lsim 0.112$
if we take into account the recent data obtained by the
WMAP satellite \cite{wmap03-1}),
have also been
imposed on the theoretical computation of the relic neutralino density,
assuming thermal production.
In addition, 
the constraints that the absence of dangerous charge
and colour breaking minima imposes on the parameter space
have also been implemented \cite{cggm03-1}.

In the usual minimal supergravity (mSUGRA) scenario, where the soft
terms of the minimal supersymmetric standard model (MSSM) are assumed
to be universal at the unification scale, $M_{GUT} \approx 2\times
10^{16}$ GeV, and radiative electroweak symmetry breaking is imposed,
the neutralino-proton cross section turns out to be constrained by
$\sigma_{\tilde{\chi}_1^0-p}\lsim 3\times 10^{-8}$ pb. Clearly, in
this case, present experiments are not sufficient and only the planned
1 tonne Ge/Xe detectors would be able to test part of the parameter
space. However, in the presence of non-universal soft scalar and
gaugino masses \cite{gr92} the cross section can be increased
significantly \cite{Bottino} in some regions with respect to the
universal scenario (see, e.g., the discussion in \cite{david}, and
references therein). Although the current upper limit on the decay
$B_s \to  \mu^+ \mu^-$ seriously affects these results, as was pointed
out in \cite{kokim}, regions of the parameter space can still be found
where the neutralino detection cross section can be within the reach
of experiments such as CDMS Soudan. An analysis, summarizing all these
results in the context of SUGRA, can be found in \cite{ko}.

On the other hand,
the low-energy limit of superstring theory is 
SUGRA, and therefore the neutralino is also a candidate for
dark matter in superstring constructions.
Let us recall that, 
in the late eighties, working in the context of the 
$E_8\times E_8$ heterotic superstring, 
a number of interesting four-dimensional vacua
with particle content not far from that of the SUSY 
standard model were found 
(see, e.g., the discussion in the introduction of \cite{viejos},
and references therein).
Such constructions
have a natural hidden sector built-in: the
complex dilaton field $S$ arising from the
gravitational sector of the theory, and the complex moduli fields
$T_i$ parametrizing the size and shape of the compactified space. 
The auxiliary fields of those gauge singlets can be the seed of SUSY
breaking, solving the arbitrariness of SUGRA where the hidden sector
is not constrained. In addition, in superstrings the
gauge kinetic function, $f_a(S, T_i)$, and the
K\"ahler potential, $K(S, S^*, T_i, T_i^*)$, 
can be computed explicitly, leading to 
interesting predictions for the soft 
parameters \cite{dilaton}. More specifically, in orbifold
constructions they show a lack of universality due to the modular
weight dependence. From these resulting SUGRA models
one can also obtain predictions for the value of the neutralino-proton 
cross section. In fact,
analyses of the detection cross section in these constructions
were carried out in the past in 
\cite{Shafi2,Drees,Mambrini}.

Our aim in this work is to study in detail the phenomenology of
these orbifold models, including the most recent experimental
constraints on low-energy observables, as well as those coming from
charge and colour breaking minima, and 
to determine how large the 
cross section for the direct detection of neutralino dark matter
can be. We therefore calculate the theoretical predictions
for the spin-independent part of the neutralino-nucleon cross section,
$\crosssec$, and compare it with the sensitivities of present and
projected experiments.
Since the soft terms in superstring
scenarios are a subset of the general soft terms studied in SUGRA
theories
we make use of previous results on departures from the mSUGRA
scenario to look for values of the orbifold soft terms 
giving rise to a large cross section accessible for experiments.

In addition, we introduce a new
ingredient in the analysis, namely the modification produced in the
soft parameters by the presence of an anomalous $U(1)$. Let us recall
that in string theory, and in particular in orbifold constructions
\cite{Dixon,Wilson} of the heterotic superstring \cite{Gross}, the
gauge groups obtained after compactification are larger than the
standard model gauge group, and contain generically extra $U(1)$
symmetries, $SU(3)\times SU(2)\times U(1)^n$ \cite{Kim}. One of these
$U(1)$'s is usually anomalous, and although its anomaly is cancelled
by the four-dimensional Green-Schwarz (GS) mechanism, it generates a
Fayet-Iliopoulos (FI) contribution to the D-term
\cite{FayetIliopoulos}. This effect is crucial for model building
\cite{Katehou} since some scalars acquire large vacuum expectation
values (VEVs) in order to cancel the FI contribution, thereby breaking
the extra gauge symmetries, and allowing the construction of realistic
standard-like models in the context of the $Z_3$
orbifold \cite{Casas1,Casas2,Font} (see also 
\cite{Giedt2}).
Recently  other interesting models in the context of the $Z_6$
orbifold \cite{Raby,Lebedev,Lebedev2}, and $Z_{12}$
orbifold \cite{Kim2,hum}, have been analysed.
Due to the FI breaking, also D-term contributions to the soft
scalar masses are generated
\cite{Nakano,Kawamura,Tatsuo1,Tatsuo2,Kawamura2,Dudas}. 
This allows more flexibility in the soft
terms
and, consequently, in the computation of the associated 
neutralino-proton cross section.

The paper is organised as follows.
In Section~\ref{departures} we briefly 
review the departures from mSUGRA
which give rise to
large values of the neutralino detection cross section.
Then, in the next sections, we use this analysis
to study several orbifold scenarios where such departures may
be present. Special emphasis is put on the effect of the various
experimental constraints on the SUSY spectrum and low-energy
observables.
We start in Section~\ref{noanomalous}
with the simplest 
(but not the most common) possibility,
where an anomalous $U(1)$ is not present.
Then, in Section~\ref{anomalous}, we discuss
the important modifications produced in the soft terms
by the presence of an anomalous  $U(1)$,
and their effects on the computation of the
neutralino-proton cross section, considering 
the effect of D-term contributions
to soft scalar masses.
The conclusions are left for Section~\ref{conclusions}.

\section{Neutralino-proton cross section and departures from mSUGRA}
\label{departures}

In this section we review possible departures from the mSUGRA
scenario, and their impact on the neutralino-proton cross section. 
This will allow us to discuss orbifold scenarios more easily.
Let us first recall that in mSUGRA
one has only four free parameters defined at the GUT scale:
the soft scalar mass, $m$, the soft gaugino mass, $M$, 
the soft trilinear coupling, $A$, and the ratio of the Higgs vacuum
expectation values, $\tan\beta\equiv\langle H^0_u\rangle/\langle
H^0_d\rangle$. In addition, the sign of the Higgsino mass parameter,
$\mu$, remains undetermined. Using these inputs the neutralino-proton
cross section has been analysed exhaustively in the literature, as
mentioned in the Introduction. Taking into account all kind of
experimental and astrophysical constraints, the resulting scalar cross
section is bounded to be $\sigma_{\tilde{\chi}_1^0-p}\lsim 3\times
10^{-8}$ pb.

Departures from the universal structure of the soft parameters in 
mSUGRA allow to increase the neutralino-proton cross section
significantly. As it was shown in the literature, it is possible to
enhance the scattering channels involving exchange of CP-even neutral
Higgses by reducing the Higgs masses, and also by increasing the
Higgsino components of the lightest neutralino. A brief analysis based
on the Higgs mass parameters, $\higgsdew$ and $\higgsuew$, at the
electroweak scale can clearly show how these effects can be achieved.

First, a decrease in the values of the Higgs masses can be
obtained by increasing $\higgsuew$ at the electroweak scale 
(i.e., making it less negative) and/or decreasing $\higgsdew$. 
More specifically, the value of the mass of the heaviest CP-even
Higgs, $H$, can be very efficiently lowered under these
circumstances. This is easily understood by analysing the (tree-level)
mass of the CP-odd Higgs $A$, which for reasonably large values of
$\tan\beta$ can be approximated as $m^2_A\approx
m_{H_d}^2-m_{H_u}^2-M_Z^2$. Since the heaviest CP-even Higgs, $H$, is
almost degenerate in mass with $A$, lowering $m^2_A$ we obtain a
decrease in $m^2_H$ which leads to an increase in the scattering
channels through Higgs exchange

Second, through the increase in the value of
$\higgsuew$ an increase in the Higgsino components of
the lightest neutralino can also be achieved. 
Making $\higgsuew$ less negative, its positive contribution to 
$\mu^2$ in the minimization of the Higgs potential
would be smaller. Eventually $|\mu|$ will be of the order of $M_{1}$,
$M_{2}$ and $\tilde{\chi}_1^0$ will then be a mixed Higgsino-gaugino
state. Thus scattering channels through Higgs exchange become more
important than in mSUGRA, where $|\mu|$ is large and
$\tilde{\chi}_1^0$ is mainly bino. It is worth emphasizing however
that the effect of lowering the Higgs masses is typically more
important, since it can provide large values for the
neutralino-nucleon cross section even in the case of bino-like
neutralinos.

Non-universal soft parameters can produce the above mentioned
effects. Let us consider in particular the non-universality in the
scalar masses, which will be the most interesting possibility in
orbifold scenarios. We can parametrize these in the Higgs sector, at
the high-energy scale, as follows:
\begin{equation}
  m_{H_{d}}^2=m^{2}(1+\delta_{H_{d}})\ , \quad m_{H_{u}}^{2}=m^{2}
  (1+ \delta_{H_{u}})\ .
  \label{Higgsespara}
\end{equation}
Concerning squarks and sleptons we will assume
that the three generations have the
same mass structure:
\begin{eqnarray}
  m_{Q_{L}}^2&=&m^{2}(1+\delta_{Q_{L}})\ , \quad m_{u_{R}}^{2}=m^2
  (1+\delta_{u_{R}})\ , 
  \nonumber\\
  m_{e_{R}}^2&=&m^{2}(1+\delta_{e_{R}})\ ,  \quad m_{d_{R}}^{2}=m^{2}
  (1+\delta_{d_{R}})\ , 
  \nonumber\\
  m_{L_{L}}^2&=&m^{2}(1+\delta_{L_{L}})\ .    
  \label{Higgsespara2}
\end{eqnarray}
Such a structure avoids potential problems with flavour changing
neutral currents\footnote{
  Another possibility would be to assume that the first
  two generations have the common scalar mass $m$, and
  that non-universalities are allowed only for the third 
  generation (as it occurs for the models analysed in
  Ref.\,\cite{Ko:2007dz}).  
  This would not modify our analysis since, as we will see below, 
  only the third generation is relevant in our
  discussion.} 
(FCNC), and arises naturally e.g. in $Z_3$ orbifold constructions with
two Wilson lines, where realistic models have been obtained. 
Note also that whereas all $\delta$'s in (\ref{Higgsespara2}) have to
satisfy $\delta\geq -1 $ in order to avoid an unbounded from below
(UFB) direction breaking charge and colour\footnote{
If we allow metastability of our vacuum, tachyonic masses 
for some sfermions, $\delta < -1$, at the high-energy scale might be allowed.
However, we do not consider such a possibility.}, $\delta_{H_{u,d}} \leq -1$
in (\ref{Higgsespara}) is possible as long as
$m_1^2\,=\,m_{H_{d}}^2+\mu^2>0$ and $m_2^2\,=\,m_{H_{u}}^2+\mu^2>0$
are fulfilled.

An increase in $\higgsu$ at the electroweak scale can be obviously
achieved by  increasing its value at the high-energy scale, i.e., with
the choice $\delta_{H_{u}}>0$. In addition, this is also produced when
$m_{Q_{L}}^2$ and $m_{u_{R}}^2$ at the high-energy scale decrease,
i.e. taking $\delta_{Q_{L},u_{R}} < 0$, due to their (negative)
contribution proportional to the top Yukawa coupling in the
renormalization group equation (RGE) of $m_{H_u}^2$.

Similarly, a decrease in the value of $\higgsd$ at the electroweak
scale can be obtained by decreasing it at the high-energy scale with 
$\delta_{H_{d}}<0$. The same effect is obtained when $m_{Q_{L}}^2$ and
$m_{d_{R}}^2$ increase at the high-energy scale, due to their
(negative) contribution proportional to the bottom Yukawa coupling in
the RGE of $m_{H_d}^2$. Thus one can deduce that $m^2_A$ will also be
reduced by choosing $\delta_{Q_{L},d_{R}} > 0$.

In fact non-universality in the Higgs sector gives the most important
effect, and including the one in the sfermion sector the cross
section only increases slightly.

Taking into account this analysis, several scenarios were discussed
in Ref.~\cite{cggm03-1}, obtaining that large values for the 
cross section are possible. For example, with
$\delta_{H_{d}}=0,\, \delta_{H_{u}}=1;\ 
\delta_{H_{d}}=-1,\, \delta_{H_{u}}=0;\
\delta_{H_{d}}=-1,\, \delta_{H_{u}}=1$,
regions of the parameter space are found which are accessible for
experiments such as CDMS Soudan \cite{ko}. Interestingly, it was also
realised that these choices of parameters were helpful in order to
prevent the appearance of UFB minima in the Higgs potential.

The different UFB directions were classified in Ref.\,\cite{clm1}. 
Among these, the one labelled as UFB-3, which involves VEVs for the
fields $\{H_u,\tilde\nu_{L_i},\tilde e_{L_j},\tilde e_{R_j}\}$ with
$i\neq j$, yields the strongest bound.
After an
analytical minimization of the relevant terms of the
scalar potential the
value of the {$\tilde\nu_{L_i},\tilde e_{L_j},\tilde e_{R_j}$} 
fields can be written in
terms of $H_u$. Then, for any value of $|H_u|<M_{GUT}$ satisfying
\begin{eqnarray}
  |H_u| > \sqrt{ \frac{\mu^2}{4\lambda_{e_j}^2}
    + \frac{4m_{L_i}^2}{g'^2+g_2^2}}-\frac{|\mu|}{2\lambda_{e_j}} \ , 
  \label{SU6}
\end{eqnarray}
the potential along the UFB--3 direction reads 
\begin{eqnarray}
  V_{\rm UFB-3}=(m_{H_u}^2
  + m_{L_i}^2 )|H_u|^2
  + \frac{|\mu|}{\lambda_{e_j}} ( m_{L_j}^2+m_{e_j}^2+m_{L_i}^2 )
  |H_u| -\frac{2m_{L_i}^4}{g'^2+g_2^2} \ .
  \label{ufb3a}
\end{eqnarray}
Otherwise
\begin{eqnarray}
  V_{\rm UFB-3}= m_{H_u}^2
  |H_u|^2
  + \frac{|\mu|} {\lambda_{e_j}} ( m_{L_j}^2+m_{e_j}^2 ) |H_u| +
  \frac{1}{8} (g'^2+g_2^2)\left[
  |H_u|^2+\frac{|\mu|}{\lambda_{e_j}}|H_u|\right]^2  \ .  
  \label{ufb3b}
\end{eqnarray}
In these expressions $\lambda_{e_j}$ denotes the
leptonic Yukawa coupling of the $j$th generation, the deepest
direction corresponding to $e_j=\tau$. The UFB-3 condition  
is then $V_{\rm UFB-3}(Q=\hat Q) > V_{\rm real\ min.}(Q=M_{SUSY})$,
where $V_{\rm real\ min.}=-\frac{1}{8}\left(g'^2 + g_2^2\right)
\left(v_u^2-v_d^2\right)^2$, with $v_{u,d}=\langle H_{u,d}\rangle$,
is the value of the potential at the realistic minimum. 
$V_{real\ min}$ is evaluated at
the typical scale of SUSY masses, $M_{SUSY}$, and $V_{\rm UFB-3}$ at 
the renormalization
scale, $\hat Q$, which is chosen to be $\hat Q\sim
\rm{Max}(\lambda_{\rm 
  top}|H_u|,M_{SUSY})$, in order to minimize the effect of one-loop
corrections to the scalar potential.

As we see from Eqs.\,(\ref{ufb3a}) and (\ref{ufb3b}), the potential
along this direction can be lifted when $m_{H_u}^2$ increases
(becomes less negative) and for large values of the stau mass
parameters, thereby making the UFB-3 condition less restrictive. 
In this sense, non-universal soft terms, like the ones discussed
above, can  be very helpful.

The question now is whether it is possible to find explicit
realisations of these scenarios within orbifold models. In the
following sections we will study this issue in detail.

\section{Orbifold scenarios}
\label{noanomalous}

Let us recall first the structure of the SUGRA theory in
four-dimensional constructions from the heterotic superstring. The
tree-level gauge kinetic function is independent of the moduli sector
and is simply given by 
\begin{equation}
  {f_a} = k_a S\ , 
  \label{kahler3}
\end{equation}
where $k_a$ is the Kac-Moody level of the gauge factor. Usually (level
one case) one takes $k_3=k_2=\frac{3}{5}k_1=1$ for the MSSM. In any
case, the values $k_a$ are irrelevant for the tree-level computation
since they do not contribute to the soft parameters. On the other
hand, the K\"ahler potential has been computed for six-dimensional
Abelian orbifolds, where three moduli $T_i$ are generically present. 
For this  class of models the K\"ahler potential has the form 
\begin{eqnarray}
  K &=& -\log(S+S^*) - \sum _i \log(T_i+T_i^*) 
  + \sum _{\alpha }|C^{\alpha }|^2\Pi_i(T_i+T_i^*)^{n_{\alpha }^i}\ . 
  \label{orbi}
\end{eqnarray}
Here $n_{\alpha }^i$ are (zero or negative) fractional numbers usually 
called `modular weights' of the matter fields $C^{\alpha }$.

In order to determine the pattern of soft parameters it is crucial to
know which fields, either $S$ or $T_i$, play the predominant role in
the process of SUSY breaking. Thus one can introduce a
parametrization for the VEVs of dilaton and moduli auxiliary fields
\cite{Brignole}. A convenient one is given by 
\cite{Brignole,Tatsuo3,Scheich} 
\begin{eqnarray}
  F^S&=& \sqrt{3}\, (S+S^*)\, m_{3/2} \sin \theta\;, \nonumber \\
  F^i&=& \sqrt{3}\, (T_i+T^*_i)\, m_{3/2} \cos \theta\; \Theta_i\;,
  \label{parameterize}
\end{eqnarray}
where $i=1,2,3$ labels the three complex compact dimensions, $m_{3/2}$
is the gravitino mass, and the angles $\theta$ and $\Theta_i$, with
$\sum_{i} |\Theta_i|^2=1$, parametrize the Goldstino direction in the
$S$, $T_i$ field space. Here we are neglecting phases and the
cosmological constant vanishes by construction.

Using this parametrization and Eqs.~(\ref{kahler3}) and (\ref{orbi})
one obtains the following results for the soft terms
\cite{Brignole,Tatsuo3,Scheich}: 
\begin{eqnarray}
  M_a &=& \sqrt{3}\,m_{3/2} \sin\theta\ , \nonumber\\
  m_{\alpha }^2 &=& m_{3/2}^2\left(1 + 3\cos^2\theta\ 
  \sum_i n^i_{\alpha } {\Theta }_i^2 \right) \ ,\nonumber\\
  A_{\alpha \beta \gamma } &=& -\sqrt{3}\, m_{3/2} \left(
  \rule{0pt}{17pt} \sin\theta \right.\nonumber\\ 
  &&\left. + \cos\theta \sum_{i} {\Theta }_i 
  \left[1 +\
    n^i_{\alpha }+n^i_{\beta
    }+n^i_{\gamma}-
    (T_i+T_i^*) \partial_i \log \lambda_{\alpha \beta \gamma}\!
    \right]\right)\ .
  \label{masorbi}
\end{eqnarray}
Although in the case of the $A$ parameter an explicit $T_i$-dependence
may appear in the term proportional to $\partial_i \log
\lambda_{\alpha \beta \gamma }$, where $\lambda_{\alpha \beta
  \gamma}(T_i)$ are the Yukawa couplings and
$\partial_i\equiv\partial/\partial_{T_i}$, it disappears in several
interesting cases \cite{Brignole,Scheich}. 
For example, the $A$-term which is
relevant to electroweak symmetry-breaking is the one associated to the
top-quark Yukawa coupling. Thus, in order to obtain the largest
possible value of the coupling, the fields should be untwisted or
twisted associated to the same fixed point. In both cases $\partial_i
\lambda_{\alpha \beta \gamma }\to 0$, and we will only consider this
possibility here.

Using the above information, one can analyse the structure of
soft parameters available in Abelian orbifolds. 
In the dilaton-dominated SUSY-breaking case ($\cos\theta =0$) the 
soft parameters are universal, and fulfil \cite{ibalu,kaplulouis}
\begin{equation}
  m\ =\ m_{3/2}\ , \ M\ =\ \pm\sqrt{3}\, m\ ,\ A\ =\ -M\ ,
  \label{cuatro}
\end{equation}
where the positive (negative) sign for $M$ corresponds to
$\theta=\pi/2$ ($\theta=3\pi/2$). Of course, these are a subset of the
parameter space of mSUGRA, and as a consequence one should expect
small dark matter detection cross sections, as discussed in the
previous Section.

However, in general, the soft terms (scalar masses and trilinear
parameters) given in Eq.~(\ref{masorbi}) show a lack of universality
due to the modular weight dependence. For example, assuming an overall
modulus (i.e., $T=T_i$ and $\Theta_i=1/\sqrt 3$), one obtains 
\begin{eqnarray}
  m_{\alpha }^2 &=& m_{3/2}^2\left(1 + n_{\alpha }\cos^2\theta 
  \right)\ , 
  \label{scalars} \\
  A_{\alpha \beta \gamma } &=& -\sqrt{3}\, m_{3/2} \sin\theta
  - m_{3/2} \cos\theta \left( 3 + n_{\alpha }+n_{\beta }+n_{\gamma } 
  \right)\ , 
  \label{masorbi2}
\end{eqnarray}
where we have defined the overall modular weights $n_{\alpha}=\sum_i
n^i_{\alpha }$. In the case of $Z_n$ Abelian orbifolds, these can take
the values $-1,-2,-3,-4,-5$. Fields belonging to the untwisted sector
of the orbifold have $n_{\alpha}=-1$. Fields in the twisted sector but
without oscillators have usually modular weight $-2$, and those with
oscillators have $n_{\alpha}\leq -3$.
Of course, if all modular weights of the standard model fields are
equal, one recovers the universal scenario. For example, taking all 
$n_{\alpha}= -1$ one has \cite{Brignole} $m=m_{3/2}\sin\theta$, $M=
\sqrt{3} m$, $A= -M$.

Using notation (\ref{Higgsespara}) and (\ref{Higgsespara2}), the
degree of non-universality in the scalar masses is therefore given by 
\begin{equation}
  \delta_\alpha=n_\alpha\cos^2\theta\ .
\end{equation}
It is worth noticing here that $\delta_\alpha\le0$
as a consequence of the negativeness of the modular weights. As we
will see, this has important phenomenological implications.

On the other hand, the apparent success of the joining of gauge
coupling constants at, approximately, $2\times 10^{16}$ GeV in the
MSSM is not automatic in the heterotic superstring, where the natural
unification scale is $M_{GUT}\simeq  g_{GUT}\times 5.27\times 10^{17}$
GeV, where $g_{GUT}$ is the unified gauge coupling.  
Therefore unification takes place at energies around a factor
$10$ smaller than expected in the heterotic superstring. This problem
might be solved with the presence of large string threshold
corrections which explain the mismatch between both scales
\cite{Choi,Ross}. 
In a sense, what would happen is that the gauge coupling constants
cross at the MSSM unification scale and diverge towards different
values at the heterotic string unification scale. These different
values appear due to large one-loop string threshold corrections.

It was found that these corrections can be obtained for restricted
values of the modular weights of the fields \cite{Ross}. In fact,
assuming generation independence for the $n_{\alpha}$ as well as
$-3\leq n_{\alpha} \leq-1$, the simplest possibility corresponds to
taking the following values for the standard model fields:
\begin{eqnarray}
  &&n_{Q_L}=n_{d_R}=-1,\quad n_{u_R}=-2,\quad n_{L_L}=n_{e_R}=-3\ , 
  \nonumber\\
  &&n_{H_u}+n_{H_d}=-5,\,-4\ ,
  \label{mod_higgs}
\end{eqnarray}
where, e.g., $u_R$ denotes the three family squarks $\tilde{u}_R$, 
$\tilde{c}_R$, $\tilde{t}_R$.
The above values together with Re\,$T\simeq 16$ lead to good agreement
for $\sin^2\theta_W$ and $\alpha_3$ \cite{Ross}. The associated soft
sfermion masses are given by \cite{Brignole}:
\begin{eqnarray}
  m_{Q_L}^2,\,m_{d_R}^2&=\,&m_{3/2}^2\,(1-\cos^2\theta)\ ,\nonumber\\
  m_{u_R}^2&=\,&m_{3/2}^2\,(1-2\,\cos^2\theta)\ ,\nonumber\\
  m_{L_L}^2,\,m_{e_R}^2&=\,&m_{3/2}^2\,(1-3\,\cos^2\theta)\ ,
  \label{weights}
\end{eqnarray}
whereas for the soft Higgs masses, choosing $n_{H_u}=-1\,,\
n_{H_d}=-3$, one obtains: 
\begin{eqnarray}
  m_{H_u}^2&=\,&m_{3/2}^2\,(1-\cos^2\theta)\ ,\nonumber\\
  m_{H_d}^2&=\,&m_{3/2}^2\,(1-3\,\cos^2\theta)\ .
  \label{higgs_b}
\end{eqnarray}
For convenience, this set of modular weights is summarised in
Table\,\ref{tablemodular} and labelled as case A).

For example, with $\cos^2\theta=1/3$, using notation
(\ref{Higgsespara}) and (\ref{Higgsespara2}), the non-universalities
in the Higgs and sfermion sectors correspond to $\delta_{H_{u}}=-1/3$,
$\delta_{H_{d}}=-1$, $\delta_{Q_{L}}=\delta_{d_{R}}=-1/3$,
$\delta_{u_{R}}=-2/3$, and $\delta_{L_L}=\delta_{e_R}=-1$.

\begin{table}[!t]\begin{center}
    \begin{tabular}{|c|ccccccc|}
      \hline
      &$n_{Q_L}$& $n_{{u_R}_{1,2,3}}$& $n_{d_R}$& $n_{{L_L}}$ 
      &$n_{{e_R}_{1,2,3}}$& $n_{H_d}$&$n_{H_u}$\\ 
      \hline
      A)&-1& -2& -1& -3& -3& -3& -1\\
      \hline
      B)&-1& -2& -1& -3& -3& -3& -2\\ 
      \hline
      C)&-1& -2& -2& -1& -1& -2& -1\\
      \hline
      D)&-2& -1& -1& -2& -1& -2& -2\\ 
      \hline
      E)&-1& -1, -3, -3& -1& -3& -1, -3, -3& -2& -3\\
      \hline\end{tabular}\end{center}
  \captions{Modular weights for the scalar fields of
    heterotic orbifold scenarios with an overall modulus 
    that can reproduce gauge unification
    \cite{Ross,ibalu}. 
    Note that cases B) and C) present extra massless
    chiral fields.} 
  \label{tablemodular}
\end{table}

Concerning the soft gaugino masses,
they are given by:
\begin{eqnarray}
  M_3&\simeq& 1.0\sqrt 3 m_{3/2}\sin\theta\ ,
  \nonumber\\
  M_2&\simeq& 1.06\sqrt 3 m_{3/2}\sin\theta\ ,
  \nonumber\\
  M_1&\simeq& 1.18\sqrt 3 m_{3/2}\sin\theta\ .
  \label{gauginoss}
\end{eqnarray}
The small departure from universality is 
due to the effect of the string threshold
corrections on the gauge kinetic function \cite{Brignole}.

Finally, for the above modular weights, and using (\ref{masorbi2}),
the expressions for the trilinear parameters read
\begin{eqnarray}
  A_\tau&=&-m_{3/2}(\sqrt 3 \sin\theta-6\cos\theta)\ ,\nonumber\\
  A_b&=&-m_{3/2}(\sqrt 3 \sin\theta-2\cos\theta)\ ,\nonumber\\
  A_t&=&-m_{3/2}(\sqrt 3 \sin\theta-\cos\theta)\ .
  \label{trilinear}
\end{eqnarray}
The $A$-term which is relevant to radiative symmetry breaking
is the one associated to the top-quark Yukawa coupling $A_t$.

These soft terms serve as an explicit model for the 
study of the neutralino detection cross section. Since they are
completely determined in terms of just the gravitino mass and the
Goldstino angle, we are left with three free parameters, namely
$m_{3/2}$, $\theta$, and $\tan\beta$, plus the sign of $\mu$. Note,
however, that the absence of negative mass-squared of the sleptons at
the GUT scale implies the constraint $\cos^2\theta\leq
\frac13$. Besides, the shift $\theta\to\theta+\pi$ implies in the
above equations $m_{\alpha}\to m_{\alpha}$, $M_{a}\to -M_{a}$ and
$A_t\to-A_t$. This fact makes it unnecessary to consider both signs of
the $\mu$ parameter. The reason is that the RGEs are symmetric under
the change $\mu,\,M,\,A\to-\mu,\,-M,\,-A$. Consequently, in the
remainder of this paper we will assume $\mu>0$. Notice in this sense
that we will always have $\mu M_i>0$ for $\theta<\pi$ whereas  $\mu
M_i<0$ for $\theta>\pi$. This will have important implications, as we
will soon see, on the effect of the experimental constraints on the
rare decays \bsg\ and $\bmumu$, and on the SUSY contribution
to the muon anomalous magnetic moment, $\asusy$.

\begin{figure}[!t]
  \epsfig{file=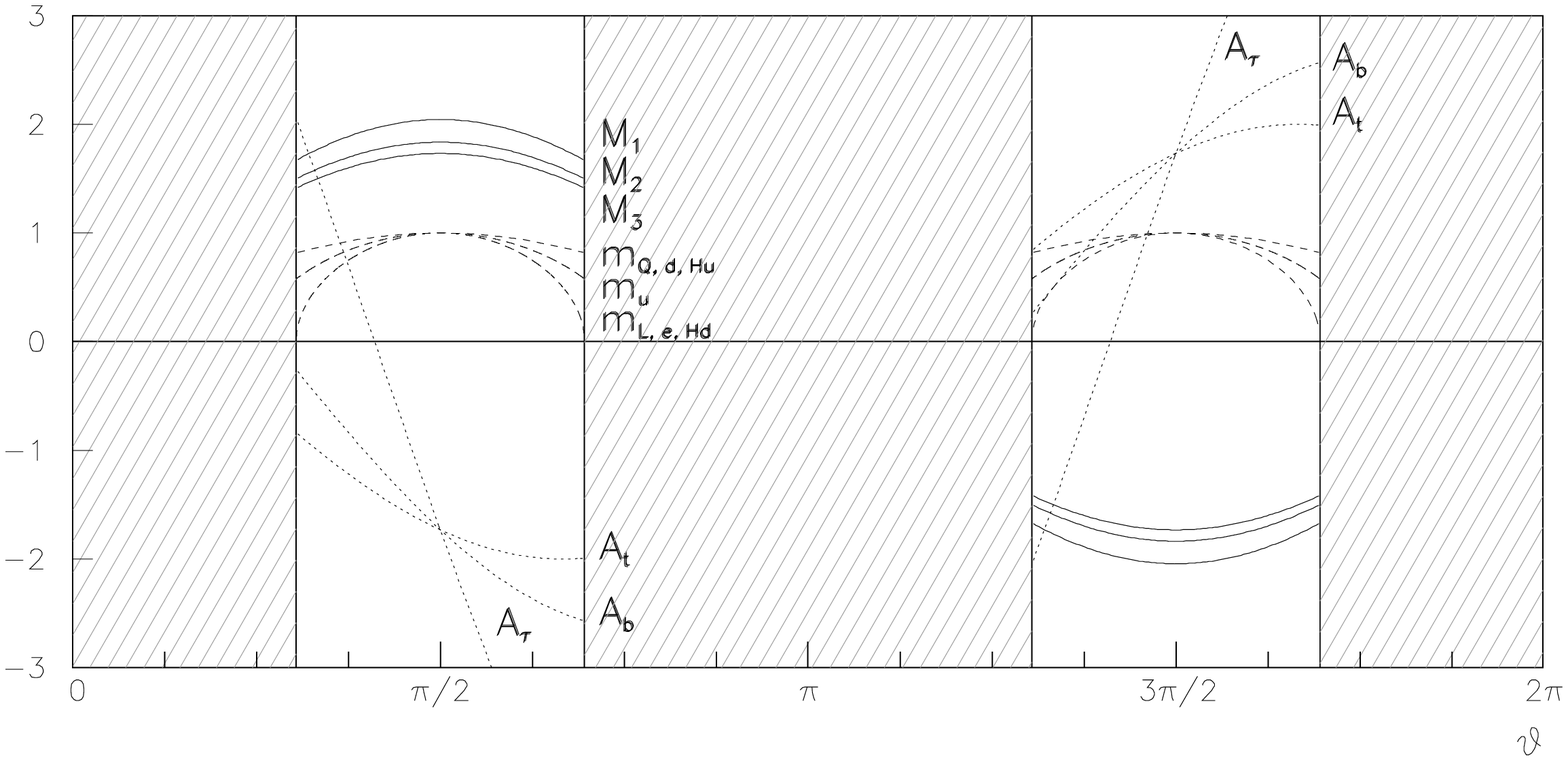,width=16cm}
  \vspace*{-1cm}
  \captions{
    Soft terms at the string scale in units of $m_{3/2}$ as a
    function of the Goldstino angle, $\theta$. 
    Solid lines represent, from top to bottom, the bino, wino and
    gluino mass parameters. The various scalar masses are depicted by 
    means of dashed lines. Finally, dotted lines correspond to the
    trilinear terms. The oblique ruled areas are excluded due
    to the occurrence of negative mass-squared parameters.
    \label{10agutsp200}}
\end{figure}

The resulting structure of the soft parameters for case A), given at
the GUT scale, is represented in 
Fig.\,\ref{10agutsp200} as a function of the Goldstino angle in
units of the gravitino mass.
Two generic features of this kind of orbifold constructions are
evidenced by the plot, namely, the fact that scalar masses are always
smaller than gaugino masses, and the presence of regions which are
excluded because some scalar masses-squared become negative. In the
present example, as already mentioned,
the strongest bound is set by 
slepton masses, for which (\ref{weights}) implies
$\cos^2\theta\le1/3$.
The ruled areas 
correspond to those where this bound is not fulfilled. 
This reduces the allowed parameter space to two
strips in $\theta$, around the dilaton-dominated case, $\theta=\pi/2,
\,3\pi/2$.

\begin{figure}[!t]
  \epsfig{file=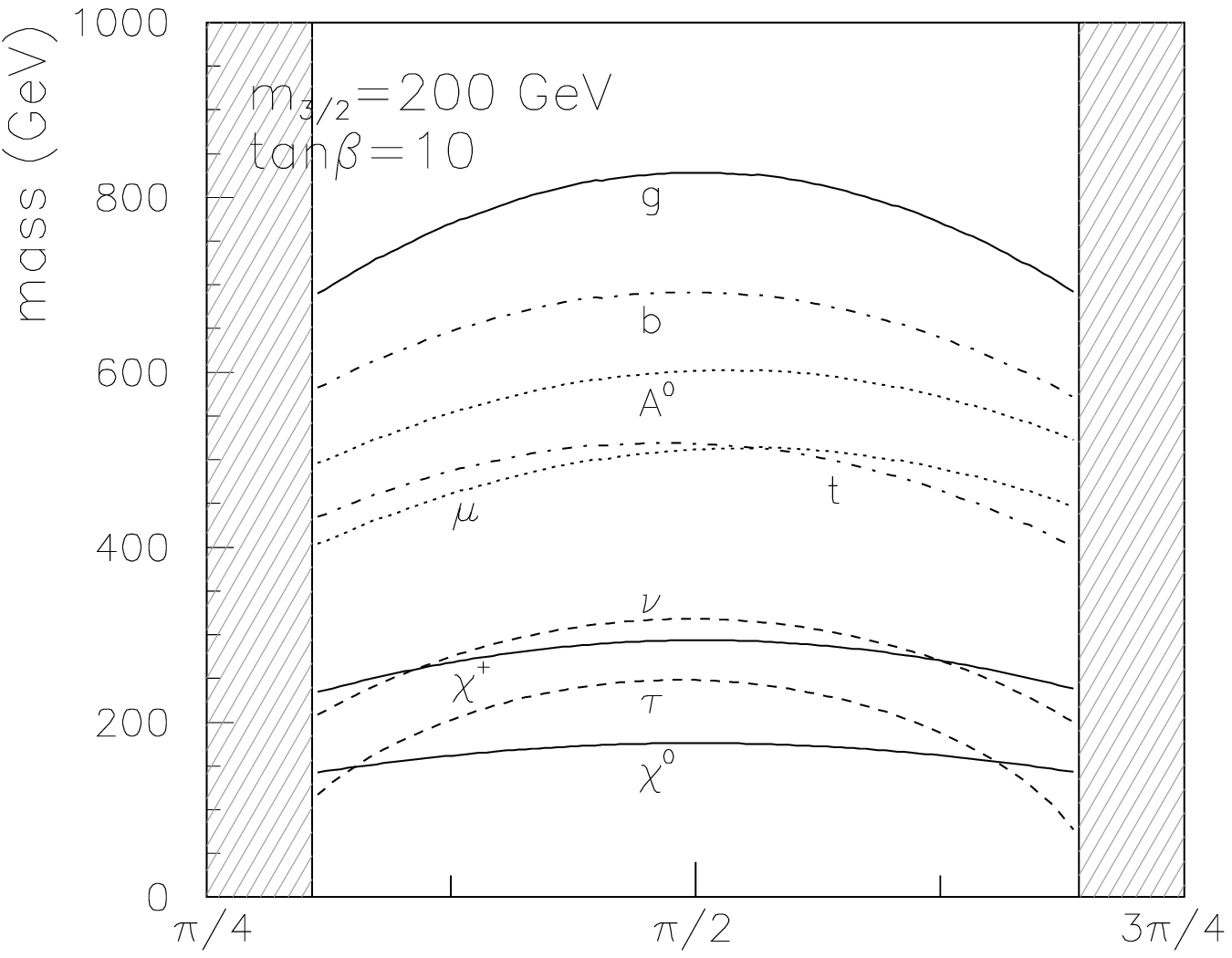,width=8.7cm}
  \hspace*{-0.9cm}  
  \epsfig{file=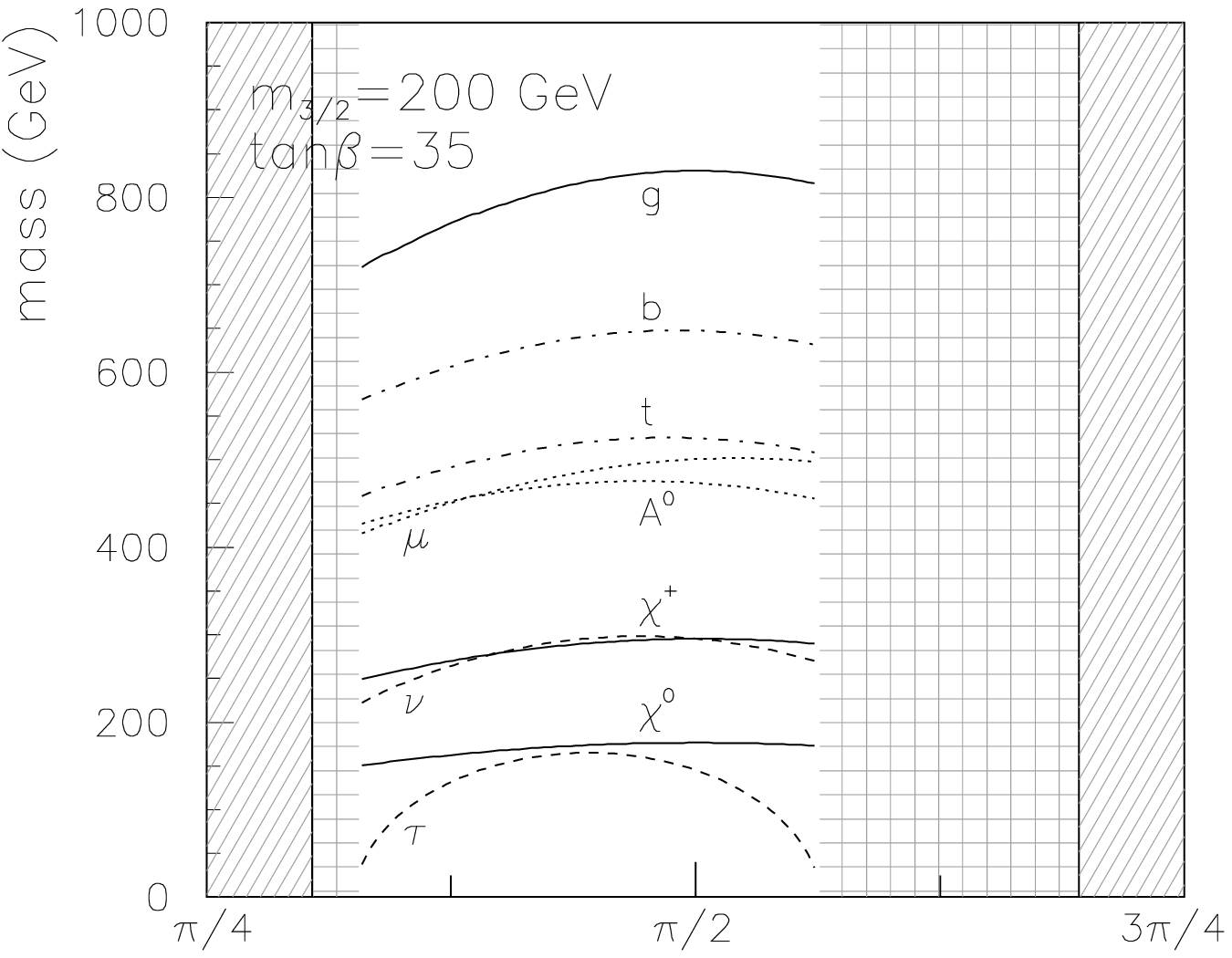,width=8.7cm}
  \vspace*{-1cm}
  \captions{
    Supersymmetric spectrum at low-energy as a function of 
    the Goldstino angle, $\theta$, for $m_{3/2}=200$ GeV and
    $\tan\beta=10$ and $35$. Only the region around $\theta=\pi/2$ is
    represented. 
    From bottom to top, the solid lines correspond to the lightest
    neutralino, lightest chargino and gluino masses. The dashed lines 
    represent the lightest stau and lightest sneutrino masses. The
    lightest stop and sbottom masses are plotted by means of
    dot-dashed lines. Finally, the dotted lines show the mass of the
    CP-odd Higgs and the resulting $\mu$ parameter.
    The oblique ruled areas are excluded due
    to the occurrence of negative mass-squared parameters 
    at the GUT scale, whereas the gridded regions
    correspond to those where tachyons appear after solving the RGEs.
    \label{10asp}}
\end{figure}

With this information, 
the RGEs are numerically solved and the low-energy
supersymmetric spectrum is calculated.
Fig.\,\ref{10asp} shows the resulting particle spectrum 
as a function of the Goldstino angle for $m_{3/2}=200$ GeV and
$\tan\beta=10$ and $35$. 
As we can see, although
slepton masses-squared are positive at the GUT scale for
$\cos^2\theta\le1/3$, the RGEs can still drive 
them negative, or lead to tachyonic mass eigenstates. 
This is typically the case of the lightest stau, $\tilde\tau_1$, and
lightest sneutrino, $\tilde\nu_1$ (the latter 
only for low values of the gravitino
mass), due to their small mass parameters (\ref{weights}). 
This is more likely to happen for large $\tan\beta$, 
since the lepton Yukawas (which are proportional
to $1/\cos\beta$) increase and induce a larger negative contribution
to the slepton RGEs. 
In such a case, the lightest stau can be the lightest SUSY particle
(LSP) in larger regions of
the parameter space, thus potentially reducing the allowed areas for
neutralino dark matter, as we see in the example with $\tan\beta=35$. 
The supersymmetric spectrum also displays a heavy squark sector, due
to the gluino contribution on the running of their mass
parameters. Similarly, the heavy Higgs masses (represented here only
with the pseudoscalar, $A^0$) are also sizable. For reference, the
value of the $\mu$ term is also displayed and found to be large.

Notice at this point that there are regions of the parameter space
where the lightest neutralino is the LSP and the stau, being the
next-to-lightest SUSY particle (NLSP),
has a very similar mass. As we will soon see, this allows reproducing
the correct dark matter relic density by means of a coannihilation
effect. On the other hand, one can readily see that in these examples
$m_{A^0}>2\,\neumass$ and therefore there is no enhancement in the
annihilation of neutralinos mediated by the CP-odd Higgs.

Finally, it is worth emphasizing that in these scenarios 
the gravitino is never the LSP. 
Despite the bino mass being larger than $m_{3/2}$ at the string scale,
its RGE always leads to $M_1<m_{3/2}$ at the electroweak scale (even
in the dilaton-dominated limit for which $M_1$ is at its maximum) so
that, at least, the neutralino mass is always lighter than $m_{3/2}$.

Having extracted the supersymmetric spectrum, we are ready to
determine the
implications for low-energy observables and study how the associated
bounds further restrict the allowed parameter space. 
In our analysis the most recent experimental and 
astrophysical constraints have been included. In particular, the lower
bounds on the masses of the supersymmetric particles and on the
lightest Higgs have been implemented, as well as the experimental
bound on the branching ratio of the \bsg\ process,
$2.85\times10^{-4}\le$B(\bsg)$\le 4.25\times10^{-4}$. 
The latter has been calculated taking into account the most recent
experimental world average for the branching ratio 
reported by the Heavy Flavour Averaging Group \cite{hfag,cleo,belle},
as well as the new re-evaluation of the SM value
\cite{misiak-newbsg}, with errors combined in quadrature.  
We also take into account the improved experimental
constraint on the $\bmumu$ branching ratio,
B$(\bmumu)<1.5\times10^{-7}$, obtained from a combination of the
results of CDF and D0, \cite{bmumuexp,bmumuexp2,bmumuexp3}.  
The evaluation of the neutralino relic density is carried
out with the program {\tt micrOMEGAs} \cite{micro}, and, due to its
relevance, the effect of the WMAP constraint will be shown
explicitly. Finally, dangerous charge and colour breaking minima of
the Higgs potential will be avoided by excluding unbounded from below
directions.

Concerning $\asusy$, we have taken into account the
experimental result for the muon anomalous magnetic moment
\cite{g-2}, as well as the most recent theoretical evaluations of the
Standard Model contributions \cite{g-2_SM,newg2,kino}. It is found
that when $e^+e^-$ data are used the experimental excess in
$a_\mu\equiv(g_\mu-2)/2$ would constrain a possible supersymmetric
contribution to be $\asusy=(27.6\,\pm\,8)\times10^{-10}$, where
theoretical and experimental errors have been combined in
quadrature. However, when tau data are used, a smaller discrepancy
with the experimental measurement is found. Due to this reason, in our
analysis we will not impose this constraint, but only indicate the
regions compatible with it at the $2\sigma$ level, this is, 
$11.6\times10^{-10}\le\asusy\le43.6\times10^{-10}$.

\begin{figure}[!t]
  \epsfig{file=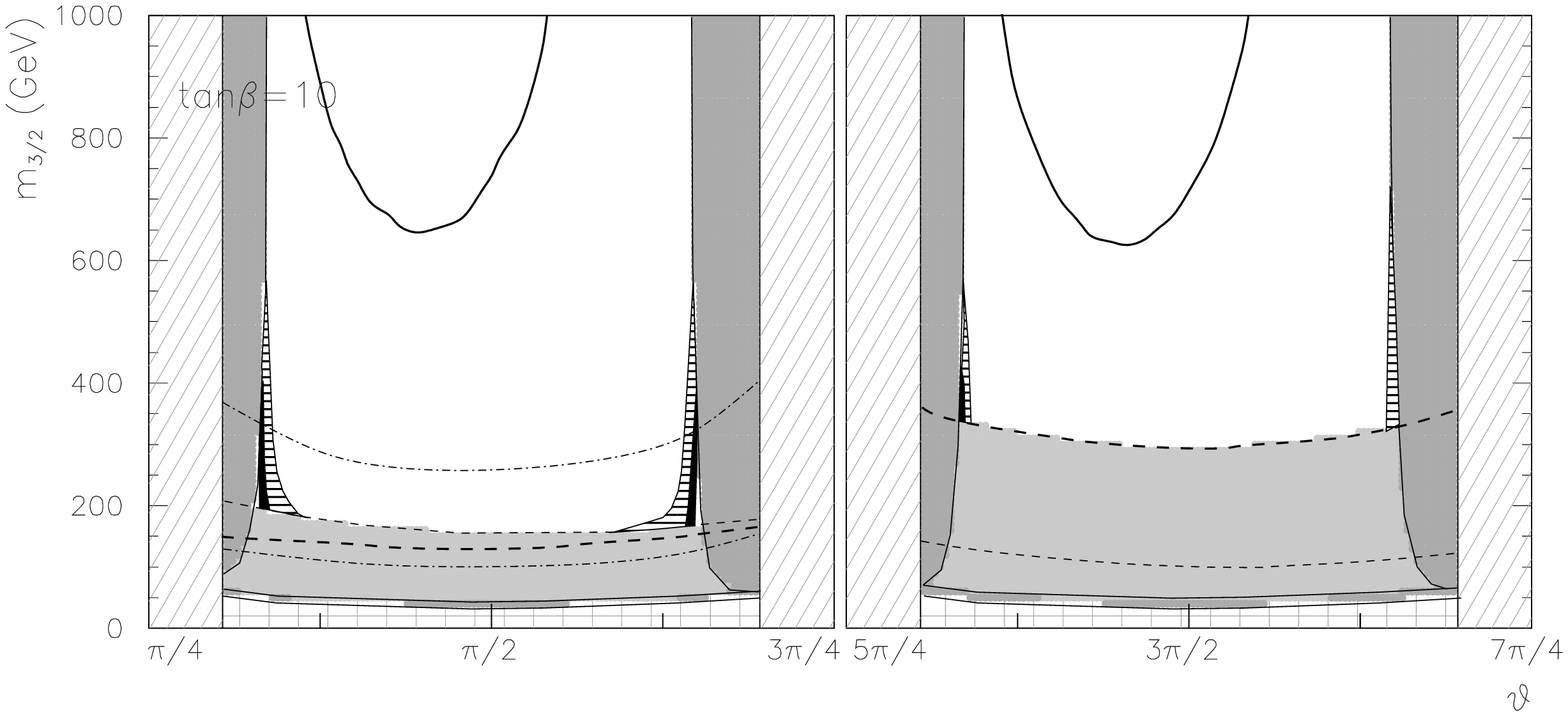,width=16cm}
  \vspace*{-1cm}
  \captions{
    Effect of the different experimental constraints on the parameter
    space  
    ($m_{3/2}$,\,$\theta$) for $\tan\beta=10$.
    The oblique ruled areas are excluded due to the occurrence of
    tachyons at the GUT scale, whereas the gridded regions
    correspond to those where tachyons appear after solving the RGEs.
    Only the two areas centred around $\theta\approx\pi/2,\,3\pi/2$
    are free from tachyons in the slepton sector.
    Dark grey areas represent those where the lightest neutralino is
    not the LSP. Among these
    regions, the narrow 
    vertical areas contained within solid lines are 
    those where the stau is the LSP, whereas in the 
    thin horizontal region at very small gravitino masses, 
    also bounded by solid lines, the LSP is the
    lightest sneutrino.
    Light grey areas stand for those not fulfilling one or more
    experimental bounds. 
    In particular,  
    the region below the thin dashed line 
    is excluded by the lower bound 
    on the Higgs mass. 
    The area below the
    thick dashed line is excluded by $b\to s\gamma$.
    The regions excluded by the experimental constraints on the
    masses of the chargino and stau are always contained within those
    ruled out by other reasons and are therefore not shown.
    The region bounded by thin dot-dashed lines is favoured by
    $\asusy$ (notice that the whole allowed area around
    $\theta\approx3\pi/2$ 
    always has $\asusy<0$ and is therefore disfavoured), although this
    constraint has not been explicitly applied.
    The remaining white area is favoured by all the experimental 
    constraints. Within it the ruled region fulfils in addition
    $0.1\leq \Omega_{\tilde{\chi}_1^0}h^2\leq 0.3$, and 
    the black area on
    top of this indicates the WMAP range
    $0.094<\Omega_{\tilde{\chi}_1^0}h^2<0.112$.
    Finally, the UFB
    constraints are only fulfilled in the area above the thick solid
    line. 
    \label{10a}}
\end{figure}

\begin{figure}[!t]
  \epsfig{file=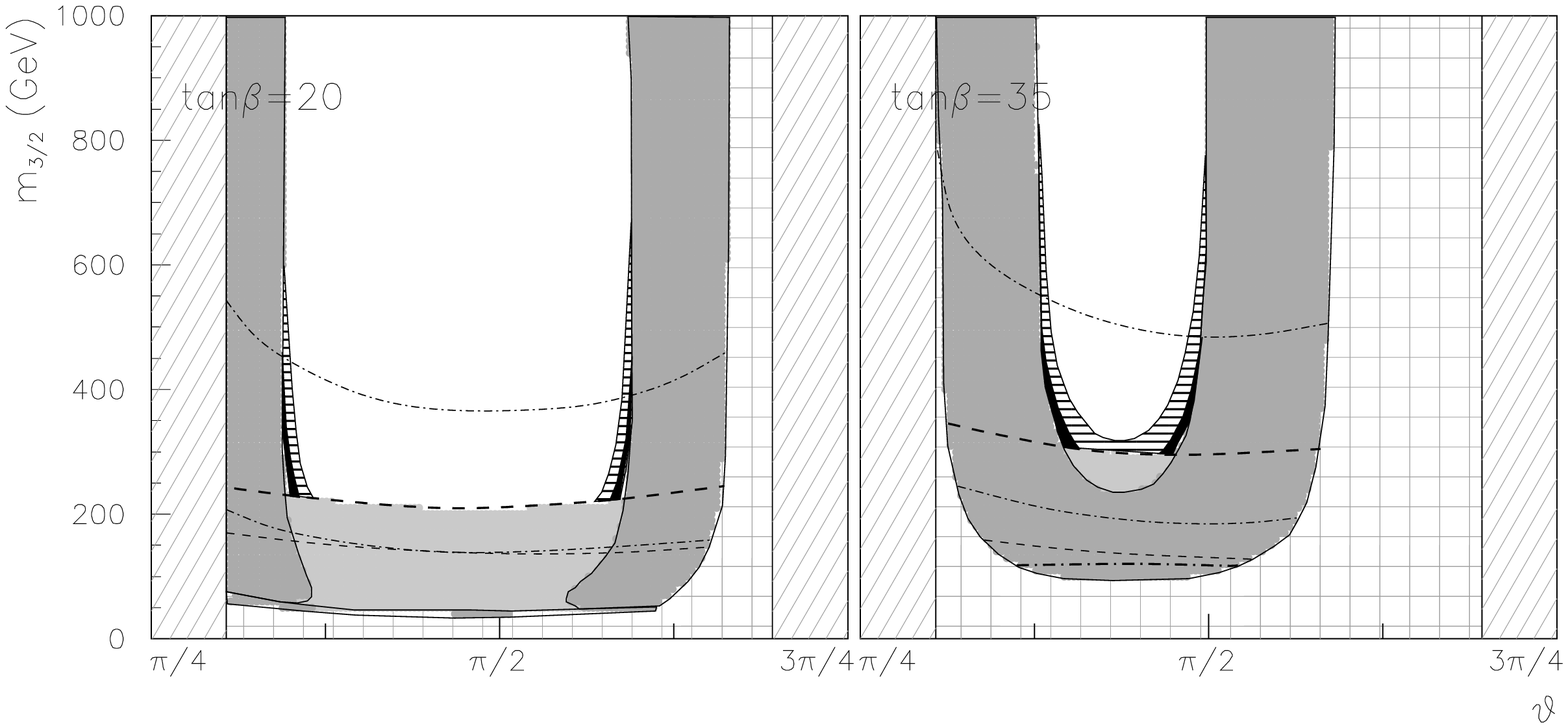,width=16cm}
  \vspace*{-1cm}
  \captions{The same as Fig.\,\ref{10a} but for $\tan\beta=20$ and
    $35$. Notice that the UFB constraints are violated in the whole
    represented parameter space and therefore no thick solid line is 
    shown. 
    In the plot for $\tan\beta=35$, the area
    below the thick dot-dashed line on the region 
    is excluded due to the constraint on the 
    $\bmumu$ branching ratio}
  \label{2035a}
\end{figure}

For a better understanding of all these constraints, 
we have represented in
Fig.\,\ref{10a} their effect on the $(m_{3/2},\theta)$ plane for
$\tan\beta=10$. For comparison, the cases with $\tan\beta=20,\,35$ are
also shown in Fig.\,\ref{2035a}.

The first thing to notice is that
extensive regions are excluded due to the occurrence of tachyonic
masses for sleptons. As already discussed, the area excluded for this
reason becomes larger when $\tan\beta$ increases, an
effect which is clearly displayed in Figs.\,\ref{10a} and \ref{2035a}.
This implies an increase in the lower bound of the gravitino
mass. Whereas for $\tan\beta=10,\,20$ the smallest allowed value is
$m_{3/2}\approx35$ GeV, in the case with $\tan\beta=35$ one needs
$m_{3/2}\gsim90$ GeV.

The above mentioned smallness of the slepton mass parameters, together
with the fact that gaugino masses are always larger than scalar masses
($M_a>m_{\alpha}$), also imply that the areas in the parameter
space where the lightest neutralino is the LSP are not very extensive. 
These regions occur for small values of $\cos\theta$ (they are
centered around $\theta\approx\pi/2,\,3\pi/2$), since the
ratio $|M|/m_{L_L,e_R}$ increases with\footnote{
  The lack of a complete mirror symmetry at $\theta=\pi/2$ and
  $3\pi/2$ is due to the trilinear terms (\ref{masorbi2}) being a
  combination of $\sin\theta$ and $\cos\theta$.
} 
$\cos\theta$. Note that such values of the Goldstino angle mean that
the breaking of SUSY is mainly due to the dilaton auxiliary term.
Once more, the allowed areas shrink for large values of $\tan\beta$
and eventually disappear for $\tan\beta\approx45$. In the rest of the
parameter space the role of the LSP
is mainly played by the lightest stau. Although, as already mentioned,
for small values of $\tan\beta$ the sneutrino can also be the LSP in a
very narrow band for small gravitino masses, this area is always
excluded by experimental bounds.

The relevance of the experimental constraints is also evidenced by
Figs.\,\ref{10a} and \ref{2035a}.
Reproducing the experimental result of the branching ratio of \bsg\ is
much easier in the region around $\theta=\pi/2$, since it has $\mu
M>0$. On the contrary, it poses a stringent lower bound on the 
value of $m_{3/2}$ for the region around $\theta=3\pi/2$, for which
$\mu M<0$. As expected, the area excluded for this reason also
increases for larger values of $\tan\beta$. Thus, 
whereas this constraint implies $m_{3/2}\gsim 150$~GeV for
$\tan\beta=10$ in the area around $\theta=\pi/2$, $m_{3/2}\gsim
250\,(300)$~GeV is necessary for $\tan\beta=20\, (35)$.

Having $\mu M<0$, the whole region around $\theta=3\pi/2$ also fails
to fulfil the experimental constraint on $\asusy$, and is therefore
further disfavoured.

The bound on the lightest Higgs mass also rules out some regions for
small gravitino masses. This is only relevant for small values of 
$\tan\beta$ and in the region around $\theta=\pi/2$. Already for
$\tan\beta\gsim15$ this bound becomes less important than the \bsg\ or
$\asusy$ constraints. The areas not fulfilling the experimental
constraints on sparticle masses are always contained within those
already excluded by other bounds and are therefore not shown
explicitly.

\begin{figure}[!t]
  \epsfig{file=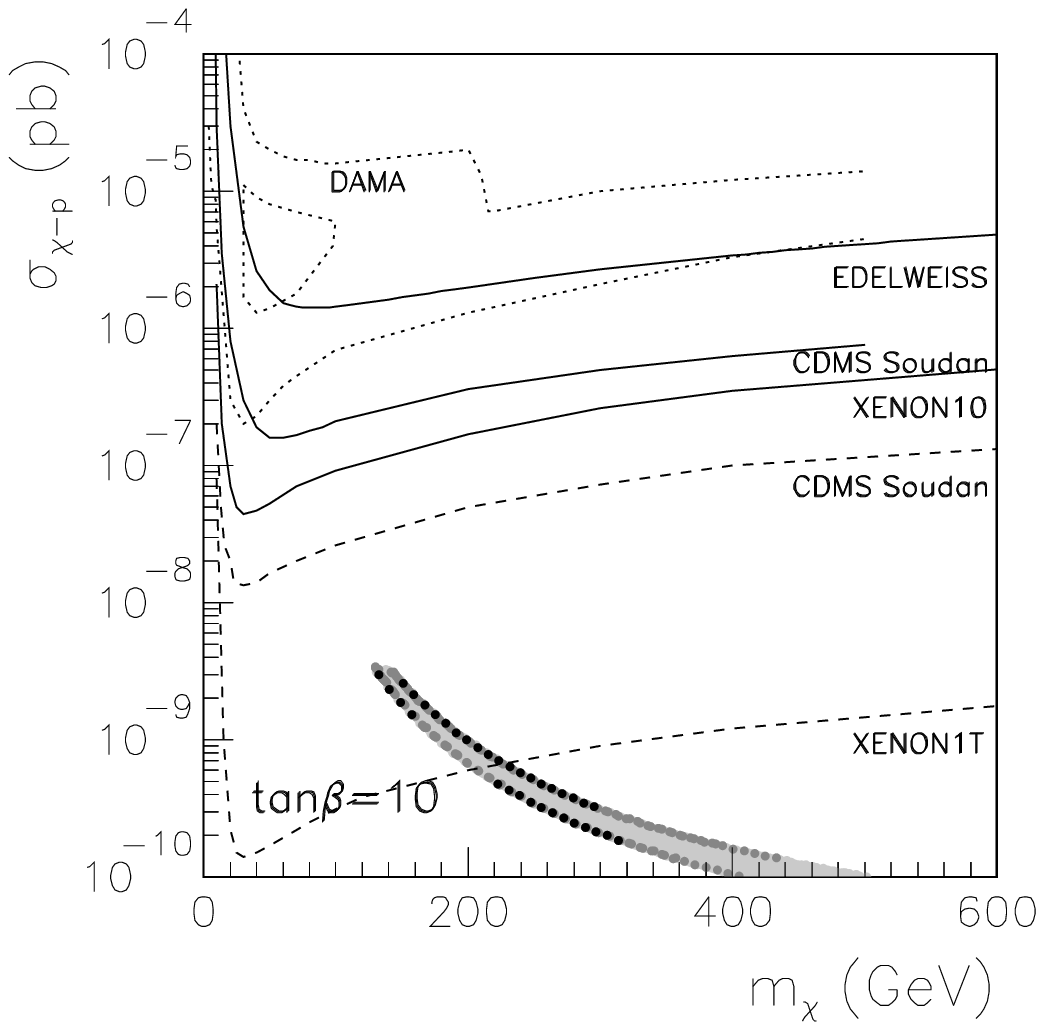,width=8cm}
  \epsfig{file=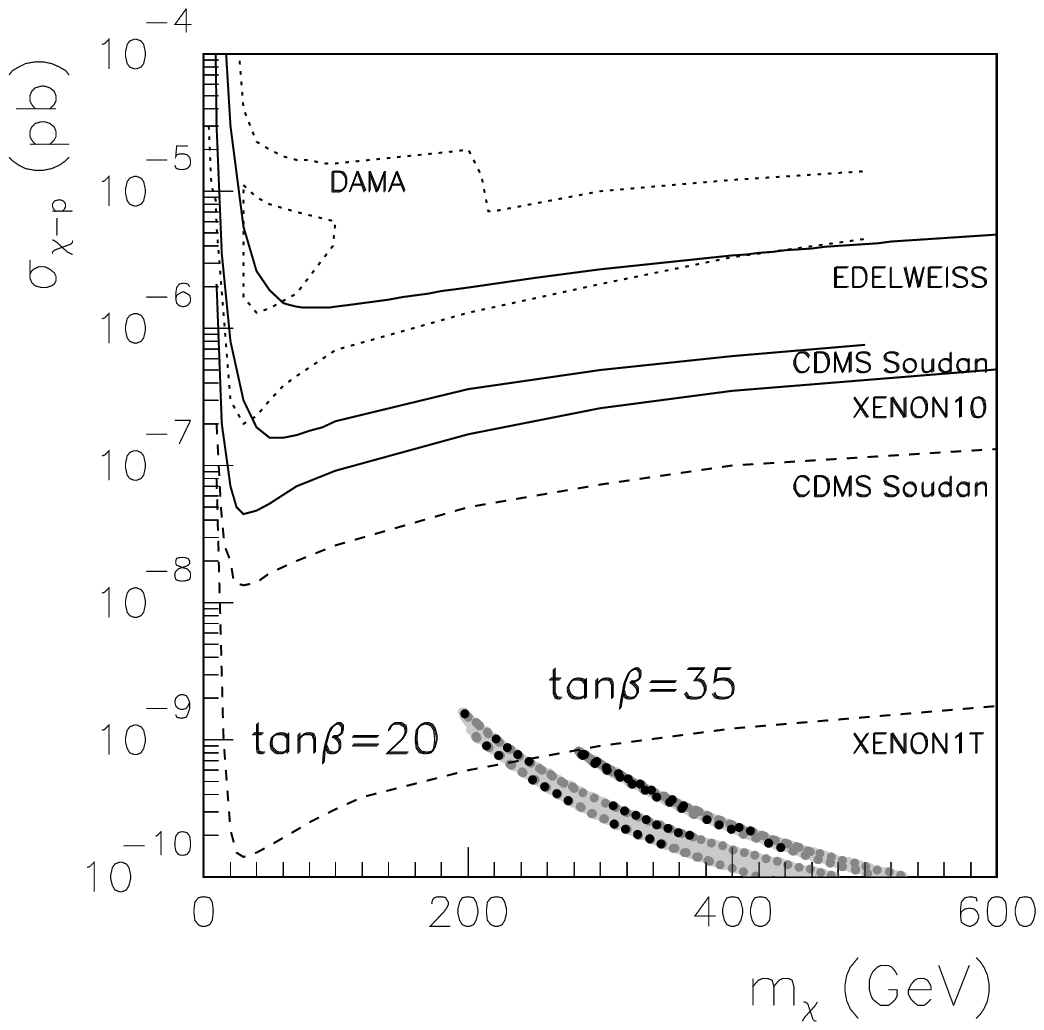,width=8cm}
  \vspace*{-1cm}
  \captions{
    Scatter plot of the scalar neutralino-proton cross section,
    $\sigma_{\tilde{\chi}_1^0-p}$, as a function of the neutralino
    mass, $m_{\tilde{\chi}_1^0}$, for $\tan \beta=10,\,20$ and
    $35$. The light grey dots represent points fulfilling all the
    experimental constraints. The dark grey dots represent points
    which satisfy in addition $0.1\leq \Omega_{\tilde{\chi}_1^0}h^2\leq
    0.3$ and the black dots on top of these indicate those in agreement
    with the WMAP range $0.095<\Omega_{\tilde{\chi}_1^0}h^2<0.112$. The
    sensitivities of present and projected experiments are also depicted
    with solid and dashed lines, respectively, in the case of an
    isothermal spherical halo model. The large (small) area bounded by
    dotted lines is allowed by the DAMA experiment when astrophysical
    uncertainties to this simple model are (are not) taken into account.  
    \label{102035across}
  }
\end{figure}

The allowed parameter space is further reduced when the constraint on
the relic density is imposed. The WMAP result is only reproduced along
the narrow regions close to the area where the stau becomes the LSP. 
This is due to the well known coannihilation effect that takes place
when the neutralino mass is close to the stau mass. The equivalent of
the ``bulk region'' in the mSUGRA parameter space is here excluded by
the experimental constraints. 
Finally, no regions are found where $2\neumass\approx m_A$, 
and consequently resonant annihilation of neutralinos does not play
any role in this case.

\begin{figure}[!t]
  \epsfig{file=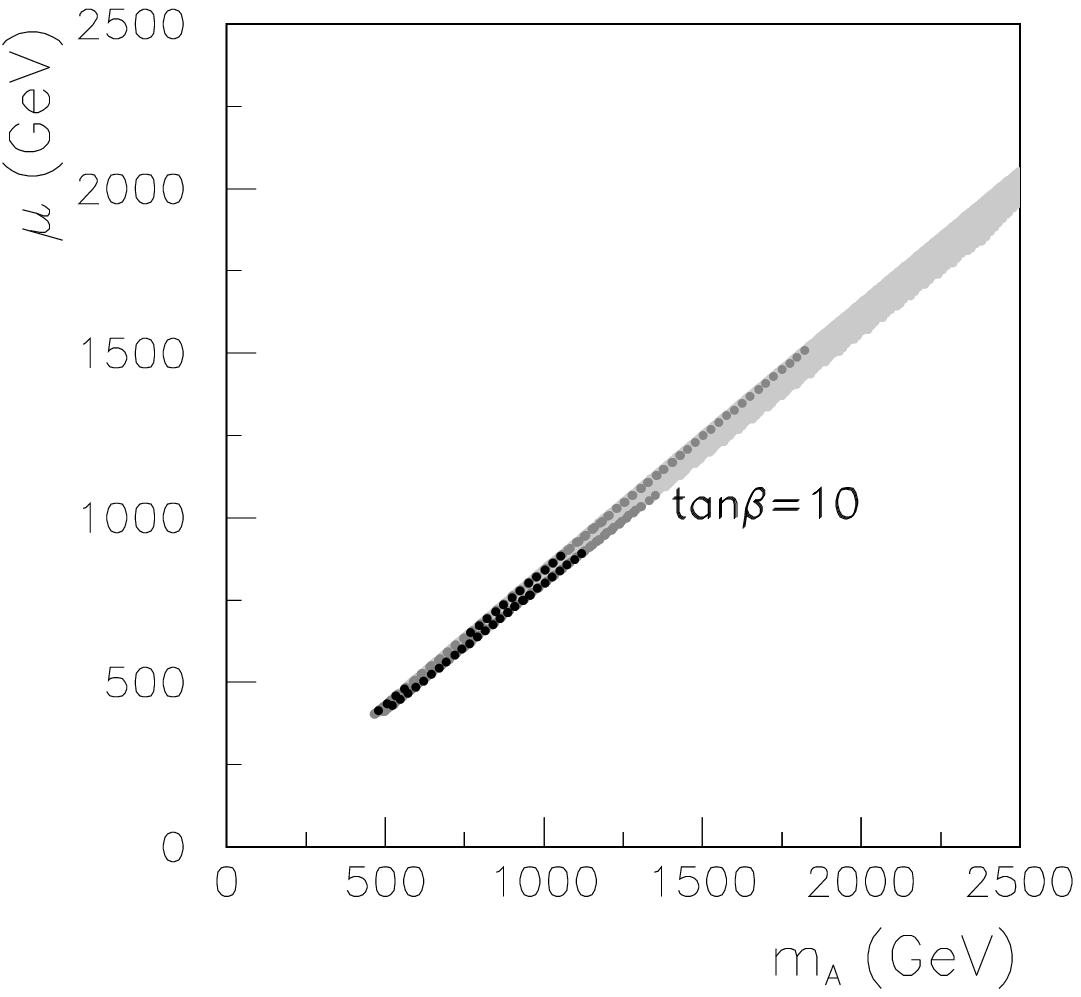,width=8cm}
  \epsfig{file=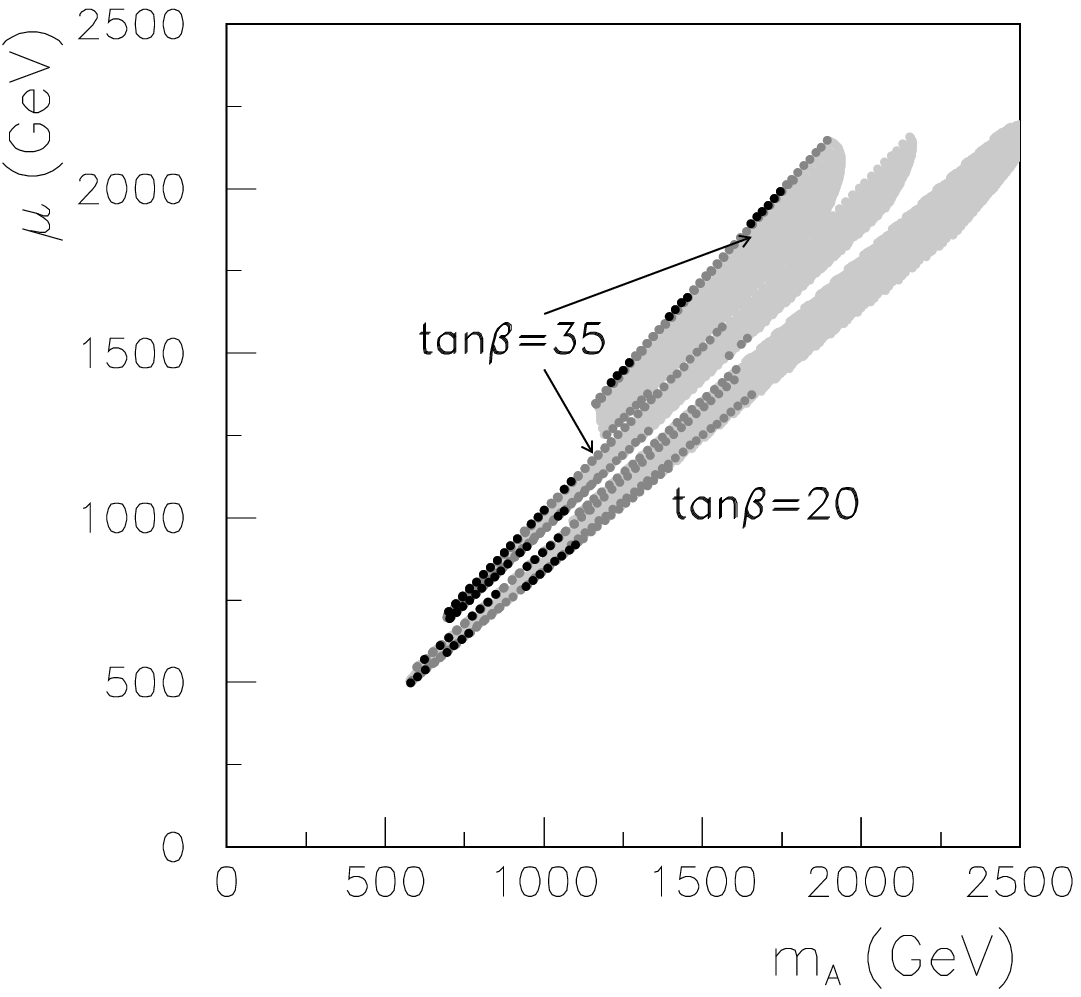,width=8cm}
  \vspace*{-1cm}
  \captions{
    Scatter plot of the resulting $\mu$ parameter as a function of the 
    pseudoscalar Higgs mass, $m_A$, for $\tan \beta=10,\,20$ and
    $35$. The colour convention is as in Fig.\,\ref{102035across}.
    \label{102035mamua}
  }
\end{figure}

Having shown that there are regions with viable neutralino dark
matter, let us now turn our attention to its possible direct
detection. Following the discussion of Section~\ref{departures}, the
Higgs modular weights giving rise to the soft masses (\ref{higgs_b}),
could induce an increase of the neutralino detection cross section
with respect to the universal case. In order to investigate this
possibility, the theoretical predictions for the spin-independent part
of the neutralino-nucleon cross section have 
been calculated in the accepted regions of the parameter space. 
They are represented versus the neutralino mass in
Fig.\,\ref{102035across} for $\tan\beta=10,\,20$ and $35$, where the
sensitivities of present and projected dark matter experiments are
also shown. These results resemble those of mSUGRA, as no high values
are obtained. As in mSUGRA, in this scenario the $\mu$ parameter and
the heavy Higgs masses are sizable (see Fig.\,\ref{10asp}), 
thus implying bino-like
neutralinos and a suppressed contribution to $\crosssec$ from
Higgs-exchanging processes. This is illustrated in
Fig.\,\ref{102035mamua}, where the resulting values of the $\mu$
parameter are plotted as a function of the pseudoscalar Higgs
mass. After analysing the range $\tan\beta=10$ to $50$ we found that
$\crosssec\lesssim5\times 10^{-9}$~pb, 
the maximum values corresponding to
$\tan\beta\approx15$. These results are therefore beyond the
present sensitivities of dark matter detectors and would only be
partly within the reach of the projected 
1 tonne detectors.

So far we have not commented on the bounds 
imposed by the UFB-3 constraint
to avoid dangerous charge and colour breaking minima of the Higgs
potential. This turns out to play a crucial role in disfavouring this
scenario. Indeed, most of the parameter space is excluded on these
grounds\footnote{This is consistent with previous analyses of
charge and colour breaking minima in different superstring
and M-theory scenarios \cite{Ibarra}.}.
Only for small values of $\tan\beta$ and heavy gravitinos do allowed
regions appear (see for instance Fig.\,\ref{10a}, where
$m_{3/2}\gsim650$~GeV is necessary), but these always correspond to
areas where the neutralino relic density is too large and
exceeds the WMAP constraint. For $\tan\beta\gsim15$ the UFB-3
constraint already excludes the complete region with
$m_{3/2}<1000$~GeV. Once more, the reason for this is the low value of
the slepton masses, and more specifically, of the stau mass.
Let us recall that the smaller this value, the more negative
$V_{UFB-3}$ in (\ref{ufb3a}) or (\ref{ufb3b}) is, 
and thus the stronger the UFB-3 bound becomes.
Moreover, the fact that in this scenario the value of $m_{H_u}^2$ is
not very large (since $\delta_{Hu}$ is negative) 
also contributes in driving the potential deeper along this
direction.

Let us finally remark that 
other examples with
different choices of modular weights for the Higgs parameters
satisfying (\ref{mod_higgs}) have been investigated, such as
case B) in Table\,\ref{tablemodular}, and lead to
qualitatively similar results.

The previous analysis suggests how to modify the model to `optimise'
its behaviour under the UFB-3 constraint \cite{Ibarra}, increasing
also the regions in the parameter space where the lightest neutralino
is the LSP. The most favorable case would correspond to slepton masses
as large as possible, i.e.,
\begin{equation}
  n_{L_L}=n_{e_R}=-1\ .
  \label{mod_higgs2}
\end{equation}
For squark and Higgs mass parameters we will continue using the
modular weights of case A)\footnote{
  Of course, with such a choice of modular weights we know that
  the string threshold corrections cannot account for the joining
  of gauge couplings at the MSSM unification scale. 
  Thus we will be tacitly assuming that there is some other
  effect (e.g., the existence of further chiral fields in the spectrum 
  below the heterotic string scale \cite{mass,Giedt2,viejos}) which
  appropriately produces the correct low-energy experimental values
  for gauge couplings.}. 
Note that now the bound on $\cos^2\theta$ is less constraining,
since we only need to impose $\cos^2\theta\leq\frac12$, thus allowing 
a larger degree of
non-universality. For example, with $\cos^2\theta=\frac13$, we get
$\delta_{H_{u}}=-1/3$ and $\delta_{H_{d}}=-1$ for the Higgs 
masses.

\begin{figure}[!t]
  \epsfig{file=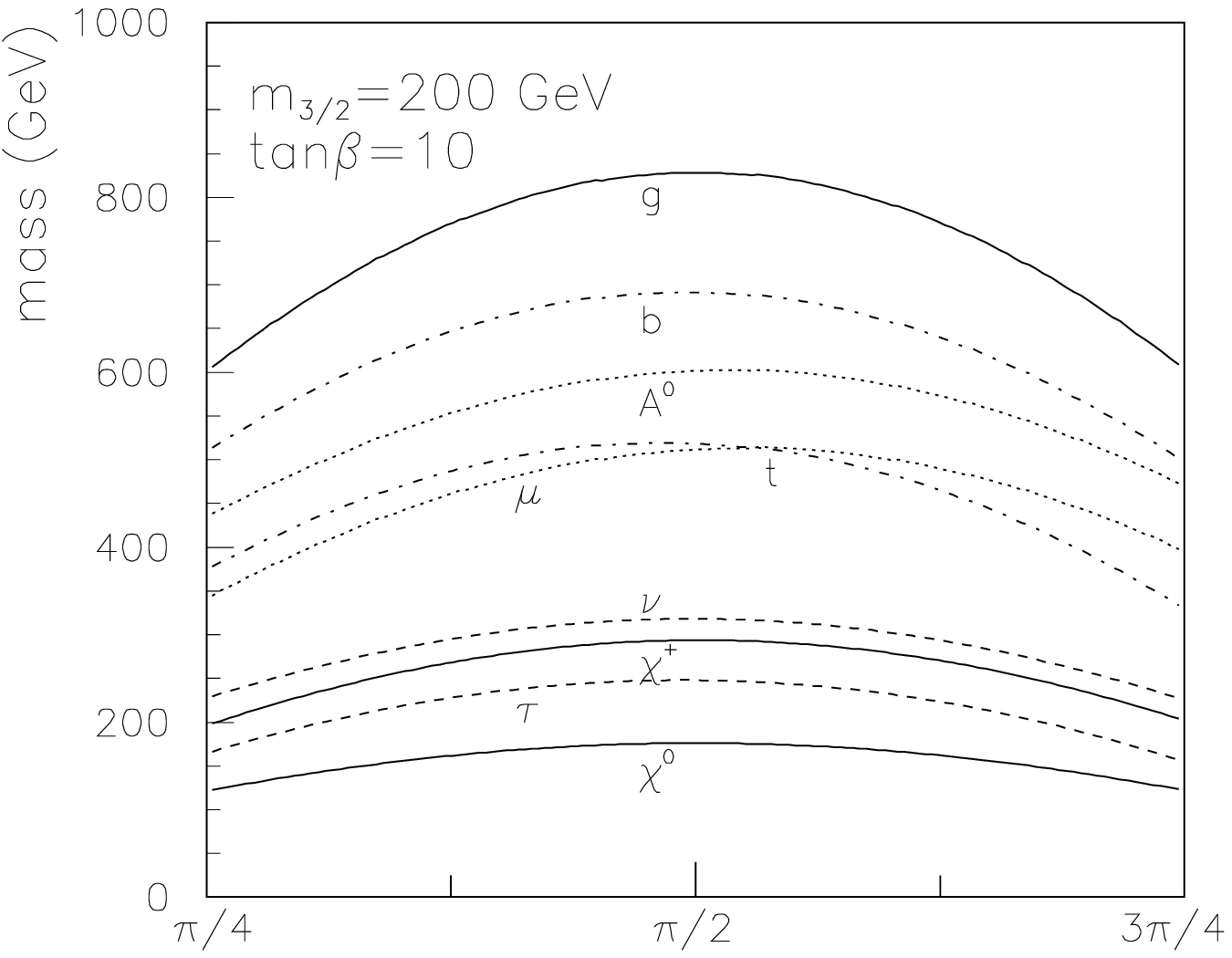,width=8.7cm}
  \hspace*{-0.9cm}  
  \epsfig{file=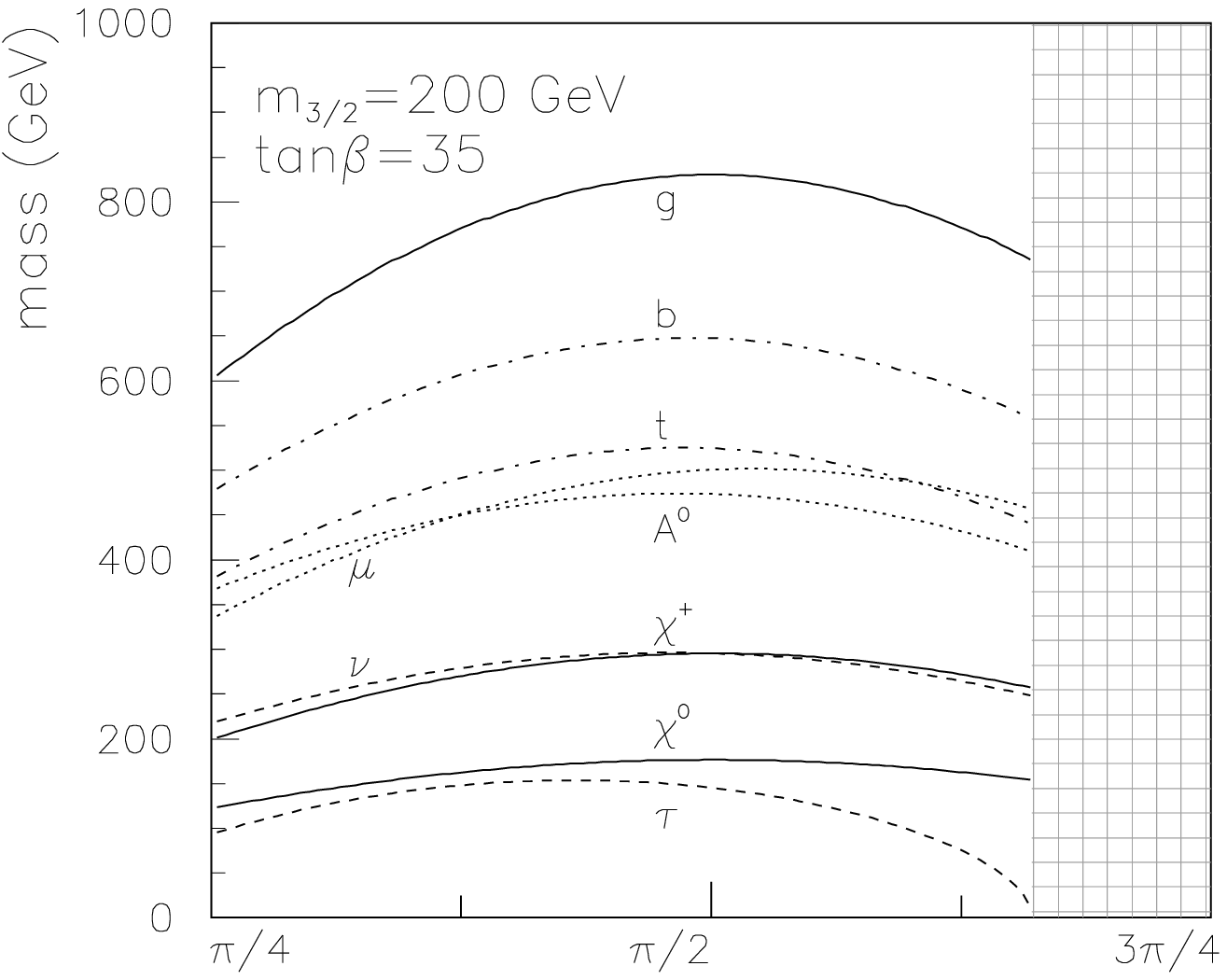,width=8.7cm}
  \captions{The same as Fig.\,\ref{10asp} but for the
    optimised example where $n_{L_L}=n_{e_R}=-1$.
    \label{1035optimsp}}
\end{figure}

The resulting supersymmetric spectrum is shown in
Fig.\,\ref{1035optimsp} as a function of the Goldstino angle for
$m_{3/2}=200$ GeV with $\tan\beta=10$ and $35$. Notice that now the
whole region with $\theta\in[\pi/4,\,3\pi/4]$ is free 
from tachyons at the GUT scale. 
For small $\tan\beta$ the increase in the slepton mass-squared
parameters leaves extensive allowed regions where the neutralino can
be the LSP. 
As expected, larger values of $\tan\beta$ lead to a reduction in the
stau mass, which now easily becomes the LSP, and gives rise to
tachyons in some regions.

\begin{figure}[!t]
  \epsfig{file=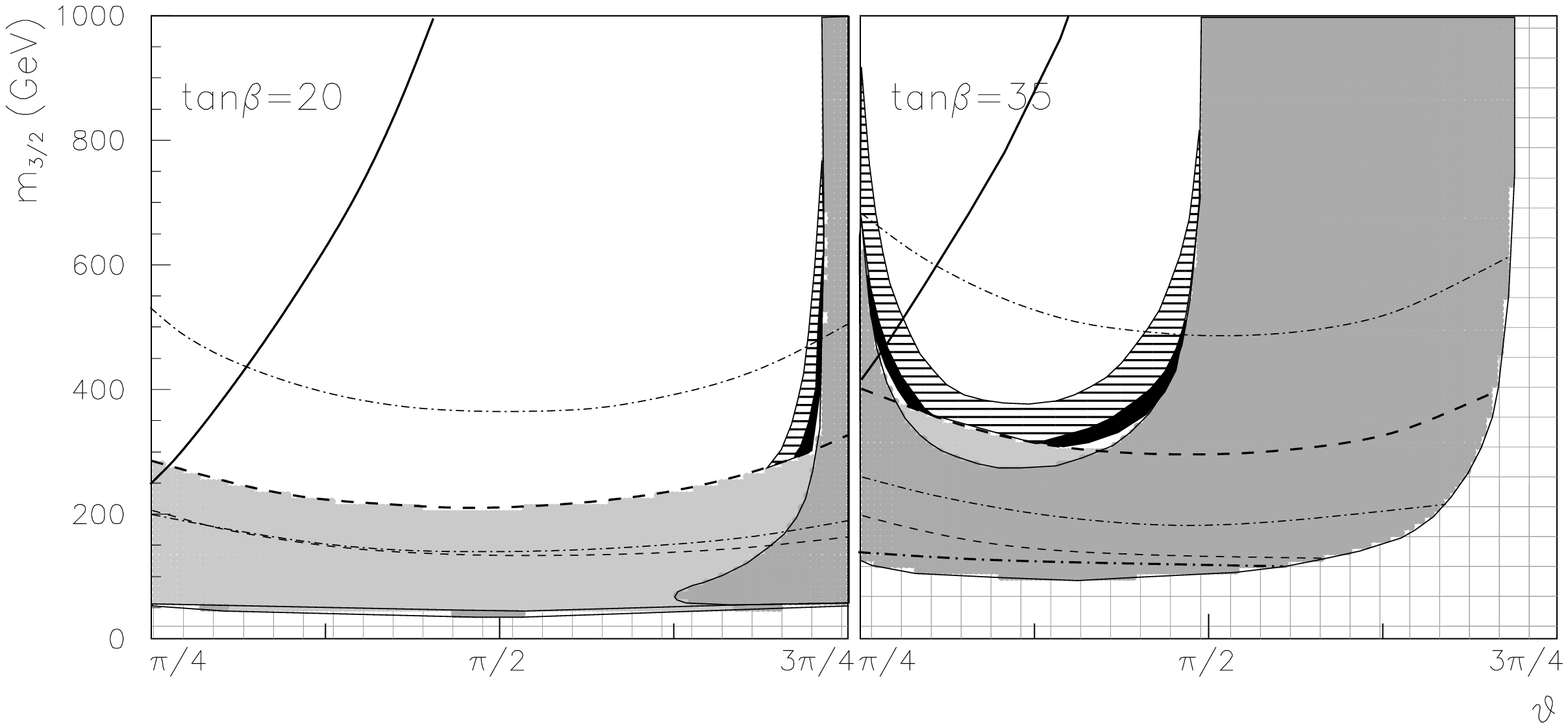,width=16cm}
  \vspace*{-1cm}
  \captions{The same as Fig.\,\ref{10a} but for the optimised example
    where $n_{L_L}=n_{e_R}=-1$ with $\tan\beta=20,\,35$.} 
  \label{1020optim} 
\end{figure}

These features are evidenced in Fig.\,\ref{1020optim}, where 
the corresponding $(m_{3/2},\theta)$ parameter space is depicted for
$\tan\beta=20,\, 35$. 
Notice that, due to the increase in the slepton mass terms, 
the stau only becomes the LSP on narrow bands on the right-hand
side of the allowed areas for $\tan\beta=20$ (for smaller values of
$\tan\beta$ the neutralino is always the LSP). As expected, these
areas with stau LSP become more sizable for $\tan\beta=35$ and
eventually dominate the whole parameter space for $\tan\beta\gsim 45$.
Also, for $\tan\beta\lsim 25$ the sneutrino can also be the LSP on a
narrow region for very light gravitinos, although this is always
excluded by experimental constraints.

\begin{figure}[!t]
  \epsfig{file=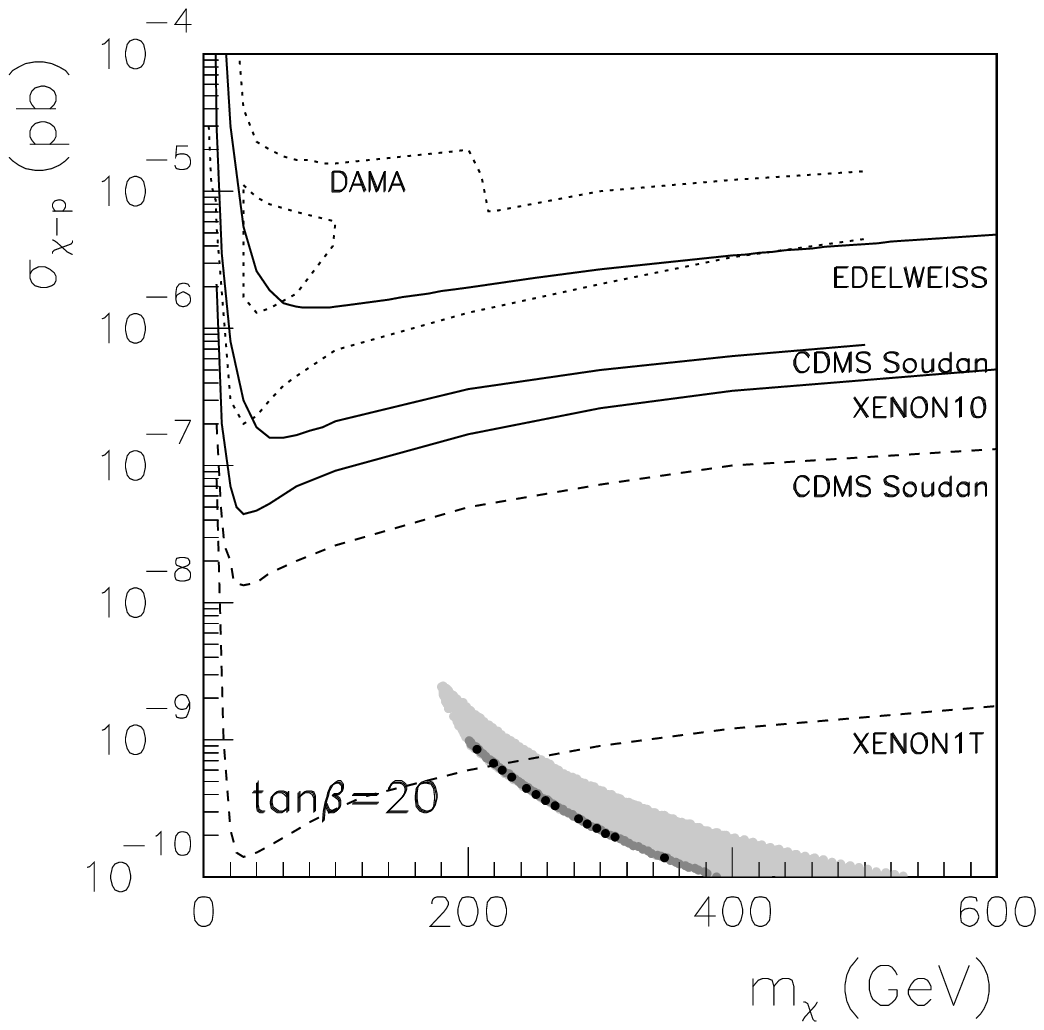,width=8cm}
  \epsfig{file=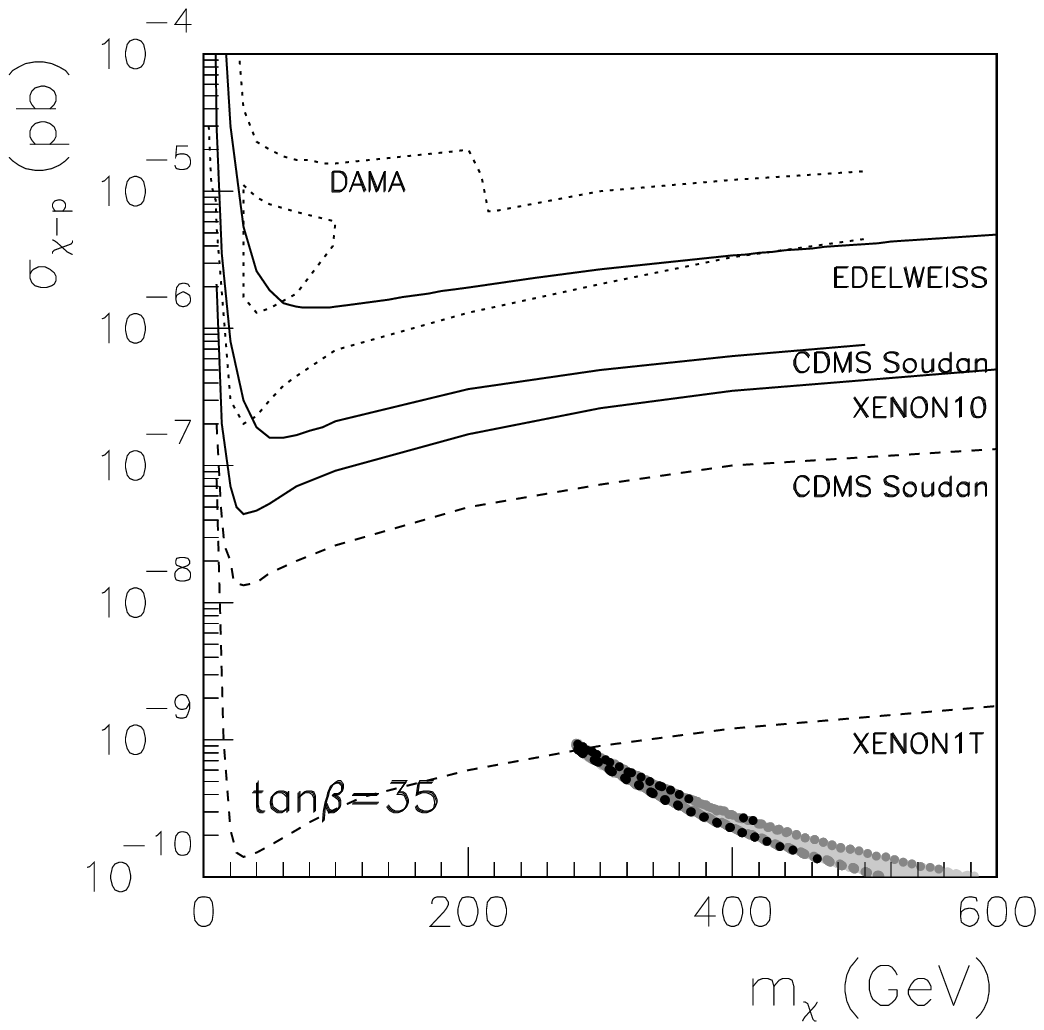,width=8cm}
  \vspace*{-1cm}
  \captions{The same as Fig.\,\ref{102035across} but for the optimised
    example where $n_{L_L}=n_{e_R}=-1$ with $\tan\beta=20,\,35$.} 
  \label{10crossoptim}
\end{figure}

The decrease of the stau mass towards the right-hand side of the
allowed areas can be understood by 
analysing the expressions for the trilinear soft terms. 
The trilinear terms associated to the top, bottom and
tau Yukawa coupling read in this example
\begin{eqnarray}
  A_t&=&-m_{3/2}\left(\sqrt{3}\sin\theta-\cos\theta\right)\
  ,\nonumber\\ 
  A_b=A_\tau&=&-m_{3/2}\left(\sqrt{3}\sin\theta-2\cos\theta\right)\ .
\end{eqnarray} 
It can be checked that for all of them
the ratio $|A/M|$ increases towards the right-hand side of
both allowed areas. In particular, $|A_{\tau,b}/M|\approx 0.05$ for
$\theta=\pi/4,\, 5\pi/4$ and becomes  $|A_{\tau,b}/M|\approx -1.8$ for
$\theta=3\pi/4,\,7\pi/4$.
The increase in $|A_{\tau,b}/M|$ leads to a larger negative correction
in the RGE for the slepton mass terms, implying lighter staus.
Large
values of $\tan\beta$ increase the corresponding Yukawas thus further
decreasing the stau mass.

The variation in the stau mass affects the area excluded by
the UFB-3 
constraint, which becomes more stringent towards the right-hand 
side of the allowed regions. 
On the left, as expected, the effect of the UFB constraints is less
severe than in the previous examples and
regions with $m_{3/2}\gsim 250\, (400)$~GeV for $\tan\beta=20\, (35)$
are allowed.
Interestingly, for $\tan\beta\gsim 30$ part of these areas can also
reproduce the correct value for the neutralino relic density.

\begin{figure}[!t]
  \epsfig{file=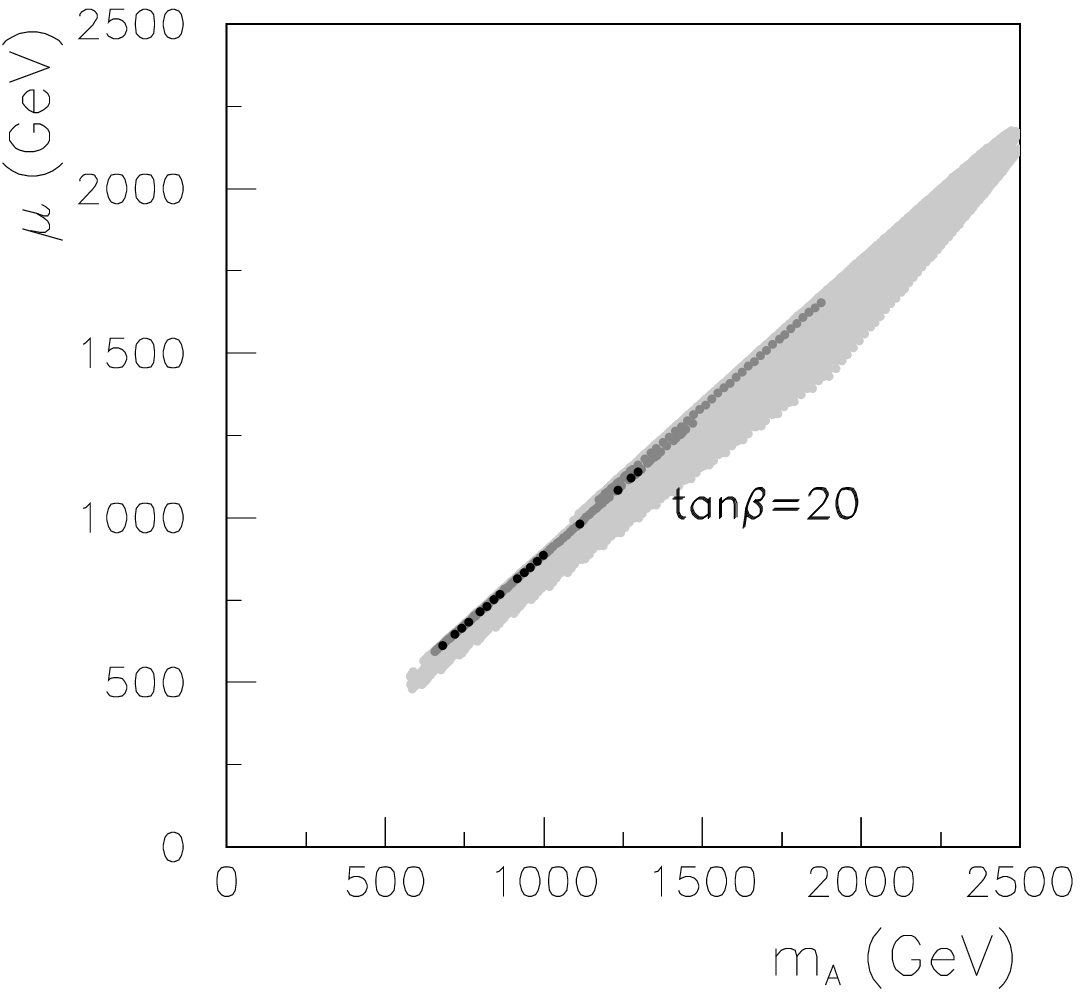,width=8cm}
  \epsfig{file=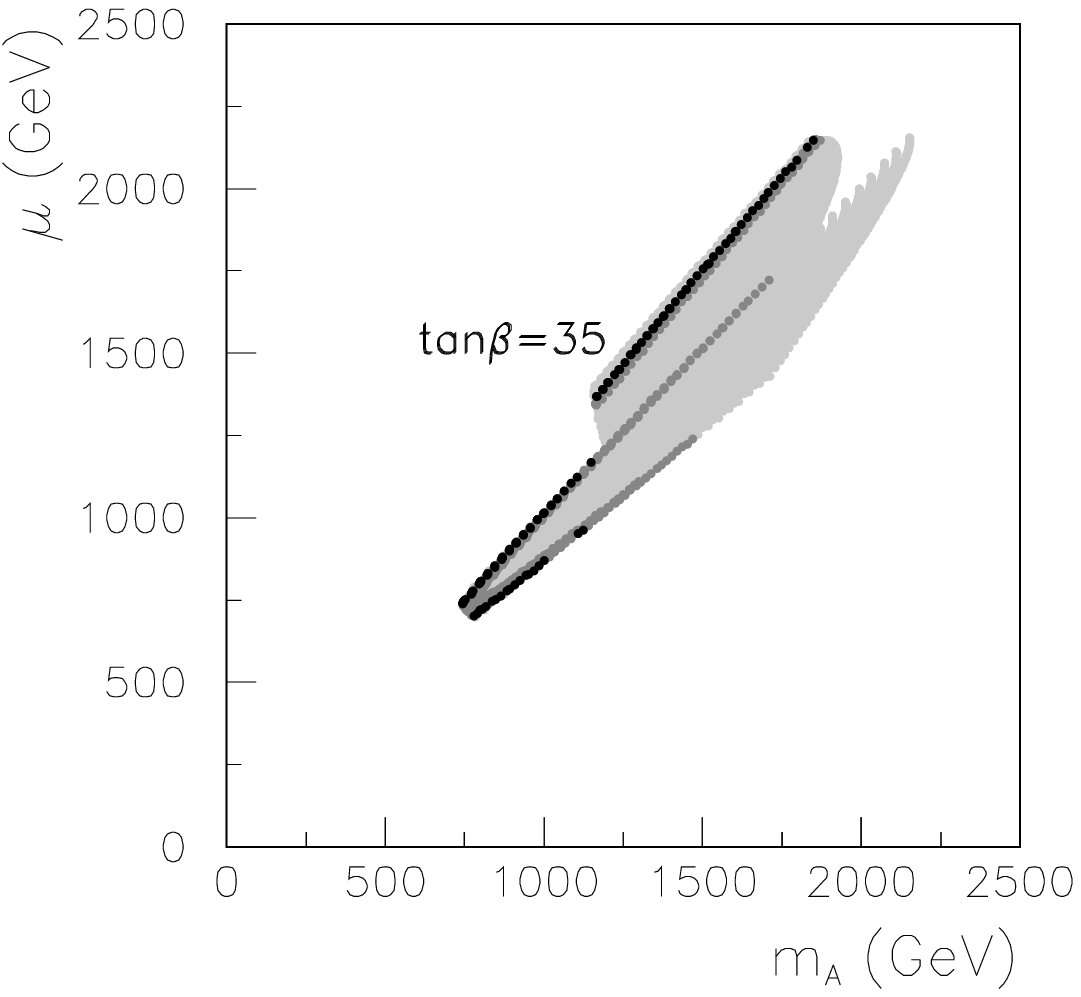,width=8cm}
  \vspace*{-1cm}  
  \captions{
    The same as Fig.\,\ref{102035mamua} but for the optimised
    example where $n_{L_L}=n_{e_R}=-1$ with $\tan\beta=20,\,35$.
    \label{102035mamuoptim} 
  }
\end{figure}

The corresponding predictions for the neutralino-nucleon cross section
are depicted in Fig.\,\ref{10crossoptim}. Although regions with the
correct relic density can appear with $\crosssec\approx10^{-8}$ pb for
$\tan\beta=20$, the points 
fulfilling the UFB-3 constraints only correspond
to those with
$\crosssec\lsim 10^{-9}$ pb for $\tan\beta\gsim 30$. 
Once more, as evidenced in Fig.\,\ref{102035mamuoptim}, where the
resulting $\mu$ parameter is represented versus the CP odd Higgs mass, 
this is due to
the large values of $\mu$ and the heavy
Higgs masses.

Notice, finally, that in this example the non-universality of the
Higgs masses, given by (\ref{higgs_b}),
was chosen to be the maximal allowed by the modular
weights ($n_{H_u}=-1$ and $n_{H_d}=-3$).  
Also, the
stau mass, for which we have $n_{L,e}=-1$, cannot be further
increased and therefore the behaviour under the UFB-3 constraint
cannot be improved.
Consequently, this
optimised scenario represents 
a good estimate of how large the neutralino
detection cross section can be in heterotic orbifolds with overall
modulus, where soft masses are given by (\ref{scalars}).
We therefore conclude that in this class of models
$\crosssec\lsim 10^{-8}$ pb. The neutralino in these scenarios
would escape detection in all present experiments and only $1$ tonne
detectors would be able to explore some small areas of the allowed
parameter space.

For completeness, we have also analysed other three scenarios,
described in \cite{ibalu}, which also give rise to gauge coupling
unification with an overall modulus. Their corresponding modular
weights are summarised in Table\,\ref{tablemodular}.

For example, in case C) unification is possible with Re\,$T=7$, but
extra massless chiral fields (one octet, one triplet, and two
multiplets transforming like right-handed electrons), with modular
weight equal to $-1$, are needed. 
This scenario seems promising, 
since the modular weights for sleptons are
less negative ($n_{L,e_R}=-1$). 
In fact, although squarks become tachyonic at
the GUT scale for $\cos^2\theta>1/2$, sleptons have a positive mass
squared in the whole remaining area
$\theta\in[\pi/4,\,3\pi/4],\, [5\pi/4,\,7\pi/4]$.
As we have learned from the optimised example, this might be helpful in
order to avoid the UFB constraints. 
The presence of extra matter
alters the running of the gauge coupling constants,
which are now dictated by the following beta functions, $b_1=-13$,
$b_2=-3$, and $b_3=0$.
As a consequence, the running of the soft masses is also
modified. In particular, all the gaugino masses
become smaller at the EW
scale, as compared to the usual running within the MSSM. Notice in
particular that the gluino mass does not run (at tree level) 
from the GUT to the EW scales.

\begin{figure}[!t]
  \epsfig{file=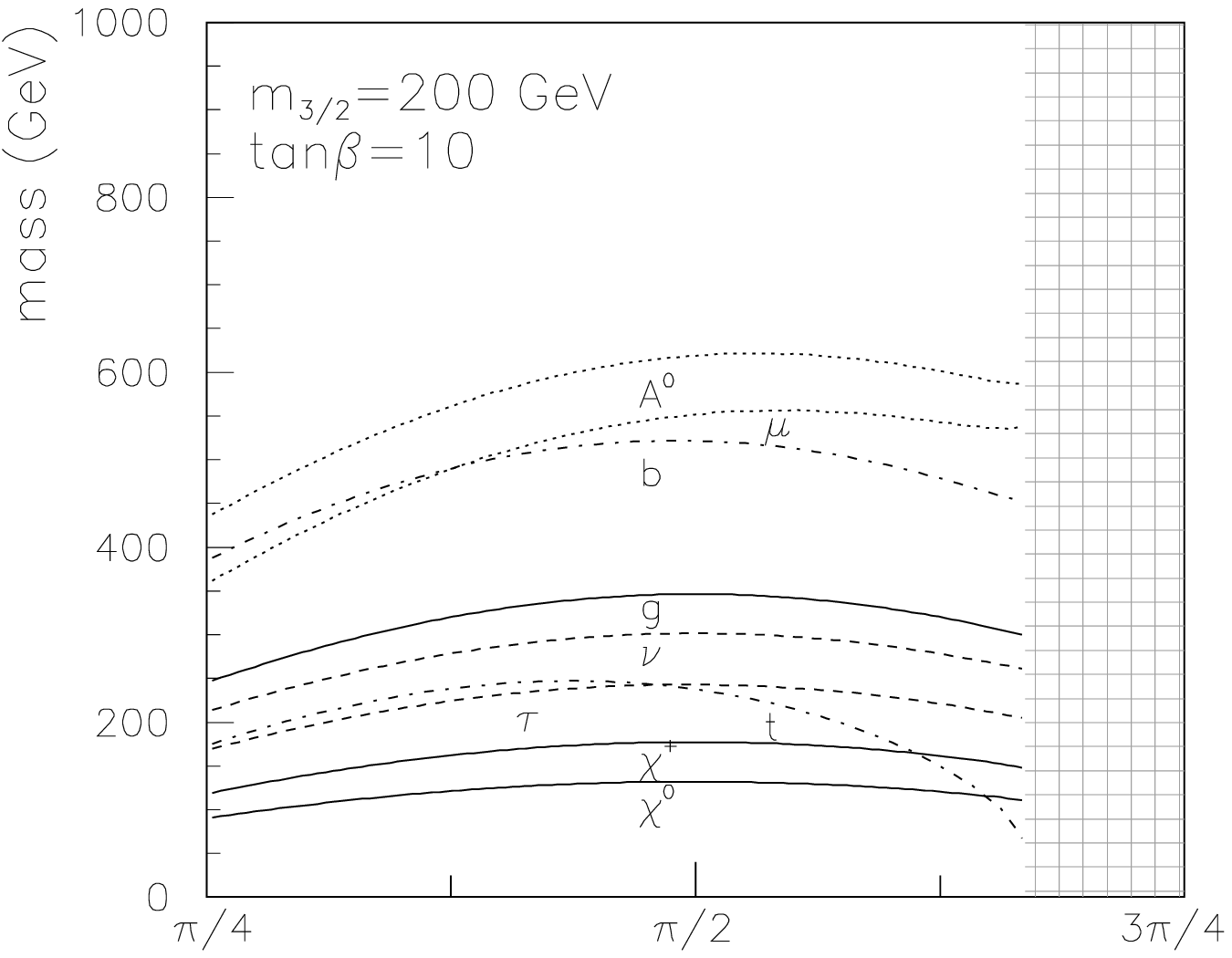,width=8.7cm}
  \hspace*{-0.9cm}  
  \epsfig{file=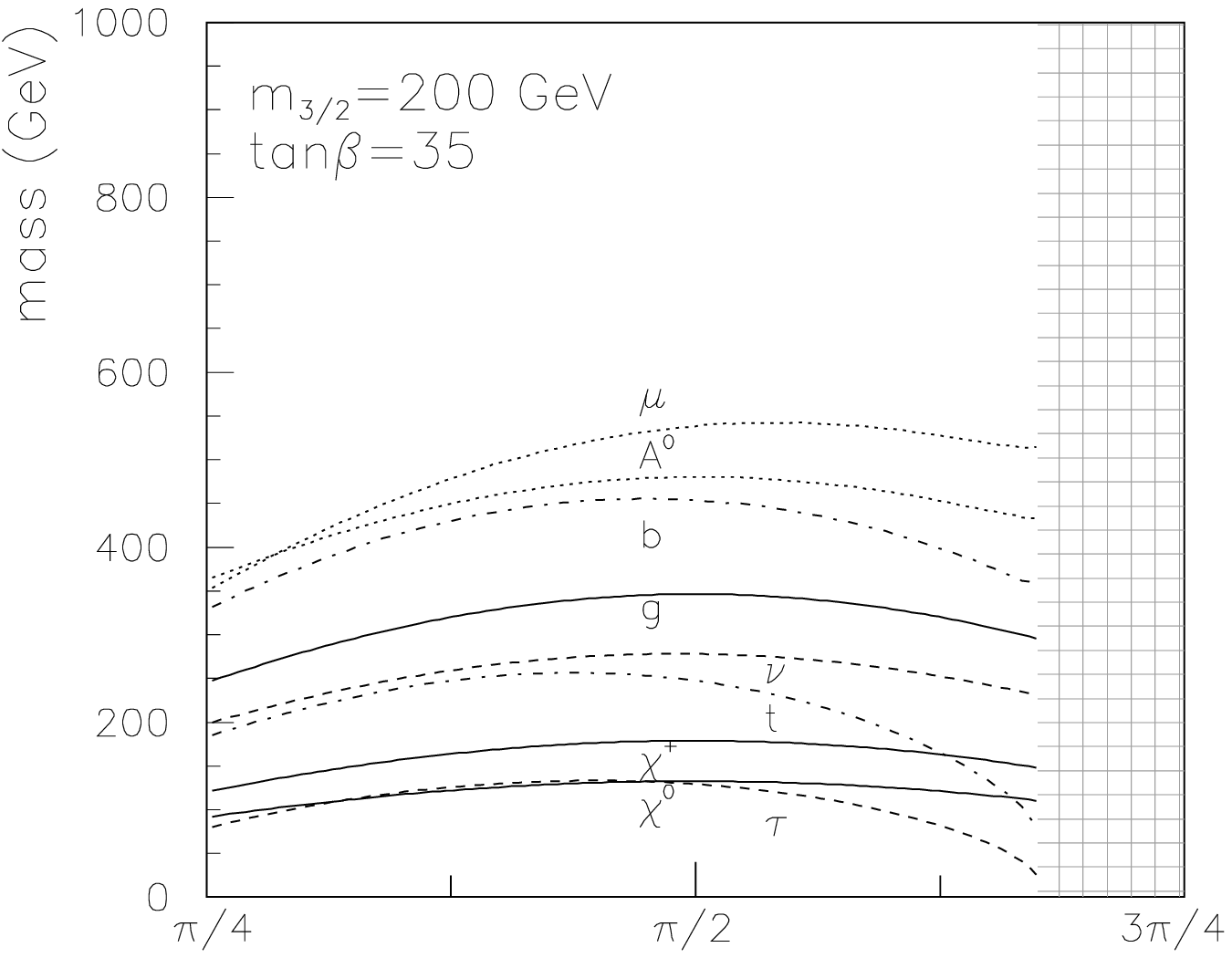,width=8.7cm}
  \captions{The same as Fig.\,\ref{10asp} but for case C) in
    Table\,\ref{tablemodular}.  
    \label{1035csp}}
\end{figure}

\begin{figure}[!t]
  \epsfig{file=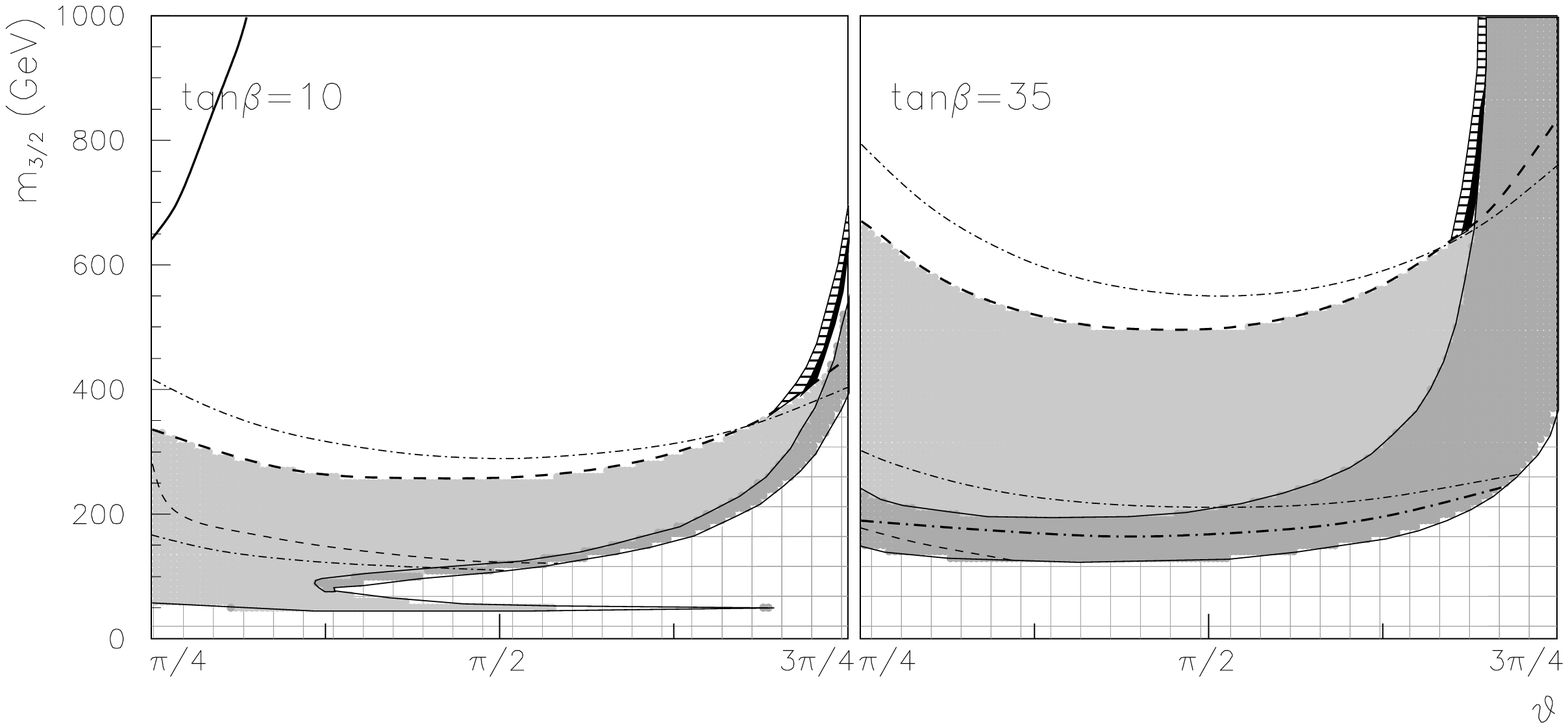,width=16cm}
  \vspace*{-1cm}
  \captions{The same as Fig.\,\ref{2035a} but for case C) in
    Table\,\ref{tablemodular} and for $\tan\beta=10$ and $35$. Notice
    that for $\tan\beta=10$ the lightest stop is the LSP on the dark
    grey area whereas for $\tan\beta=35$ it is the lightest stau.}
  \label{10c}
\end{figure}

\begin{figure}[!t]
  \begin{center}
  \epsfig{file=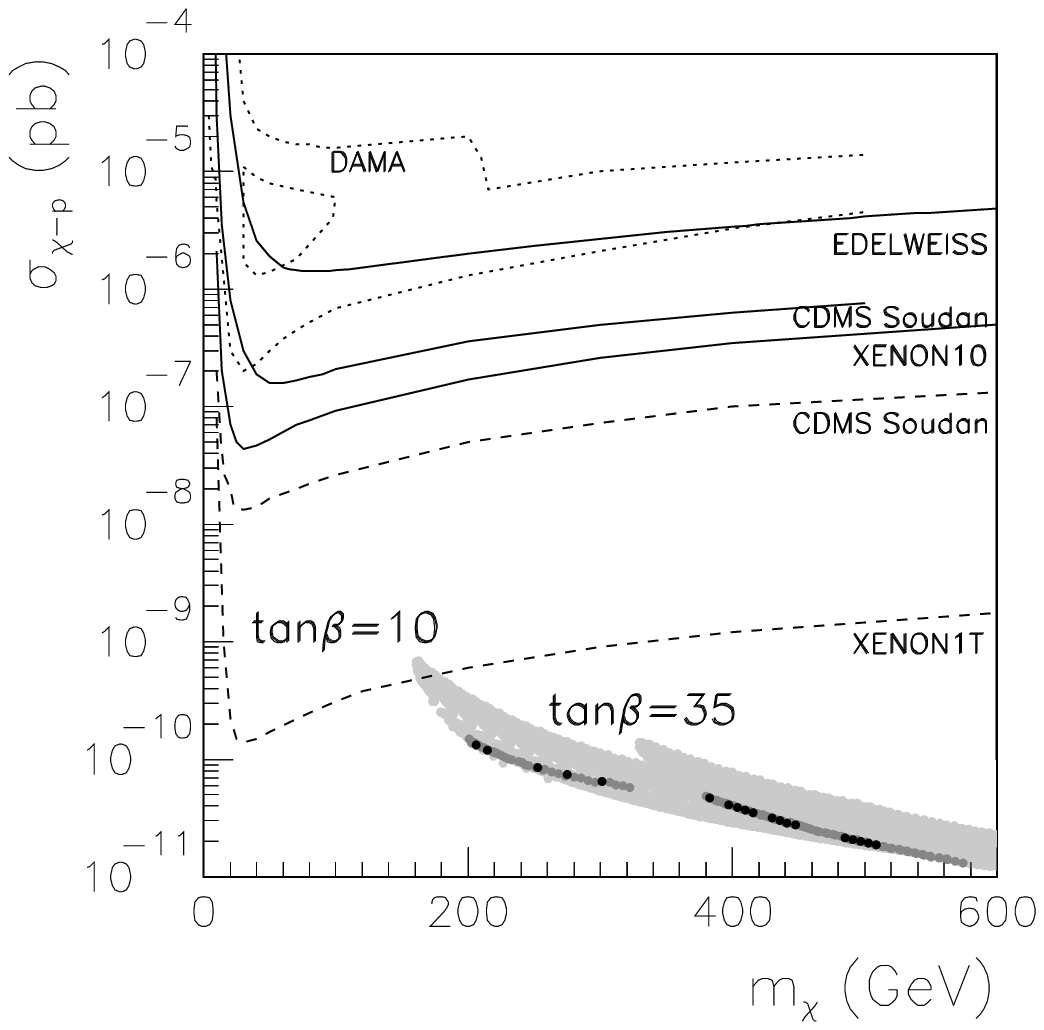,width=8cm}
  \end{center}
  \vspace*{-1cm}
  \captions{The same as Fig.\,\ref{102035across} but for case C) in
    Table\,\ref{tablemodular} and with 
    $\tan\beta=10\,,35$.}
  \label{1035ccross}
\end{figure}

The decrease in $M_1$ and (especially) $M_2$ affect the
running of the scalar mass parameters, rendering them smaller at the
EW scale. This is enough to offset the increase in $m_{L,E}^2$ due
to the smaller modular weights.
Similarly, the important decrease in the gluino mass implies a very
light squark sector. This leads to a qualitatively different structure
of the SUSY spectrum in which squarks and sleptons
have a similar mass.  
This is illustrated 
in Fig.\,\ref{1035csp}, where we have represented 
the resulting spectrum for
$m_{3/2}=200$ GeV and $\tan\beta=10$ and $35$. Unlike the previous
cases, this example displays very light gluinos and squarks. There are
even regions where the stop is the LSP
(especially for low values of
$\tan\beta$ for which the top Yukawa is larger).

The regions allowed by experimental constraints become larger in this
example, as we can see in Fig.\,\ref{10c}, where the
$(m_{3/2},\theta)$ plane is depicted for $\tan\beta=10$ and $35$. 
It is important to mention that, due to the resulting light squarks,
the supersymmetric contribution to B(\bsg) becomes sizable. Unlike in
the previous examples, the experimental bound on this observable 
becomes the most stringent constraint,
even for small values of $\tan\beta$. 
There are also areas which reproduce the correct dark matter
relic density through coannihilation effects with the stop (for small
values of $\tan\beta$) and the stau (for $\tan\beta\gsim 20$). 
Noticeably, in spite of the less negative modular weights for
sleptons, the modifications in the RGEs (especially the decrease in
$M_{1,2}$) imply smaller values for the Higgs mass parameters. In
particular, $\higgsu$ is more negative,  
making it more difficult to avoid the UFB-3 constraint. Only for
heavy gravitinos do allowed regions occur
($m_{3/2}\gsim650$ GeV is necessary 
for $\tan\beta=10$, whether for larger
$\tan\beta$ gravitinos heavier than 1 TeV are needed). As already
observed in case A), these regions
never correspond to those with the correct neutralino relic density.

Finally, it is worth mentioning that, since the Higgs mass parameters
have a smaller departure from universality, we do not expect large
neutralino detection cross sections in this example. The results for
$\crosssec$ are represented in Fig.\,\ref{1035ccross} for
$\tan\beta=10$ and $35$ and, clearly lie beyond the reach of current
and projected direct
dark matter searches.

\begin{figure}[!t]
  \epsfig{file=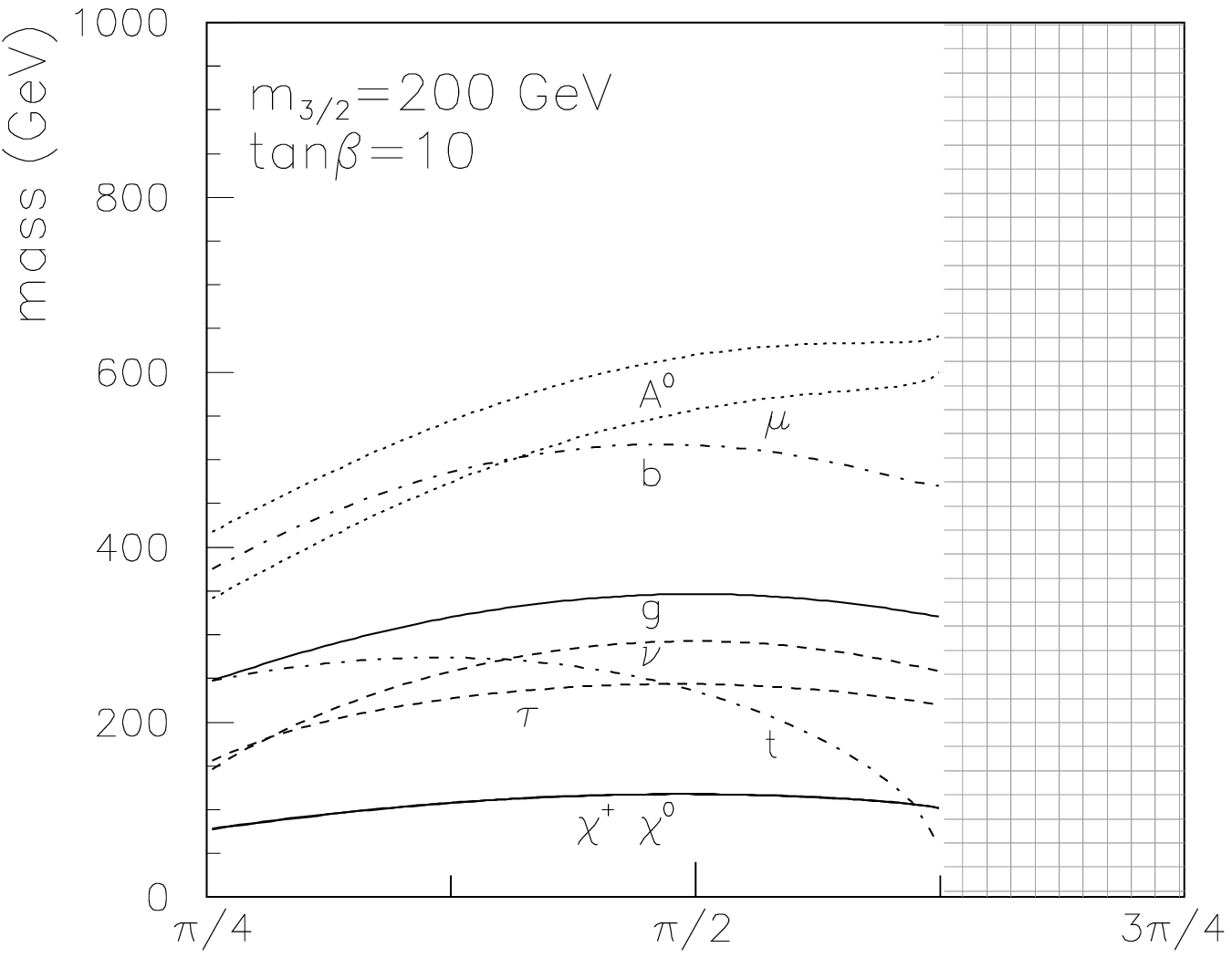,width=8.7cm}
  \hspace*{-0.9cm}  
  \epsfig{file=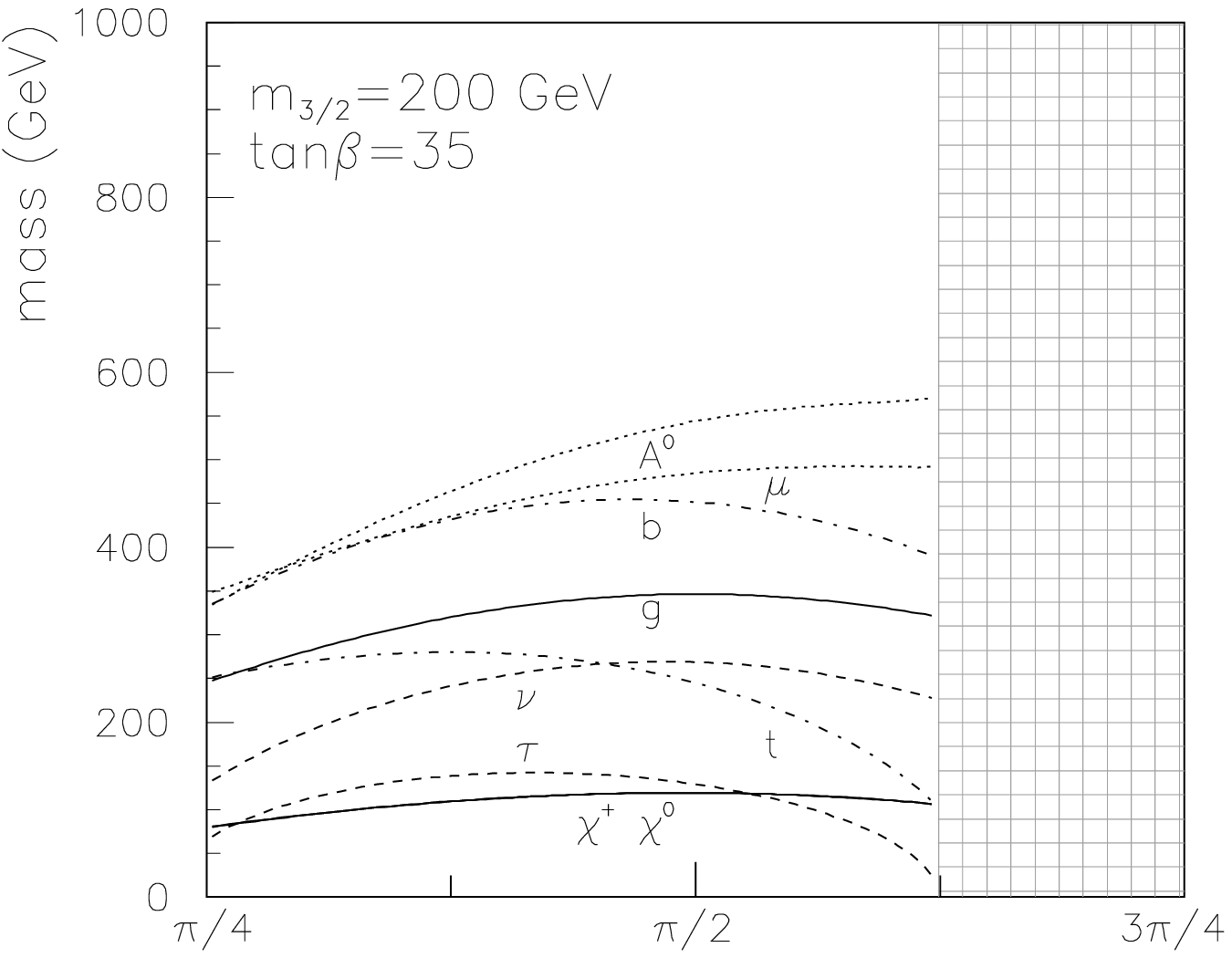,width=8.7cm}
  \captions{The same as Fig.\,\ref{10asp} but for case D) in
    Table\,\ref{tablemodular}.  
    \label{1035dsp}}
\end{figure}

\begin{figure}[!t]
  \epsfig{file=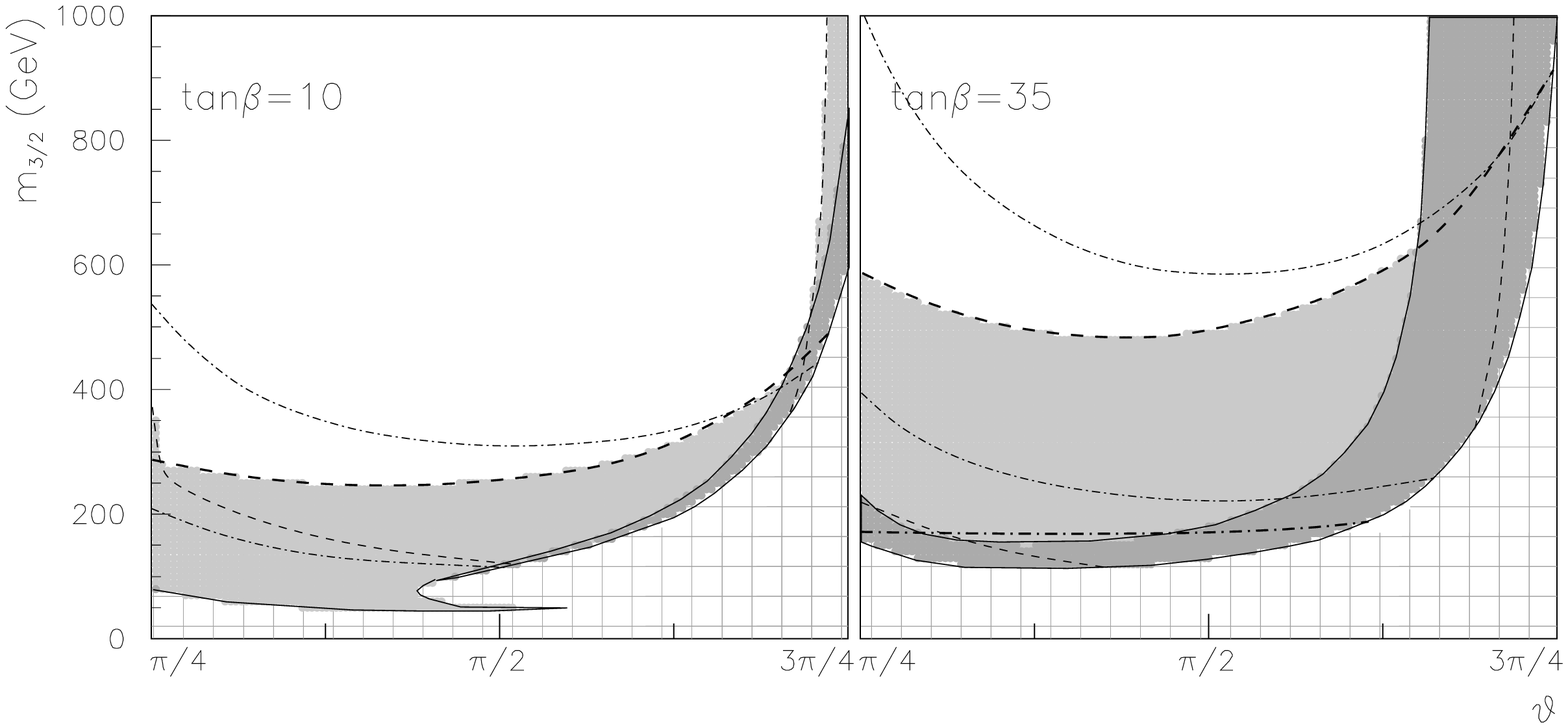,width=16cm}
  \vspace*{-1cm}
  \captions{The same as Fig.\,\ref{10a} but for case D) in
    Table\,\ref{tablemodular}. Notice
    that for $\tan\beta=10$ the lightest stop is the LSP on the dark
    grey area whereas for $\tan\beta=35$ it is the lightest stau.}
  \label{10d}
\end{figure}

Another potentially interesting scenario is case D), once more due to
the reduced modular weights for sleptons. As in the previous example,
the region allowed at the GUT scale is $\theta\in[\pi/4,\,3\pi/4],\,
[5\pi/4,\,7\pi/4]$, where $m^2_{Q_L,L_L}\ge0$. In this scenario one
needs four 
extra multiplets, transforming like $(Q,\bar Q, D,\bar D)$, and
Re\,$T=9$ for unification to take place, thus implying  $b_1=-12$,
$b_2=-4$, and $b_3=0$. The absence of running for $\alpha_3$ and
therefore for the gluino mass parameters has the same consequences as
in case C), leading to a light squark spectrum. This is illustrated in
Fig.\,\ref{1035dsp}, where the sparticle masses are plotted as a
function of the Goldstino angle for $m_{3/2}=200$ GeV and
$\tan\beta=10$ and $35$. Interestingly, the running of the wino mass
parameter is slightly enhanced and $M_1\sim M_2$ is found at the
electroweak scale. This makes the lightest neutralino a mixed
bino-wino state and almost
degenerate in mass with the lightest chargino. Although this can
lead to an increase of the resulting neutralino direct detection cross
section, it also 
implies a more efficient neutralino annihilation and, consequently, a
relic density which is too small to account for the dark matter of the
Universe. In particular, one obtains
$\relic h^2\lsim0.01$ for the whole region with gravitinos lighter than
$1$ TeV, independently of the value of $\tan\beta$. Thus, although the
area allowed by experimental constraint, represented in
Fig.\,\ref{10d} for $\tan\beta=10$ and $35$ is sizable, the
astrophysical constraint on the relic density is never
fulfilled. As in the previous examples, the presence of light squarks
induce larger contributions to B(\bsg) and the experimental constraint
on it excludes extensive regions of the parameter space,
even at low $\tan\beta$. Furthermore, the steeper running of $M_2$
renders $\higgsu$ more negative, which makes the UFB constraints even
more restrictive. All the area represented in Fig.\,\ref{10d} becomes
excluded for this reason.

Finally, case E) in Table\,\ref{tablemodular} corresponds to
a $Z_8'$ orbifold with a universal modulus, in which case Re\,$T\sim24$
is needed. Due to the small modular weights for sleptons this example
yields similar results regarding the UFB constraints as case
A), with allowed regions appearing only for very massive gravitinos
and incompatible with the astrophysical bound on the dark matter relic
density. Moreover, since the non-universality in the Higgs mass parameters
($\higgsu<\higgsd$) is not the optimal to increase the neutralino
detection cross section, the theoretical predictions for  $\crosssec$
are even smaller than those represented in Fig.\,\ref{102035across}.

The presence of different moduli can provide some extra freedom to the
non-universalities of the soft scalar masses (\ref{masorbi}), which
are then parametrized by new Goldstino angles, $\Theta_i$. 
We have analysed two scenarios of this kind, whose modular weights are
described in Table\,\ref{tablemodularmulti}, and which have also been
shown to reproduce gauge coupling unification \cite{ibalu}.
Scenario F) corresponds to a $Z_6$ orbifold with a rotated plane for
which unification is achieved with Re\,$T_1\sim 10\gg$ Re\,$T_2$. In
scenario G), a $Z_2\times Z_2$ orbifold was taken, again
non-isotropic, with $T_1\gg T_{2,3}$.

\begin{table}[!t]\begin{center}
    \begin{tabular}{|c|ccccccc|}
      \hline
      &$n_{Q_L}^i$& $n_{{u_R}}^i$& $n_{d_R}^i$& $n_{e_R}^i$&
      $n_{{L_L}}^i$&$n_{H_d}^i$& $n_{H_u}^i$\\ 
      \hline
      F)& $0, -1, 0$& $\frac{-2}3, 0, \frac{-1}3$& $0, -1, 0$&
      $\frac{-5}3, 
      0, \frac{-1}3$& $\frac{-2}3, 0, \frac{-1}3$& $0, 0, -1$&
      $0, 0, -1$\\ 
      \hline
      G)& $\frac{-1}2, \frac{-1}2, 0$& $-1, 0, 0$& $0, -1, 0$&
      $-1,0,0$& $-1, 0, 0$& $-1, 0, 0$& $0, -1, 0$\\ 
      \hline\end{tabular}\end{center}
  \captions{Modular weights for the scalar fields of two multimoduli
    heterotic orbifold scenarios that can reproduce gauge unification
    \cite{ibalu}.} 
  \label{tablemodularmulti}
\end{table}

In both examples, 
some combinations of the Goldstino angles lead to potentially
interesting non-universalities in the soft scalar masses. In
particular, it is always possible to enlarge the stau
mass, thus avoiding UFB constraints. 
For instance, in case F) this can be done by choosing $\Theta_3^2=1$
(hence $\Theta_1^2=\Theta_2^2=0$) 
obtaining the following expressions for the soft terms, 
\begin{eqnarray}
  m^2_{Q_L}=m^2_{d_R}&=&m^2_{3/2}\,,\nonumber\\
  m^2_{u_R}=m^2_{e_R}=m^2_{L_L}&=&m^2_{3/2}(1-\cos^2{\theta})
  \,,\nonumber\\    
  m^2_{H_d}=m^2_{H_u}&=&m^2_{3/2}(1-3\cos^2{\theta})\ .
\end{eqnarray}
Alternatively, in case G) one can take $\Theta_1^2=0$ (and therefore
$\Theta_2^2+\Theta_3^2=1$)  and obtain
\begin{eqnarray}
  m^2_{Q_L}&=&m^2_{3/2}\left(1-\frac32\,\Theta_2^2\,\cos^2{\theta}
  \right)  \,,\nonumber\\
  m^2_{u_R}=m^2_{e_R}=m^2_{L_L}=m^2_{H_d}&=&m^2_{3/2}
  \,,\nonumber\\    
  m^2_{d_R}=m^2_{H_u}&=&m^2_{3/2}(1-3\,\Theta_2^2\,\cos^2{\theta})\ .
  \label{massesg}
\end{eqnarray}
Notice, however, that the resulting Higgs mass parameters are not
adequate to obtain large neutralino detection cross sections. 
On the one hand, 
in case F) these are
universal by construction since they have the same modular weights. 
On the other hand, in case G), the Higgs soft masses are related by
$m_{H_u}^2/m_{H_d}^2=
(1-3\,\Theta_2^2\,\cos^2{\theta})/
(1-3\,\Theta_1^2\,\cos^2{\theta})$.
If heavy
staus are required, as in (\ref{massesg}), 
$m_{H_d}^2\ge m_{H_u}^2$ is obtained, which implies small
$\crosssec$. 
In order to reproduce 
the optimal Higgs non-universality, $m_{H_u}^2\ge
m_{H_d}^2$, one needs to take $\Theta^2>\Theta_1^2$ but then
light staus appear (since $m^2_{e_R}=m^2_{L_L}=m^2_{H_d}$) 
which therefore lead to strong UFB constraints.

As a consequence, none of these two examples can provide larger
theoretical predictions for $\crosssec$ than those obtained in the
optimised case of Fig.\,\ref{10crossoptim}.

\subsection{Non-thermal production of neutralinos from late gravitino
  decays}

Due to their extremely weak (gravitational) interactions, 
gravitinos can have a very long lifetime (typically longer than
$10^2$ sec for $m_{3/2}\lsim 1$TeV).
Consequently, gravitinos which are produced thermally in the reheating
phase after inflation and which decouple with a given relic density
$\Omega_{3/2}h^2$ will decay at late times into the LSP (or cascading
down to the LSP). This constitutes 
a source of non-thermal production of neutralinos, 
$\Omega^{NTP}_{\neut}$,
which contributes to the total relic density of dark matter, and 
is related to the gravitino relic density as
\begin{equation}
  \Omega^{NTP}_{\neut}h^2=\frac{\neumass}{m_{3/2}}\ \Omega_{3/2}h^2\ .
  \label{neutralino_ntp}
\end{equation}

It is easy, however to argue that $\Omega^{NTP}_{\neut}$ is typically
very small in the orbifold models we have just presented. It can be
shown that in the case of gravitinos that are much lighter than
gluinos ($m_{3/2}\ll m_{\tilde g}$), which generally holds in our
case, 
the gravitino abundance from
thermal production can be estimated as \cite{buchmuller}
\begin{equation}
  \Omega_{3/2}h^2 \approx 0.21\left(\frac{T_R}{10^{10}{\rm
      GeV}}\right) 
  \left(\frac{100\,{\rm GeV}}{m_{3/2}}\right)
  \left(\frac{m_{\tilde g}}{1\,{\rm TeV}}\right)^2\ .
  \label{gravitino_tp}
\end{equation}

The reheating temperature, $T_R$, is here a free parameter. The larger
the reheating temperature, the more sizable $\Omega_{3/2}h^2$
becomes. However, in order not to
reintroduce the infamous gravitino problem and spoil the results from
Big Bang nucleosynthesis, the reheating temperature   is constrained
to be \cite{kawasaki} $T_R\lesssim 10^6\, {\rm GeV}$ for a gravitino
with a mass of order 
$m_{3/2}\approx100$ GeV. This implies an upper limit on
the gravitino relic density, and thus on the non-thermal production of
neutralinos. From (\ref{neutralino_ntp}) and (\ref{gravitino_tp}), and
using typical values for the gluino and neutralino mass, we obtain  
$\Omega^{NTP}_{\neut}\lsim2\times 10^{-4}$.
Thus, even if the reheating temperature was increased to $T_R=10^8\,
{\rm GeV}$ (for heavier gravitinos, and provided their hadronic
branching ratio is small), the contribution to the neutralino relic
density is negligible.

\section{Modifications due to the presence of an anomalous $U(1)$}
\label{anomalous}

As discussed in the Introduction, an anomalous $U(1)$
\cite{FayetIliopoulos} 
is usually present after compactification\footnote{
  In \cite{Tatsuo4}, some conditions for the absence of the anomalous
  $U(1)$ are discussed, and classifications of models with anomalous 
  $U(1)$ are also attempted.}.
For example, it was found in \cite{Giedt}
that only 192 different three-generation models containing the
$SU(3)\times SU(2)\times U(1)^n$ gauge group can be constructed within the
$Z_3$ orbifold with two Wilson lines.
The matter content of 175 of them was analysed in detail and only
7 of them turn out not to have an anomalous $U(1)$ associated.
Let us recall that the presence of the anomalous $U(1)$ is crucial
for model building \cite{Katehou}.
It generates a FI contribution to the D-term \cite{FayetIliopoulos},
breaking extra $U(1)$ symmetries, and allowing the construction
of realistic standard-like models 
\cite{Casas1,Casas2,Font,Lebedev,Lebedev2,Kim2,hum}.
Let us now discuss the possible contributions to soft scalar masses
generated because of the FI breaking, and their effect on the
neutralino direct detection cross section.

The FI breaking induces additional terms to 
soft SUSY-breaking scalar masses\footnote{
  Let us remark that
  there are no additional contributions to gaugino masses and 
  A-terms when Higgs fields relevant to such symmetry breaking have 
  less F-term than those of dilaton and moduli fields.} 
due to F-terms, namely, the so-called D-term contribution
\cite{Nakano,Kawamura,Tatsuo1,Tatsuo2,Kawamura2,Dudas}. In particular,
the presence of an anomalous $U(1)$ after compactification generates
the dilaton-dependent FI term. That is, the D-term of the anomalous
$U(1)$ is given by  
\begin{equation}
  D^A=\frac{\delta^A_{GS}}{S+ S^*}+
  \sum_\alpha 
  (T + T^*)^{n_\alpha}
  q^A_\alpha |\phi_\alpha|^2\ ,
\end{equation}
where the first term corresponds to the dilaton-dependent 
Fayet-Iliopoulos term with the GS coefficient $\delta_{GS}^A$
proportional to the value of the anomaly, and 
the second one is the usual D-term with the $U(1)$ charges
$q^A_\alpha$ of the fields $\phi_\alpha$. Then, some of these fields
(with vanishing hypercharges), let us call them $C_\beta$, develop
large vacuum expectation values (VEVs) along the D-flat direction in
order to cancel the FI term, inducing the D-term contribution to the
soft scalar masses of the observable fields. The resulting scalar mass
squared is given by 
\cite{Tatsuo1}\footnote{
  Assuming for simplicity that the fields $C_\beta$ have no other
  $U(1)$ charges.}
\begin{equation}
  m^2_\alpha = m^2_{3/2}\left\{ 1 + n_\alpha \cos ^2 \theta + 
  q_\alpha^A 
  \frac{\sum_\beta (T + T^*)^{n_\beta} q_\beta^A |C_\beta|^2
    \left[ (6- n_\beta) \cos^2 \theta  - 5 \right]} 
       {\sum_\beta (T + T^*)^{n_\beta} (q^A_\beta)^2 |C_\beta|^2}
       \right\},
       \label{totally}
\end{equation}
where the first two terms are the usual contributions 
(see (\ref{scalars})), and 
the third term is the D-term contribution.
Obviously, if the observable fields have vanishing $U(1)$ charges,
$q_\alpha^A$, this contribution is also vanishing, and we recover
the situation of the previous Section.
However, the observable fields have usually non-vanishing charges 
in explicit models 
\cite{Casas1,Casas2,Font,Giedt2,Raby,Lebedev,Lebedev2,Kim2,hum},
and the effect of this contribution must be taken into account
in the analysis.

As we can see in the above formula, the D-term contribution
generates in general an additional non-universality among soft scalar
masses which depends on $q_\alpha^A$. Let us simplify the analysis
considering the case where only a single field $C$ develops a VEV in
order to cancel the FI term. Thus  the above result reduces to the
following form: 
\begin{equation}
  m^2_\alpha = m^2_{3/2}\left\{ 1 + n_\alpha \cos ^2 \theta + 
  \frac{q_\alpha^A}{q_C^A}\left[ (6- n_C) \cos^2 \theta  - 5 \right]
  \right\}\ ,
  \label{oneX}
\end{equation}
where $q_C^A$ and $n_C$ are the $U(1)$ charge and modular weight of 
the field $C$, respectively.
It is worth emphasizing here that even in the
dilaton-dominated case ($\cos\theta =0$) the 
soft scalar masses are non-universal, 
\begin{equation}
  m^2_\alpha = m^2_{3/2}\left( 1 - 
  5 \frac{q_\alpha^A}{q_C^A}
  \right)\ .
  \label{dilatond}
\end{equation}
This result should be compared with the one in (\ref{scalars}).
The contributions $-5 \,{q_\alpha^A}/{q_C^A}$
correspond to the $\delta$'s in (\ref{Higgsespara}) and
(\ref{Higgsespara2}). It is noteworthy that, 
contrary to the cases without an anomalous U(1) analysed in the
previous Section, positive values for the non-universalities are now
possible, by choosing ${q_\alpha^A}/{q_C^A}<0$.
This is welcome, as we will see, in order to enhance the stau masses
and thus avoid the UFB constraints, or increase the value of
$\higgsu$. Moreover, we may expect that, for appropriate values of 
the $U(1)$ charges, the above additional terms lead to interesting
values for the neutralino detection cross section.

\begin{table}[!t]\begin{center}
    \begin{tabular}{|c|ccc|}
      \hline
      &${q_{L_L,e_r}^A}/{q_C^A}$ &${q_{H_d}^A}/{q_C^A}$
      &${q_{H_u}^A}/{q_C^A}$\\  
      \hline
      A-I)&-2 & 0 & 0\\
      \hline
      A-II)& -2 & 0 & -2\\
      \hline
      A-III)& -2 & $\frac12$ & -2\\
      \hline\end{tabular}\end{center}
  \captions{Variations of scenario A) in
    Table\,\ref{tablemodular}. The fields have the same modular
    weights, but we have assigned the above $U(1)$ charges to the
    slepton and Higgs fields.} 
  \label{tablemodularu1}
\end{table}

Let us consider again scenario A) of the previous section, defined by
expressions (\ref{weights}) and (\ref{higgs_b}), where the sleptons
masses were too small and therefore problematic. Using (\ref{oneX}),
these are now given by
\begin{equation}
  m^2_{L_L, e_R} = m^2_{3/2}\left\{ 1 -3 \cos ^2 \theta +  
  \frac{q_{L_L, e_R}^A}{q_C^A}\left[ (6- n_C) \cos^2 \theta  - 5
    \right] 
  \right\}\ .
  \label{sleptons}
\end{equation}
For instance, assuming that $C$ is a twisted field with modular weight
$n_C=-2$, and the ratios\footnote{
  These values for modular weights, and ratios between
  $U(1)$ charges of observable and FI fields, are typically obtained
  in explicit models \cite{Casas1,Casas2,Font}.}
${q_{L_L,e_R}^A}/{q_C^A}=-2$, one has  
\begin{equation}
  m^2_{L_L, e_R} = m^2_{3/2}\left(11 -19 \cos ^2 \theta\right)
  \ .
  \label{sleptons2}
\end{equation}
This corresponds to case A-I) in Table\,\ref{tablemodularu1}.
The degree of non-universality, as defined in (\ref{Higgsespara2}), is
therefore $\delta_{L_L,e_R}=10-19\cos^2\,\theta$.
Unlike the examples in the previous section, where all the $\delta$'s
had to be negative, positive values can now be obtained.

\begin{figure}[!t]
  \epsfig{file=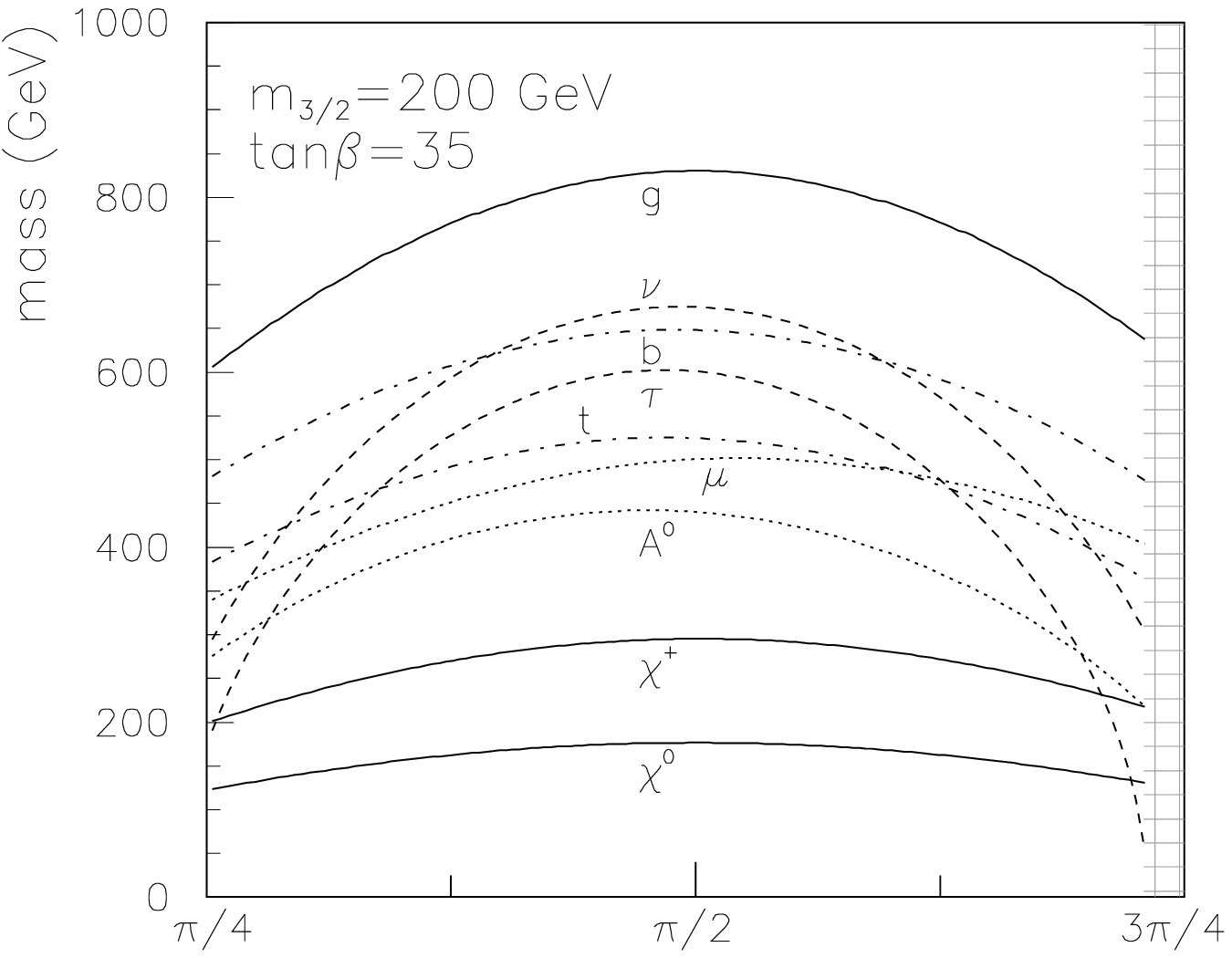,width=8.7cm}
  \hspace*{-0.9cm}  
  \epsfig{file=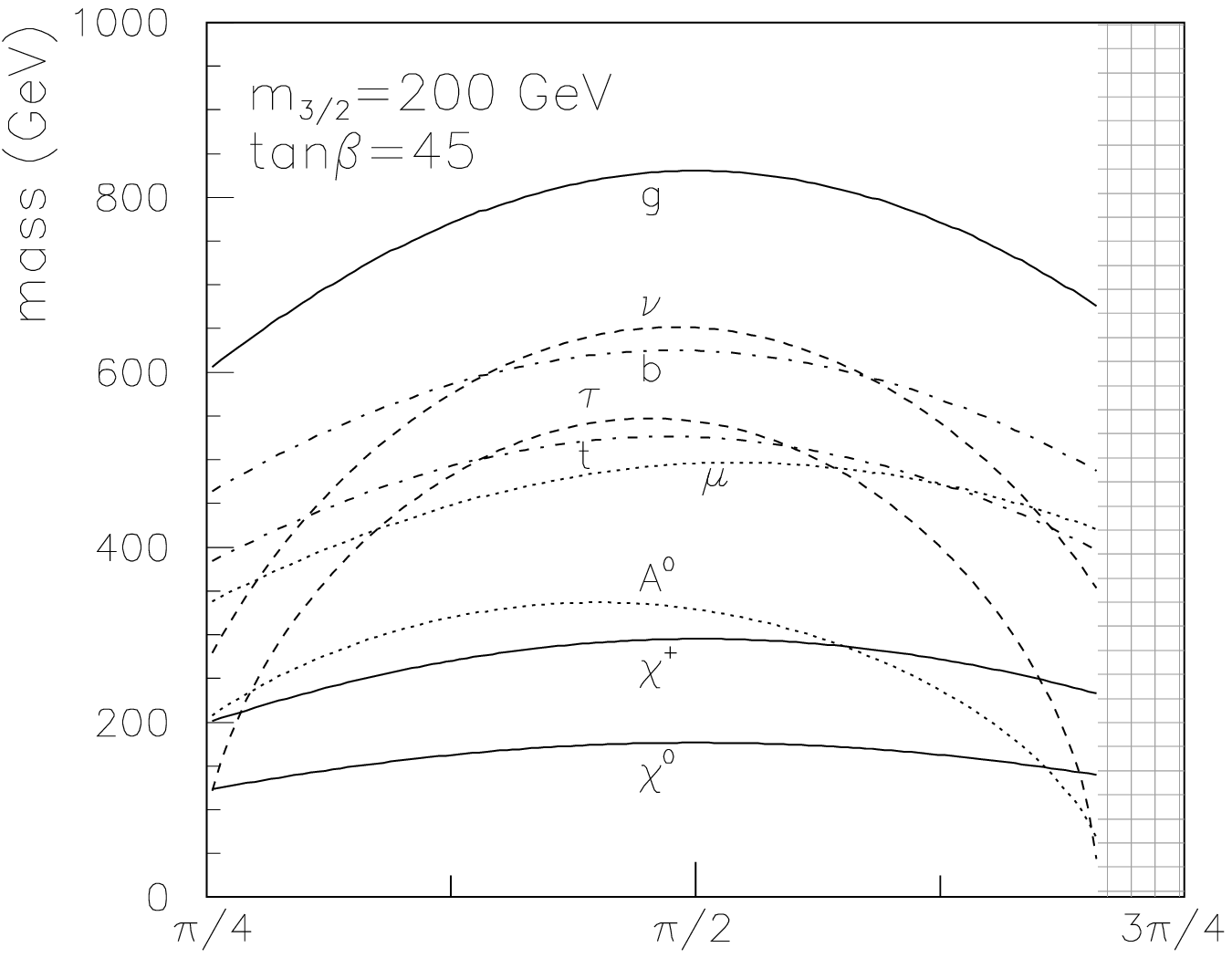,width=8.7cm}
  \captions{The same as Fig.\,\ref{10asp} but for
    example A-I) of
    Table.\,\ref{tablemodularu1} with $\tan\beta=35$ and $45$.
    Soft masses satisfy (\ref{weights}) and (\ref{higgs_b})
    with slepton masses modified according to Eq.(\ref{sleptons2}).
    \label{3545dterm1}}
\end{figure}

Condition
$\cos^2\theta\leq 1/3$ 
is no longer necessary in order to avoid tachyonic sleptons, since
$\cos^2\theta\leq 11/19$ 
is sufficient, thus implying broader allowed regions. 
Noticeably, unlike the examples in the previous Section, gaugino
masses (\ref{gauginoss}) 
can be of the same order, and even smaller than slepton
masses at the GUT scale. 
For example, with $\cos^2\theta=1/2$ one obtains 
$m^2_{L_L, e_R} = \frac{3}{2}\, m^2_{3/2}$, with gaugino masses of the
same order $M_a^2\simeq m^2_{L_L, e_R}$. 
And in the dilaton dominated limit
one obtains 
$m^2_{L_L, e_R} = 11\, m^2_{3/2}$, much larger than gaugino 
masses, for
which $M_a^2\simeq 3\,m^2_{3/2}$.

Very heavy sleptons are therefore possible and, although their masses
still decrease away from the dilaton limit, in most of the parameter
space their mass is similar to that of squarks. Such a heavy scalar
sector is clearly shown in Fig.\,\ref{3545dterm1} for $m_{3/2}=200$
GeV and $\tan\beta=35$ and $45$. As we can see, the lightest
neutralino and chargino are much lighter and the lightest stau rarely
becomes the LSP. Only along extremely narrow areas, which appear for
$\tan\beta\gsim25$ at the edge of the allowed zones, 
is the stau lighter
than the neutralino. Consequently, 
the allowed parameter space where the lightest neutralino is the LSP 
is more extensive 
and larger values of $\tan\beta$ are
allowed.
This leads to a more efficient decrease of both the pseudoscalar and
heavy scalar Higgs masses.

\begin{figure}[!t]
  \epsfig{file=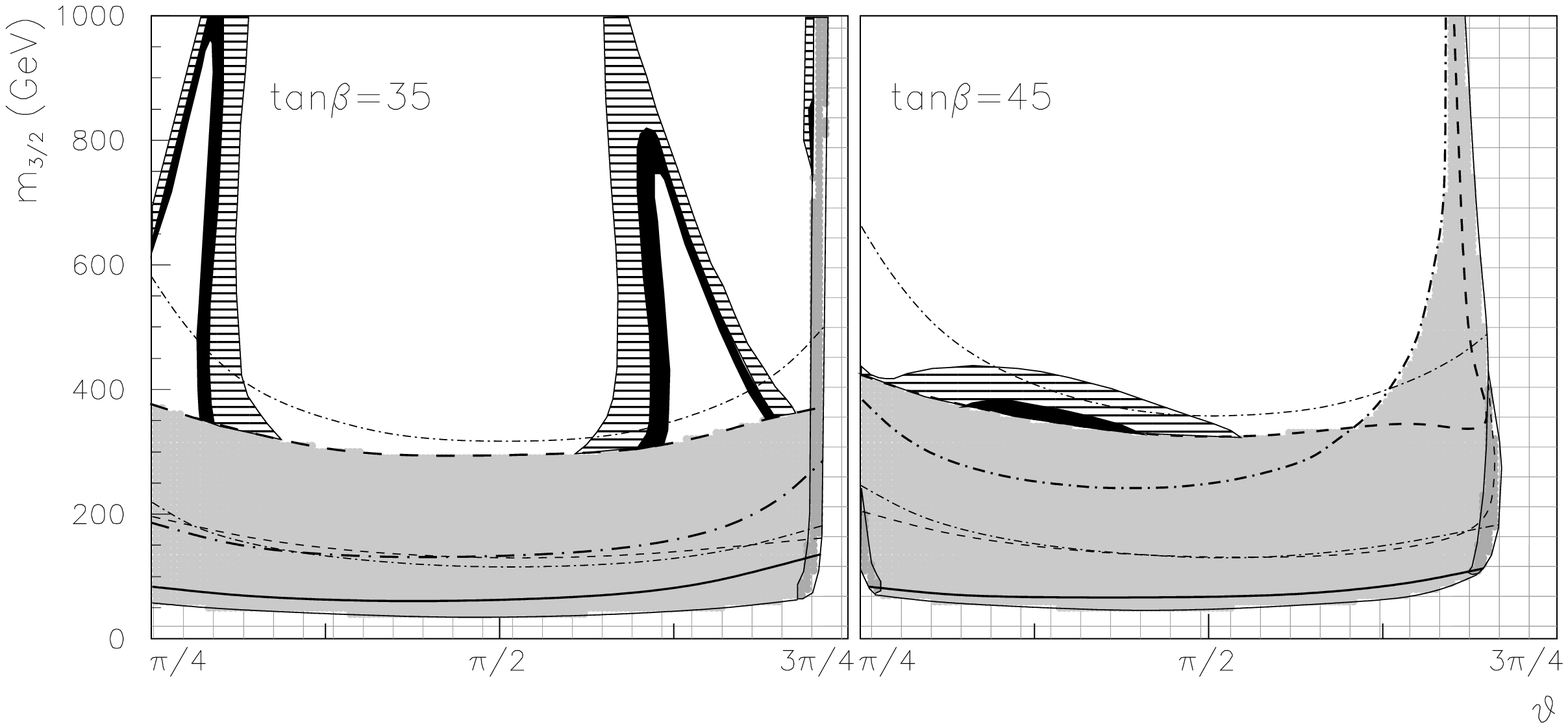,width=16cm}
  \vspace*{-1cm} 
  \captions{The same as Fig.\,\ref{10a} but for example A-I) of
    Table.\,\ref{tablemodularu1} with $\tan\beta=35\,,45$.    
    Soft masses satisfy (\ref{weights}) and (\ref{higgs_b}) 
    with slepton masses modified according to Eq.(\ref{sleptons2}).
  }
  \label{35adterm}
\end{figure}

The resulting $(m_{3/2},\theta)$ parameter space is represented in
Fig.\,\ref{35adterm} for $\tan\beta=35\, ,45$. 
The most salient consequence of the increase in the slepton masses 
is the reduction of the regions excluded by
the UFB-3 constraint (this is particularly true in the dilaton limit, as
pointed out by \cite{alejandro}).
Contrary to what happens in case A), for which the whole parameter
space was excluded (see
Fig.\,\ref{2035a}), now  only some points
with $m_{3/2}\lsim100$ GeV are disfavoured.
On the other hand, 
as a consequence of the increase in $\tan\beta$ and the associated
decrease of the pseudoscalar Higgs mass, the experimental bounds 
on $\bmumu$ and \bsg\ become much more constraining. These imply a
stronger lower limit on the gravitino mass as we can see in the plot.

\begin{figure}[!t]
  \epsfig{file=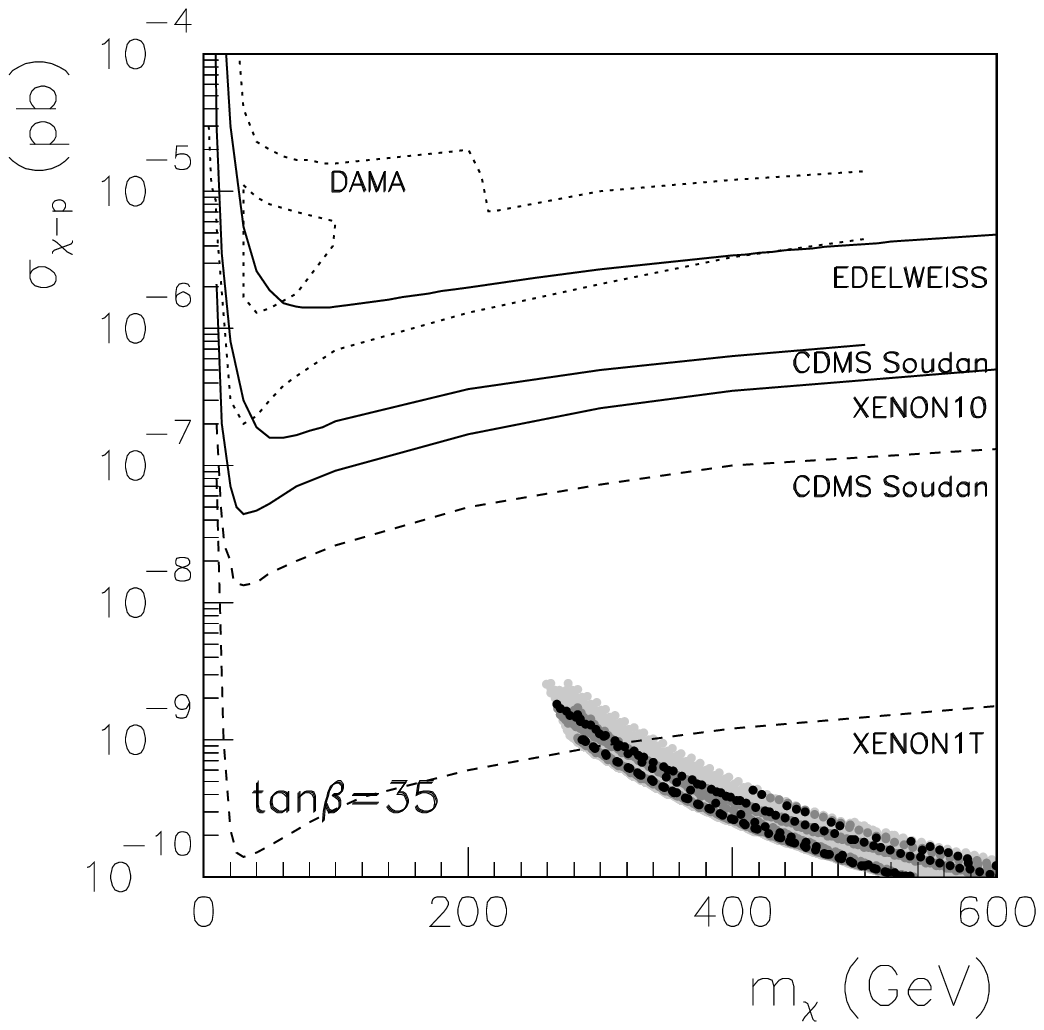,width=8cm}
  \epsfig{file=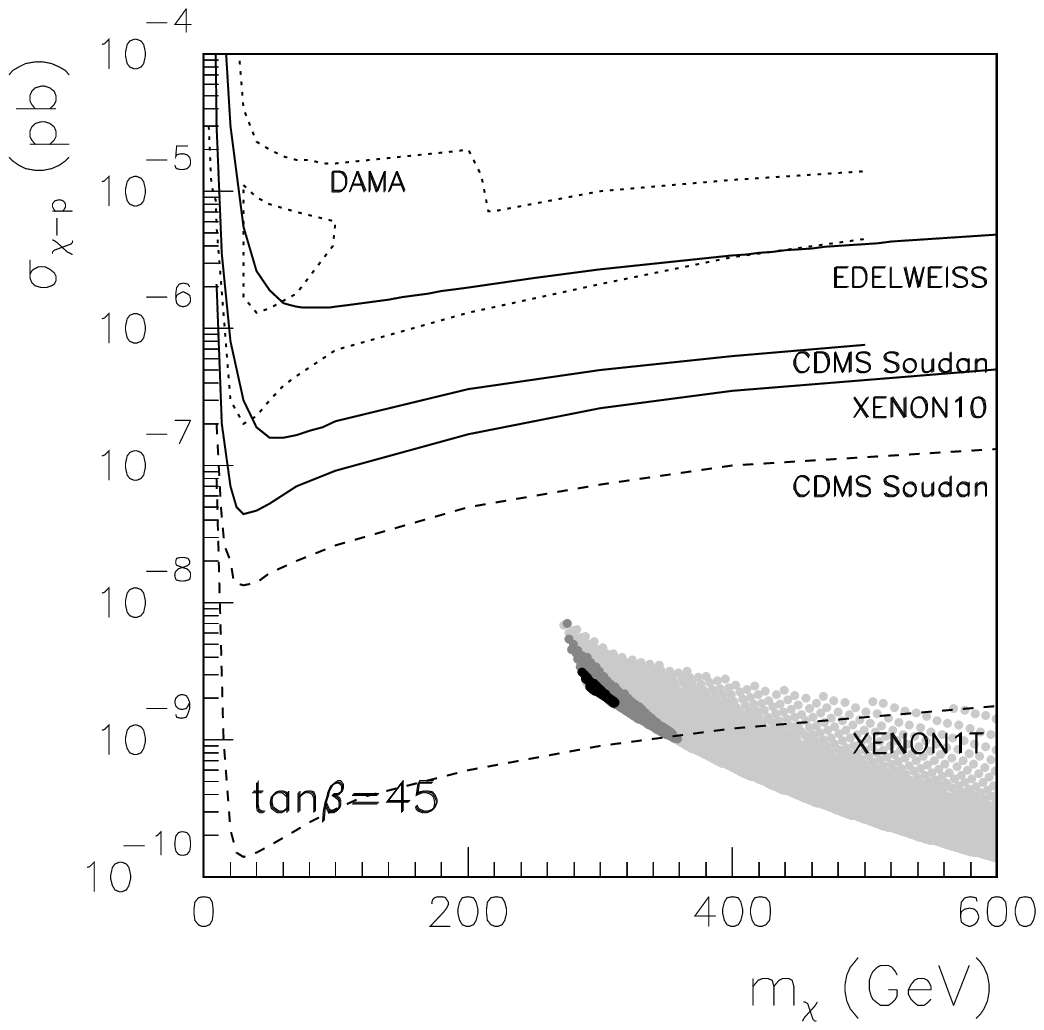,width=8cm}
  \vspace*{-1cm}
  \captions{The same as Fig.\,\ref{102035across} but for example A-I) of
    Table.\,\ref{tablemodularu1} with $\tan\beta=35\,,45$.    
    Soft masses satisfy (\ref{weights}) and (\ref{higgs_b})
    with slepton masses modified according to Eq.(\ref{sleptons2}).
  }
  \label{3545adtermcross}
\end{figure}

Regarding the neutralino relic density, for $\tan\beta\lsim 25$, the
stau coannihilation regions are not present and $\relic
h^2\gsim0.3$ is obtained in the whole $(m_{3/2},\,\theta)$
plane. The coannihilation strip only appears  
for $\tan\beta\gsim 25$, where the correct relic density can be
reproduced in extremely narrow areas. 
However, the most interesting case is $\tan\beta\gsim30$,
where thanks to the decrease in the pseudoscalar Higgs mass, 
we can find regions in the
parameter space where $2\neumass\approx m_A$, in which case the
resonant annihilation of neutralinos helps reproducing the WMAP
results. Due to this effect, the regions of the parameter space
satisfying experimental and astrophysical constraints are
now significantly larger. For instance, with $\tan\beta=35$ we find
allowed points for the whole range of gravitino masses between 350 GeV
to 1000 GeV.
Notice also that in the dilaton limit, $\theta=\pi/2$, 
the region reproducing the WMAP relic density is 
experimentally excluded for $\tan\beta\lsim47$.

\begin{figure}[!t]
  \epsfig{file=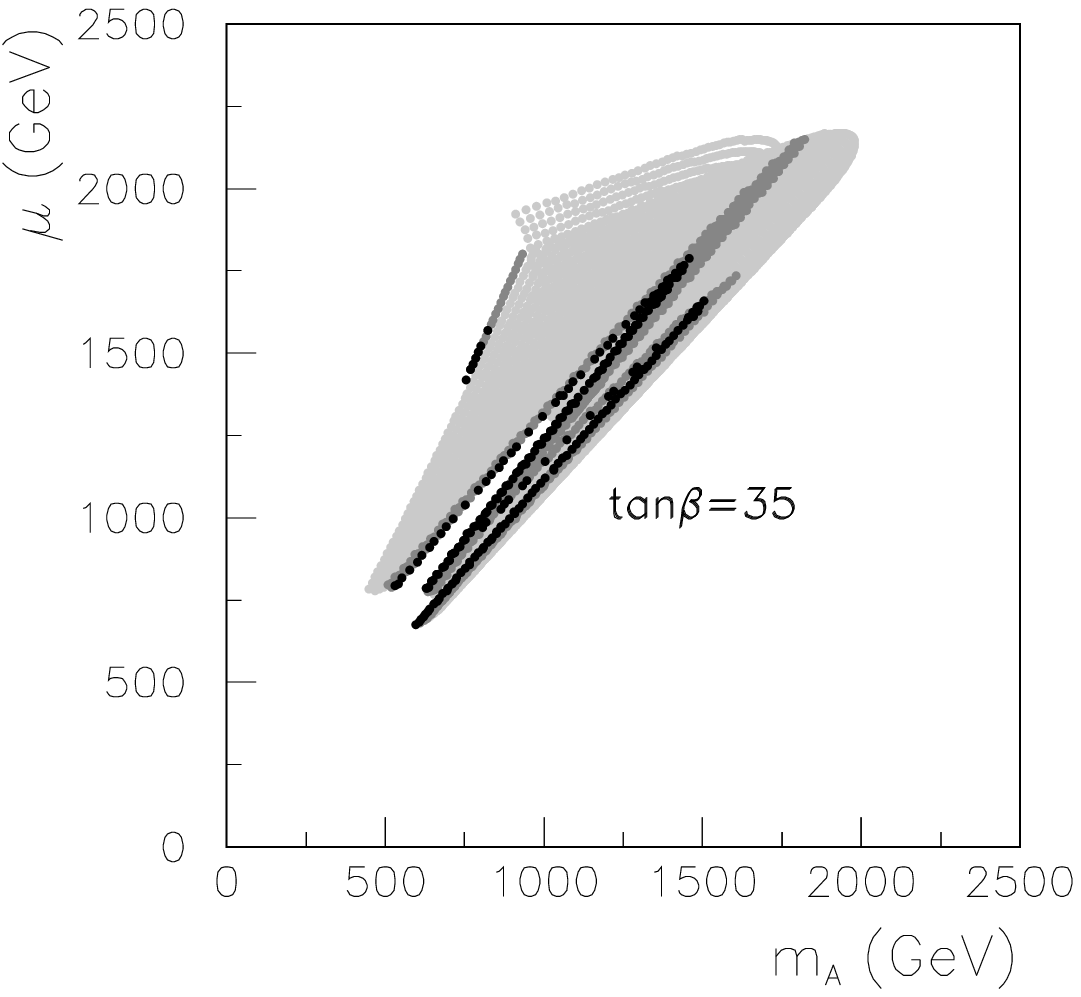,width=8cm}
  \epsfig{file=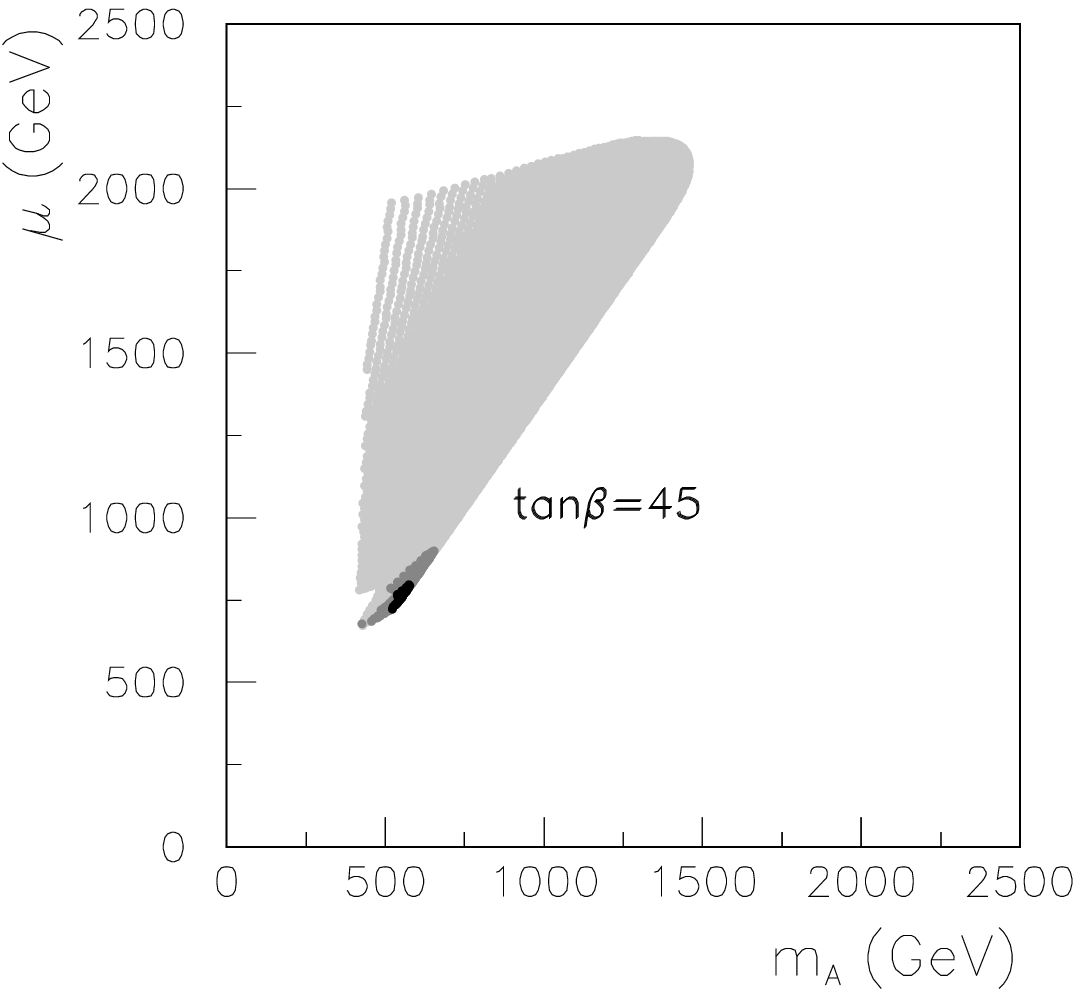,width=8cm}
  \vspace*{-1cm}
  \captions{
    The same as in Fig.\,\ref{102035mamua}  but for example A-I) of
    Table.\,\ref{tablemodularu1} with
    $\tan\beta=35\,,45$.
    Soft masses satisfy (\ref{weights}) and (\ref{higgs_b})
    with slepton masses modified according to Eq.(\ref{sleptons2}).
    \label{3545mamuadterm} 
  }
\end{figure}

The decrease on the Higgs masses for $\tan\beta\gsim 35$ 
leads to an enhancement of the
neutralino-nucleon cross section.
However, for such values of $\tan\beta$ large
branching ratios for the rare processes 
\bsg\ and $\bmumu$ are found, which therefore 
put stringent upper bounds on $\crosssec$.
This can be seen in  Fig.\,\ref{3545adtermcross}, where
the theoretical predictions for the 
neutralino detection cross section are represented as a function of
the neutralino mass for
$\tan\beta=35,\,45$. The cross section is bounded to be
$\crosssec\lsim 2\times10^{-9}$~pb by the aforementioned constraints.
The corresponding values for $\mu$ and pseudoscalar mass are shown in
Fig.\,\ref{3545mamuadterm}. Both are rather large, comparable to what
we obtained in Fig.\,\ref{102035mamuoptim} for the optimised case in
the previous Section. For such large $\tan\beta$ the small CP-odd Higgs
mass often leads to an excessive contribution to $\bmumu$,
incompatible with the present experimental bound.

In order to further increase the neutralino-nucleon cross section
while avoiding the experimental constraints, we
need to find a way to decrease the Higgs masses even for small values
of $\tan\beta$. This can be done by further exploiting the D-term
contribution if we assume that also the Higgses have
non-vanishing anomalous $U(1)$ charge. For example, setting
${q_{H_u}^A}/{q_C^A}=-2$, using result (\ref{oneX}) equation
(\ref{higgs_b}) is modified as 
\begin{eqnarray}
  m_{H_u}^2&=\,&m_{3/2}^2\,(11-17\cos^2\theta)\ ,\nonumber\\
  m_{H_d}^2&=\,&m_{3/2}^2\,\left(1-3\cos^2\theta\right)\ .
  \label{higgs_b2}
\end{eqnarray} 
We have labelled this as case A-II) in Table\,\ref{tablemodularu1}.
The degree of non-universality, using notation
(\ref{Higgsespara}), is given by $\delta_{H_{u}}=10-17\cos^2\theta$
and $\delta_{H_{d}}=-3\cos^2\theta$. In the dilaton-dominated case
this turns out to be very large, $\delta_{H_{u}}=10$.
Notice once more that, due to the presence of the anomalous U(1),
positive values for $\delta_{H_u}$ can be obtained.

\begin{figure}[!t]
  \epsfig{file=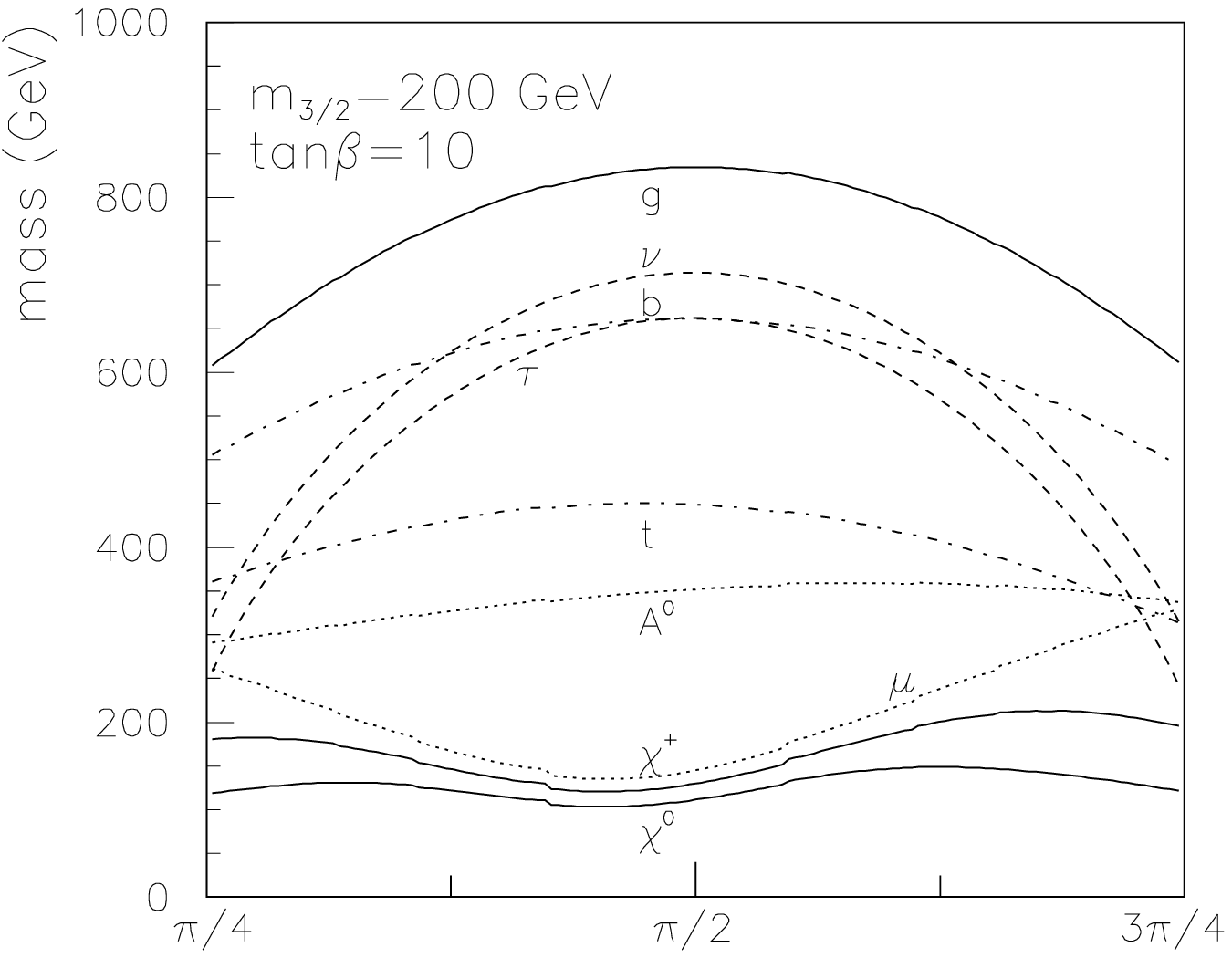,width=8.7cm}
  \hspace*{-0.9cm}  
  \epsfig{file=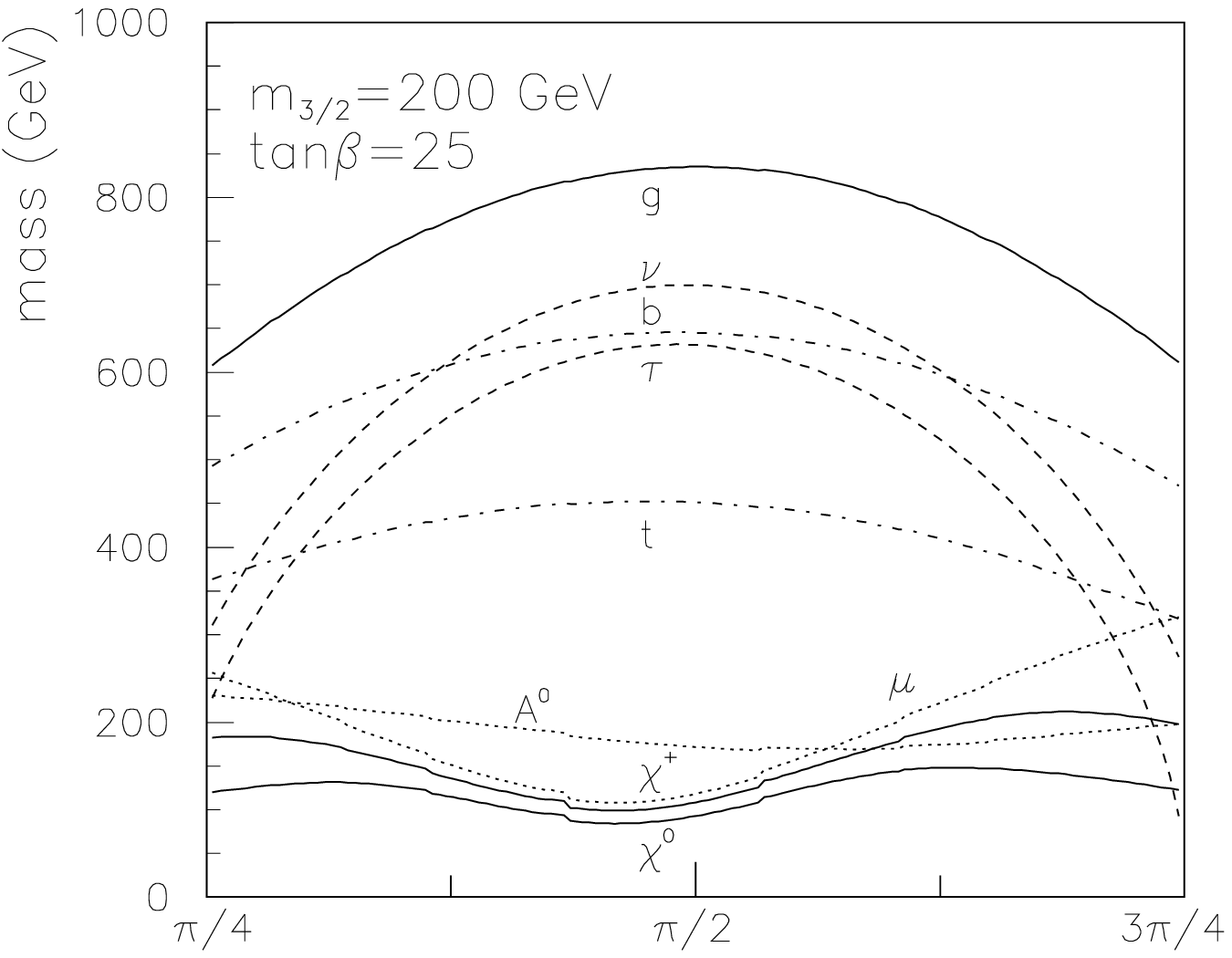,width=8.7cm}
  \captions{The same as Fig.\,\ref{10asp} but for
    example A-II) of
    Table.\,\ref{tablemodularu1} with $\tan\beta=10$ and $25$.
    Soft masses satisfy (\ref{weights})
    with slepton masses modified according to Eq.(\ref{sleptons2}) and
    Higgs mass parameters given by (\ref{higgs_b2}).
    \label{1025dterm5sp}}
\end{figure}

Due to the large non-universalities, the $\mu$ parameter and heavy
Higgs masses can be significantly reduced, even for moderate values of
$\tan\beta$, and especially in the dilaton limit. This is evidenced in
Fig.\,\ref{1025dterm5sp} where the SUSY spectrum is plotted for
$m_{3/2}=200$ GeV and $\tan\beta=10$ and $25$. The
lightest neutralino has an increased Higgsino composition in the area 
close to the dilaton-dominated limit and its mass is similar to that
of the lightest chargino. This, together with the decrease in
$m_{A^0}$, leads to a more effective neutralino annihilation in the
early Universe and hence a reduced relic density towards
$\theta=\pi/2$.

\begin{figure}[!t]
  \epsfig{file=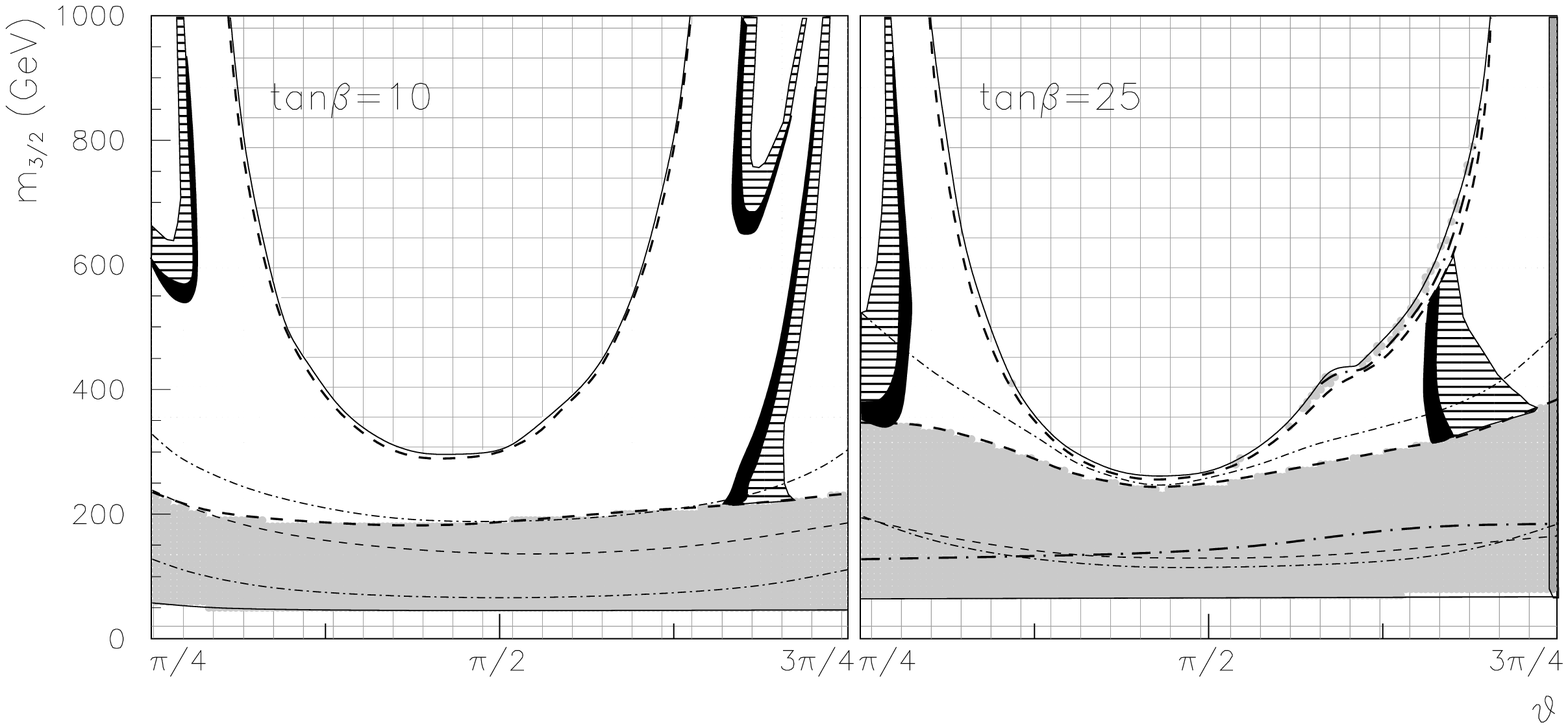,width=16cm}
  \vspace*{-1cm}
  \captions{The same as Fig.\,\ref{10a} but for
    example A-II) of
    Table.\,\ref{tablemodularu1} with $\tan\beta=10$ and $25$.
    Soft masses satisfy (\ref{weights})
    with slepton masses modified according to Eq.(\ref{sleptons2}) and
    Higgs mass parameters given by (\ref{higgs_b2}).}
  \label{all_adterm5}
\end{figure}

The associated $(m_{3/2},\,\theta)$ plane is represented in
Fig.\,\ref{all_adterm5} for $\tan\beta=10$ and $25$. 
Extensive areas of the parameter
space become excluded since $m_A^2$ becomes negative. These regions
correspond to the ruled areas above the experimentally allowed
ones. They are specially constraining in the dilaton limit, for which
$m_{3/2}\lsim300$ GeV is needed. As $\tan\beta$ increases the dilaton
limit eventually becomes excluded and the allowed regions shrink
towards $\theta=\pi/4$ and $3\pi/4$.
As mentioned above, the resulting
neutralino relic density is too small in the dilaton limit, but away
from it, regions appear where the WMAP result is reproduced.
Once more, due to the increase
of the stau mass and $\higgsu$, the UFB constraints pose no problem in
this scenario, and all the points depicted in Fig.\,\ref{all_adterm5}
satisfy them.

\begin{figure}[!t]
  \epsfig{file=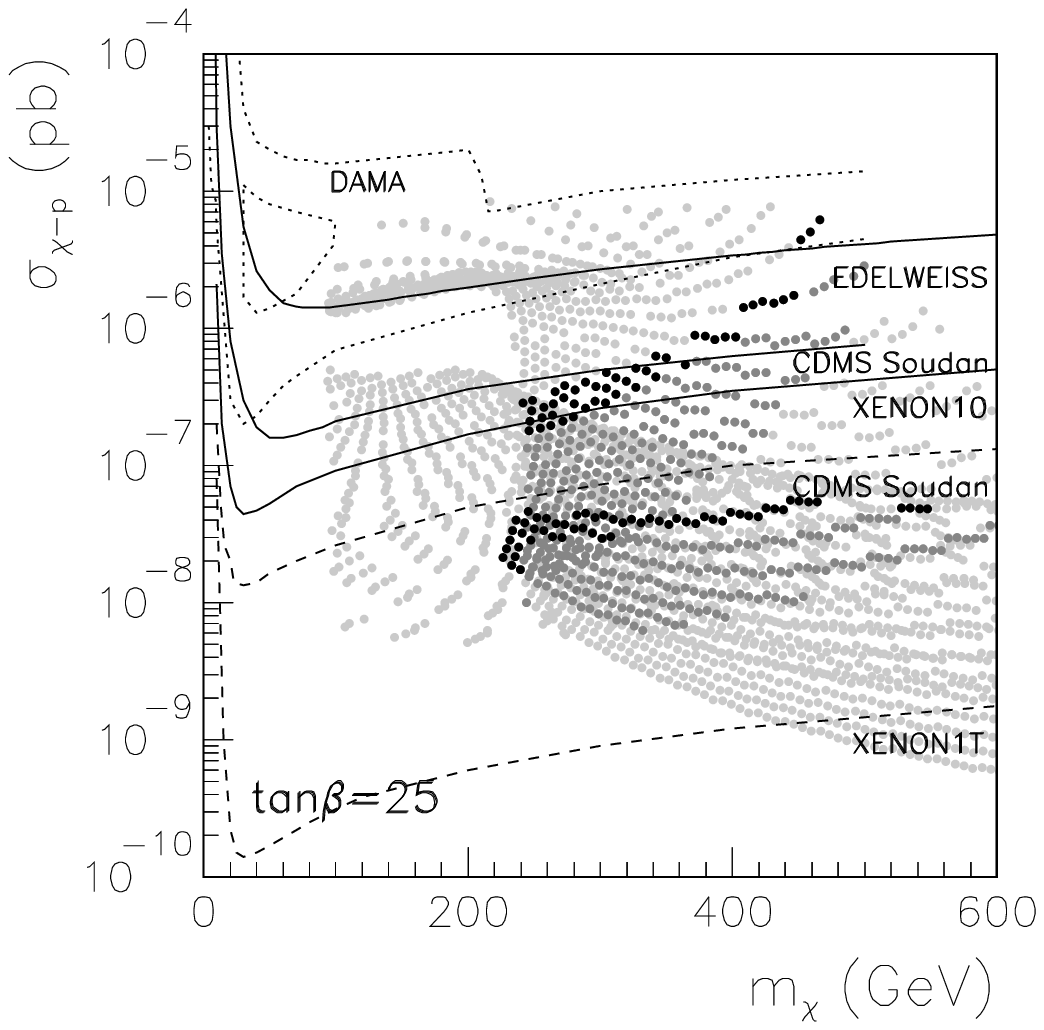,width=8cm}
  \epsfig{file=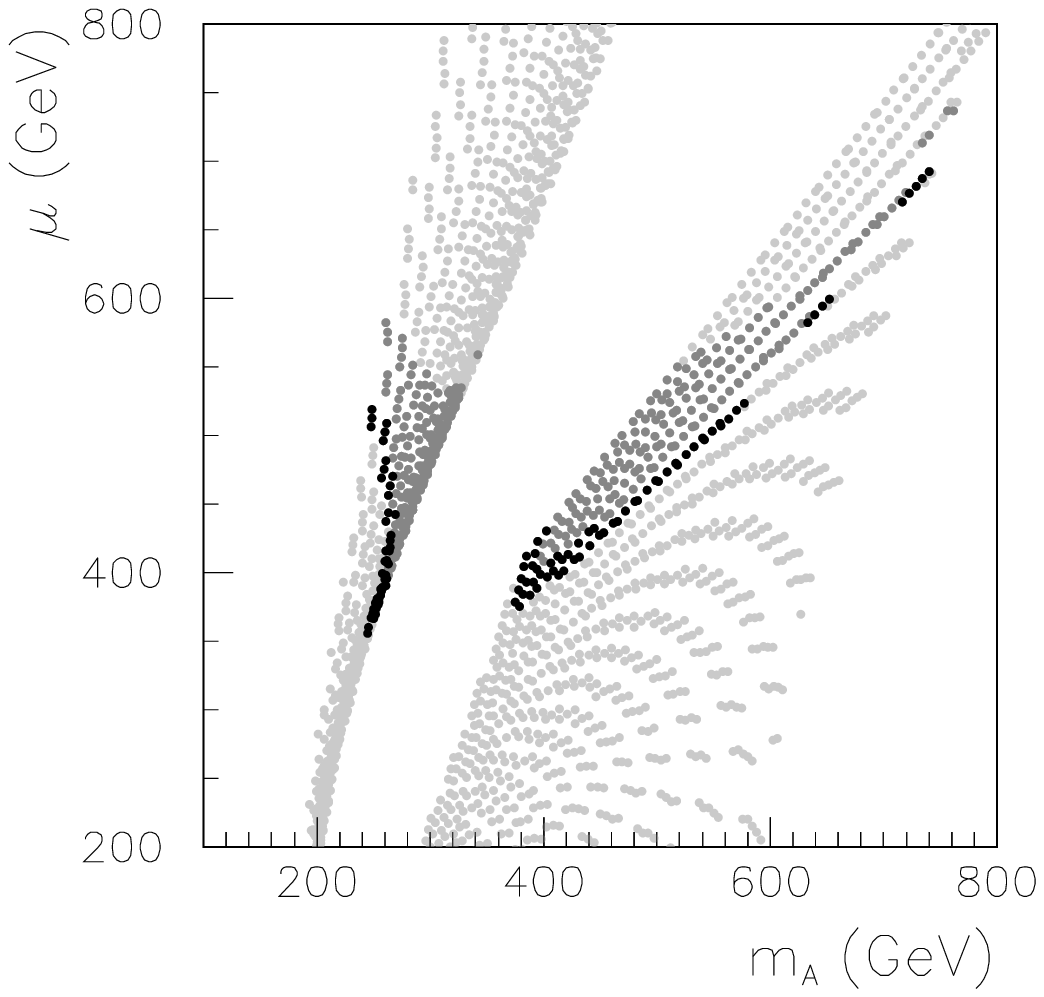,width=8cm}
  \vspace*{-1cm}
  \captions{The same as Fig.\,\ref{102035across} but for
    example A-II) of
    Table.\,\ref{tablemodularu1} with $\tan\beta=25$.
    Soft masses satisfy (\ref{weights})
    with slepton masses modified according to Eq.(\ref{sleptons2}) and
    Higgs mass parameters given by (\ref{higgs_b2}).}
   \label{all_adterm5cross}
\end{figure}

The decrease of the $\mu$ parameter and the resulting larger Higgsino
component for the lightest neutralino, $N_{13}^2+N_{14}^2\approx0.4$,
contribute to the increase of $\crosssec$. The
neutralino-nucleon cross section is depicted on the left-hand side of
Fig.\,\ref{all_adterm5cross} for $\tan\beta=25$, displaying neutralinos
with $\crosssec\gsim10^{-7}$ pb and a mass in the range
$\neumass\approx200-500$ GeV that would be within the reach of the
CDMS Soudan or XENON10 experiments. On the right-hand side of
Fig.\,\ref{all_adterm5cross} the corresponding value of the $\mu$
parameter is represented as a function of the pseudoscalar Higgs mass
for the same example, evidencing the decrease in both quantities. 
The points with a larger detection cross section
correspond to those with $m_A\approx220-240$ GeV and
$\mu\approx350-500$ GeV.

Notice that, unlike the previous case, a significant increase of
$\crosssec$ is obtained while keeping the observables B(\bsg) and
B($\bmumu$) under control. As already mentioned, this owes to the fact
that large values of $\tan\beta$ are no longer needed for obtaining
sizable detection cross sections.

The Higgs non-universality can be further increased if we also
consider $H_d$ to be charged under the anomalous $U(1)$. For example,
taking ${q_{H_d}^A}/{q_C^A}=1/2$, which corresponds to case A-III) in
Table\,\ref{tablemodularu1}, the soft masses for the Higgses
become 
\begin{eqnarray}
  m_{H_u}^2&=\,&m_{3/2}^2\,(11-17\cos^2\theta)\ ,\nonumber\\
  m_{H_d}^2&=\,&m_{3/2}^2\,\left(-\frac32+\cos^2\theta\right)\ .
  \label{higgs_b3}
\end{eqnarray}
In this case the non-universalities are given by
$\delta_{H_{u}}=10-17\cos^2\theta$, and
$\delta_{H_{d}}=-\frac52+\cos^2\theta$, which implies that in the
dilaton limit they become $\delta_{H_{u}}=10$ and
$\delta_{H_{d}}=-\frac52$.

\begin{figure}[!t]
  \epsfig{file=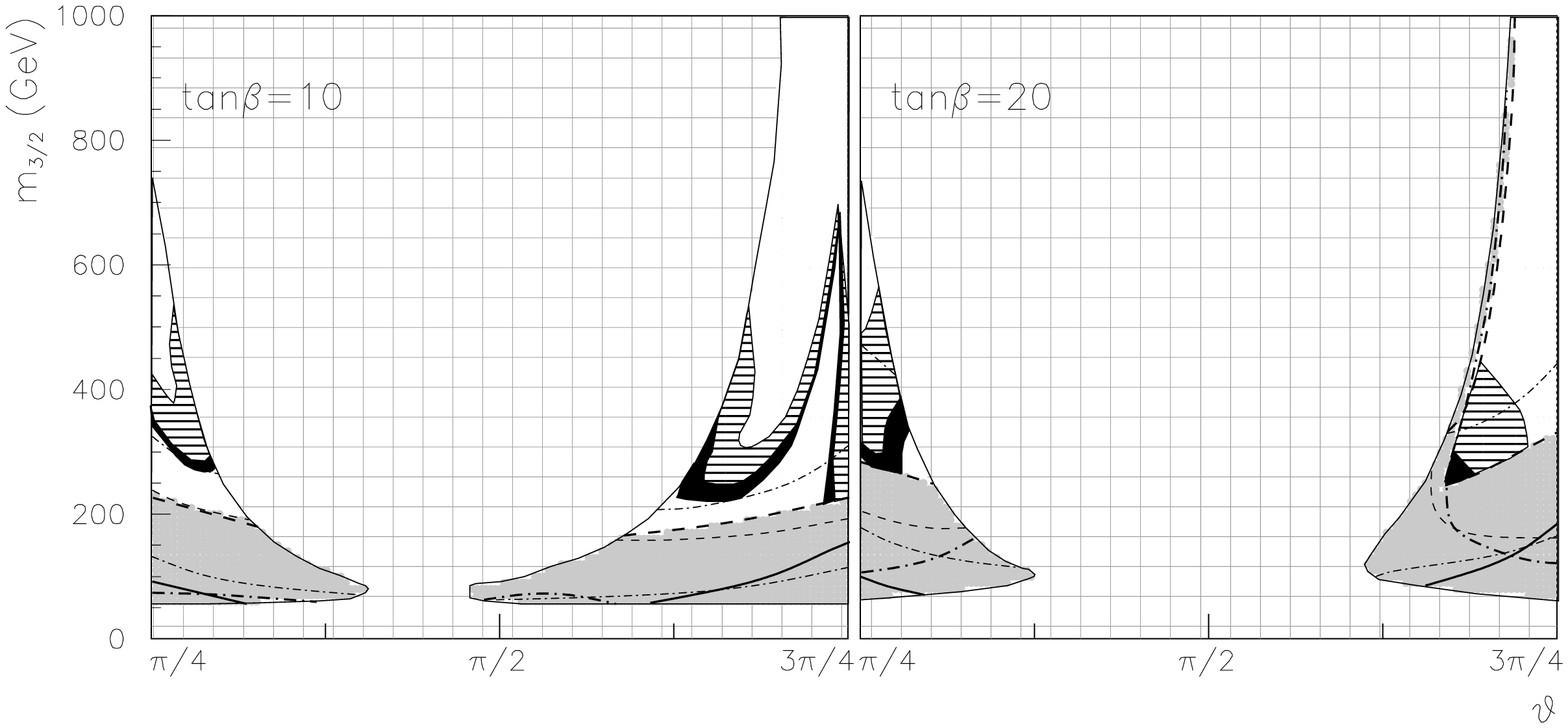,width=16cm}
  \vspace*{-1cm}
  \captions{The same as Fig.\,\ref{10a} but for
    example A-III) of
    Table.\,\ref{tablemodularu1} with $\tan\beta=10$ and $20$.
    Soft masses satisfy (\ref{weights})
    with slepton masses modified according to Eq.(\ref{sleptons2}) and
    Higgs mass parameters given by (\ref{higgs_b3}).}
  \label{all_adterm4}
\end{figure}

The associated parameter space is represented in
Fig.\,\ref{all_adterm4} for $\tan\beta=20$. As in the previous
example, extensive areas of the $(m_{3/2},\,\theta)$ plane are ruled
out because of $m_A^2$ becoming negative. Now, the region around the
dilaton-dominated case, where the non-universality is maximal, is
completely excluded for this reason. Only narrow allowed areas of the
parameter space for specific values of the Goldstino angle
survive. These allowed areas become smaller as $\tan\beta$ increases,
since the pseudoscalar mass becomes tachyonic more easily, and
eventually disappear for $\tan\beta\gsim 30$.

\begin{figure}[!t]
  \epsfig{file=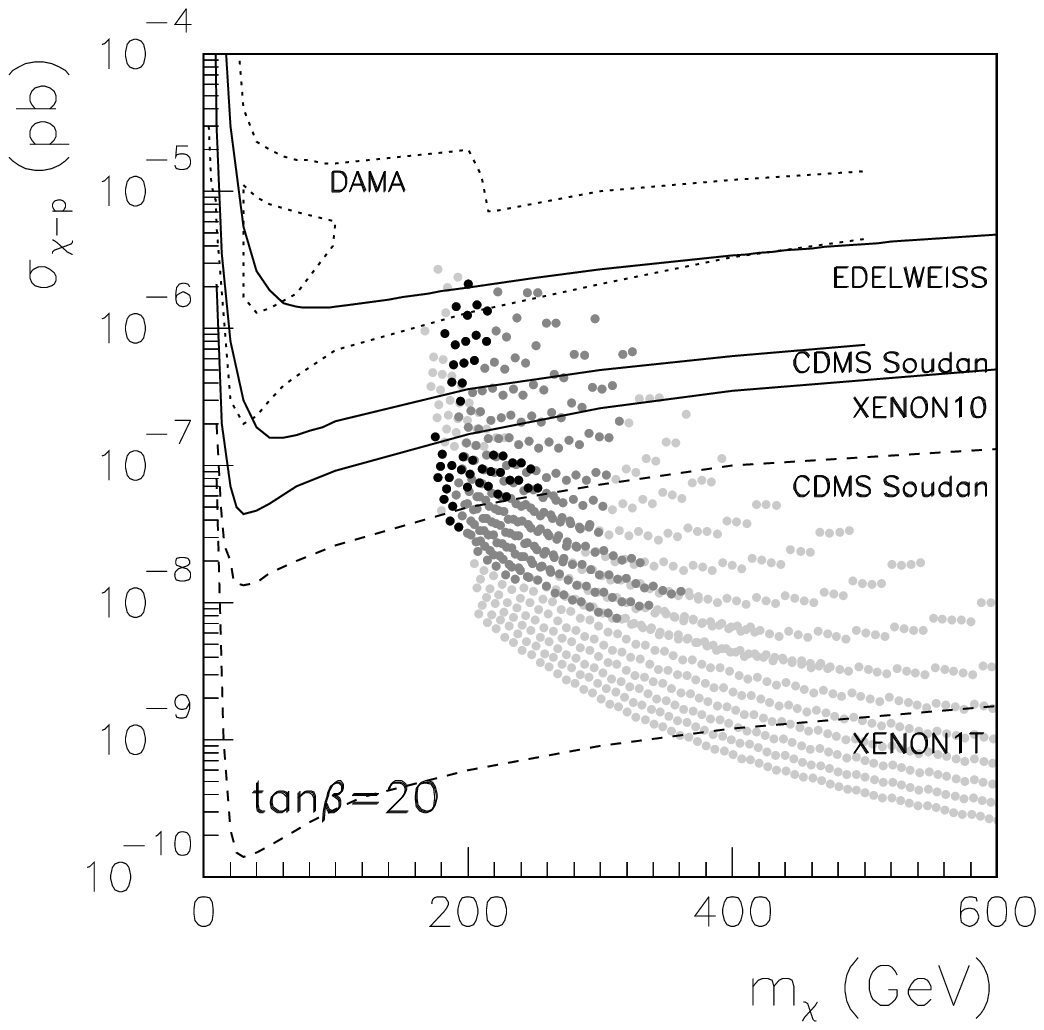,width=8cm}
  \epsfig{file=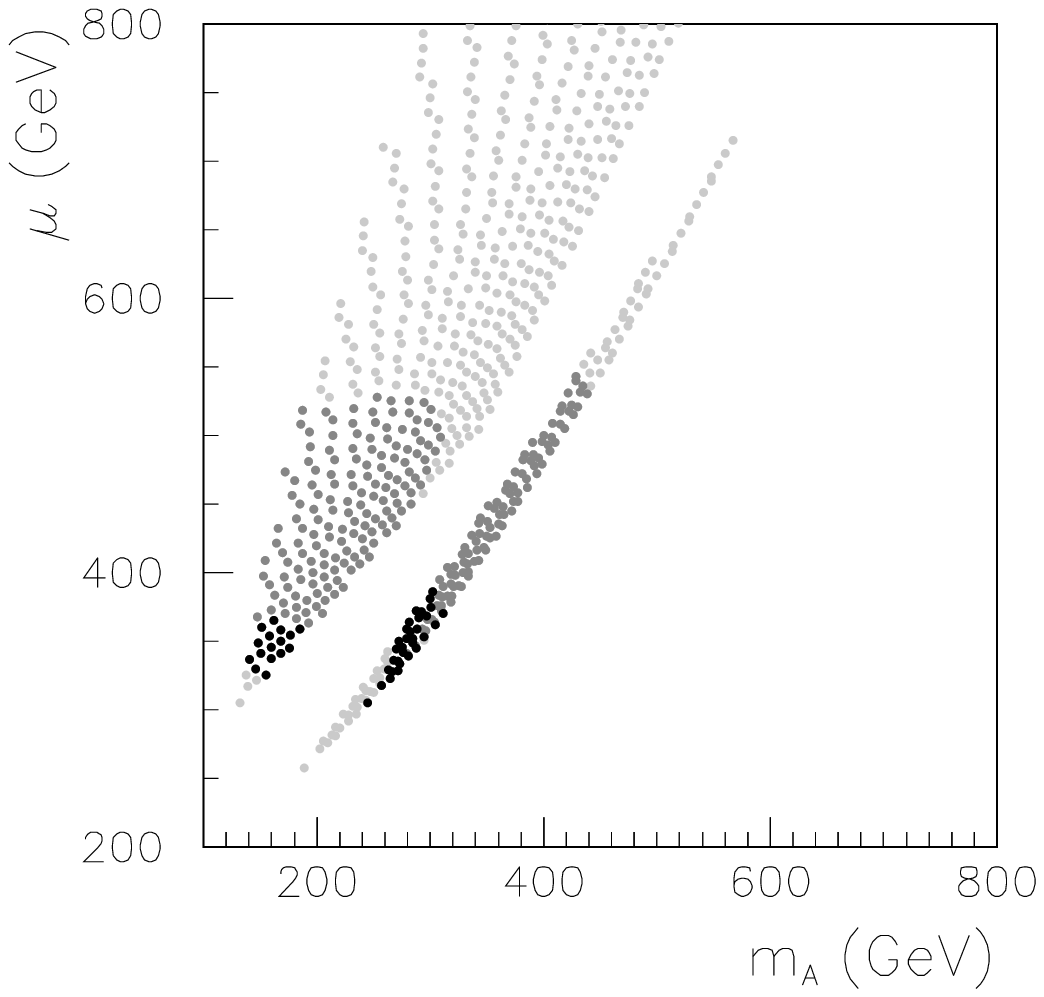,width=8cm}
  \vspace*{-1cm}
  \captions{The same as Fig.\,\ref{102035across} but for
    example A-III) of
    Table.\,\ref{tablemodularu1} with $\tan\beta=20$.
    Soft masses satisfy (\ref{weights})
    with slepton masses modified according to Eq.(\ref{sleptons2}) and
    Higgs mass parameters given by (\ref{higgs_b3}).}
  \label{all_adterm4cross}
\end{figure}

As a consequence of the further decrease in $\higgsd$, even smaller
values of $\tan\beta$ are enough in order to obtain light pseudoscalar
and scalar Higgses. Consequently, values
of $\crosssec$ within the reach of present experiments can now be
found for even smaller values of $\tan\beta$. 
These are shown as a function of the
neutralino mass on the left-hand side of Fig.\,\ref{all_adterm4cross}
for $\tan\beta=20$, which is the optimal choice for this
case. 
The pseudoscalar mass is
very small in these points, as evidenced on the right-hand side of
Fig.\,\ref{all_adterm4cross}, where the $\mu$ parameter is plotted
versus $m_A$. The points with $\crosssec\gsim10^{-7}$ pb correspond
to those with $m_A\lsim200$ GeV, and have a moderate Higgsino
composition.

Larger values of ${q_{H_d}^A}/{q_C^A}$ would further reduce
$m_{H_d}^2$ and lead to more sizable non-universalities. However, this
makes it more difficult to obtain positive $\mu^2$ or $m_A^2$, and the
allowed areas in the parameter space become more fine-tuned. For
instance, with ${q_{H_d}^A}/{q_C^A}=1$, which leads to
\begin{eqnarray}
  m_{H_u}^2&=\,&m_{3/2}^2\,(11-17\cos^2\theta)\ ,\nonumber\\
  m_{H_d}^2&=\,&m_{3/2}^2\,(-4+5\,\cos^2\theta)\ ,
  \label{higgs_b4}
\end{eqnarray} 
there are only two narrow regions left with $200$~GeV$\lsim
m_{3/2}\lsim 400$~GeV around $\theta\approx 0.8$ and $\theta\approx
2.3$, for which $\delta_{L_L,e_R}\approx 0.8$, $\delta_{H_u}\approx
1.7$, $\delta_{H_d}\approx -2.6$, and $\delta_{L_L,e_R}\approx 1.6$,
$\delta_{H_u}\approx 2.5$, $\delta_{H_d}\approx -2.8$,
respectively. The results for $\crosssec$ are qualitatively similar to
those of Fig.\,\ref{all_adterm4cross}, with points within the reach of
CDMS Soudan and XENON10 for $\tan\beta=10$ to $25$. The complete
$(m_{3/2},\theta)$ plane is ruled out for $\tan\beta\gsim25$ due to
the smallness of $m_{A}^2$.

\begin{figure}[!t]
  \epsfig{file=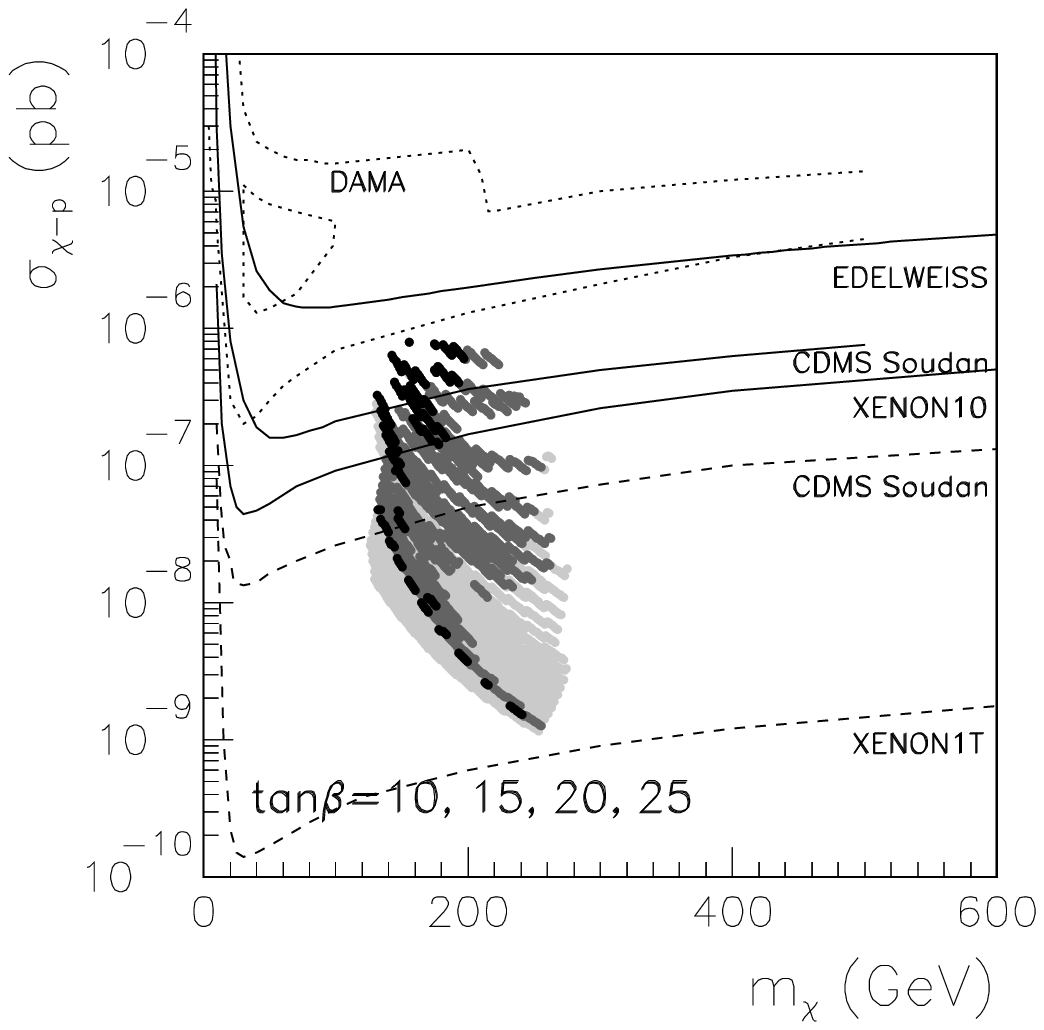,width=8cm}
  \epsfig{file=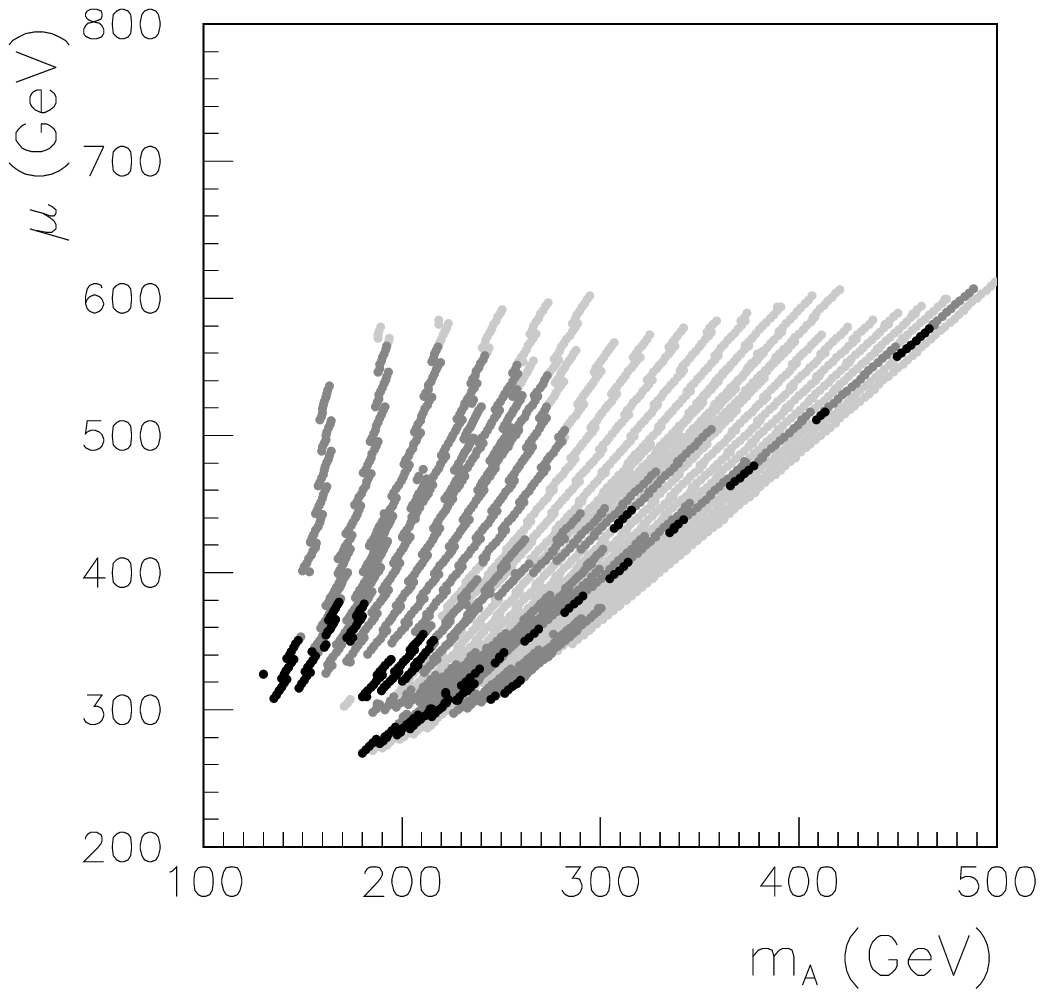,width=8cm}
  \vspace*{-1cm}
  \captions{The same as Fig.\,\ref{102035across} but for
    $\tan\beta=10,\,15,\,20$ and $25$. 
    Soft masses satisfy (\ref{weights}) and the effect of D-terms on
    the slepton and Higgs masses has been included according to 
    Eqs.(\ref{sleptons2}) and (\ref{higgs_b4}), respectively.}
  \label{all_adterm3cross}
\end{figure}

In order to illustrate this possibility, the theoretical predictions
for $\crosssec$ are represented in Fig.\,\ref{all_adterm3cross} 
for a scan of the cases with $\tan\beta=10,\,15,\,20$ and $25$.
The neutralino detection cross section reaches the
sensitivity of the CDMS Soudan and XENON10 experiments 
for $\neumass\approx150-200$
GeV. On the right-hand side of Fig.\,\ref{all_adterm3cross} the
corresponding values of $\mu$ are plotted as a function of the CP-odd
Higgs mass, evidencing the important decrease in both as a consequence
of Higgs non-universalities.

If the ratio of $U(1)$ charges is further increased to 
${q_{H_d}^A}/{q_C^A}=2$, the whole parameter
space becomes excluded.

Notice finally that,
although we have exemplified 
the relevance of the contribution due to
an anomalous $U(1)$ with a variation of case A) of
Table\,\ref{tablemodular}, the rest of the
examples in that table can also improve their behaviour under UFB
constraints and lead to detectable neutralino dark matter by using
this procedure. This owes to the fact that 
the non-universality introduced by the D-term
contribution typically dominates over the modular weights, especially 
in the dilaton limit, as we noted in Eq.~(\ref{dilatond}).

\section{Conclusions}
\label{conclusions}

We have studied the phenomenology of supergravity
theories which result from orbifold scenarios of the
heterotic superstring and the resulting theoretical predictions
for the direct detection of neutralino dark matter. 
These scenarios are specially interesting, since the soft
terms can be computed explicitly and have, in general, a non-universal
structure.

We have studied the parameter space of these
constructions, computing the low-energy spectrum and taking into
account the most recent experimental and 
astrophysical constraints. In addition, we have imposed the absence of
dangerous charge and colour breaking minima. In the remaining
allowed regions the spin-independent part of the neutralino-proton
cross section has been calculated and compared with the sensitivity of
current and projected dark matter detectors.

In the absence of an anomalous $U(1)$ the non-universality of the soft
scalar mass parameters is
always negative with respect to the gravitino mass. 
The smallness of the stau mass implies that the UFB-3 direction in the
parameter space becomes very deep and the realistic minimum is no
longer the global one.
In most of the cases it is not possible to satisfy this constraint
in the regions permitted by astrophysical and experimental bounds. 
Also, the non-universality on the Higgs masses is not sufficient to
produce any significant increase of the neutralino-nucleon cross
section. Consequently, the theoretical predictions for $\crosssec$ are
typically beyond the sensitivity of dark matter experiments.

The presence of an anomalous $U(1)$ gives more flexibility to the 
non-universality in the scalar masses. These can even be heavier than
the gravitino mass at the GUT scale. This  
allows to increase the slepton masses, thus avoiding the
UFB constraints. Moreover, the lightest neutralino becomes the LSP in
most of the parameter space and regions fulfilling all the
experimental and astrophysical constraints can be found. 
Furthermore, the non-universality of the 
Higgs mass parameters can be tailored to favour the
presence of light Higgses and the 
increase in the Higgsino composition of the lightest neutralino.
This leads to
a sizable increase in the theoretical predictions of $\crosssec$
and compatibility with present experiments can be achieved.

\noindent{\bf Acknowledgements}

D.G. Cerde\~no is supported by the program ``Juan de la Cierva'' of
the Ministerio de Educaci\'on y Ciencia of Spain.
T. Kobayashi is supported in part by the
Grand-in-Aid for Scientific Research \#17540251 and  
the Grant-in-Aid for
the 21st Century COE ``The Center for Diversity and
Universality in Physics'' from the Ministry of Education, Culture,
Sports, Science and Technology of Japan.
The work of C. Mu\~noz was supported
in part by the Spanish DGI of the
MEC under Proyecto Nacional FPA2006-05423,
by the European Union under the RTN program
MRTN-CT-2004-503369, and under the ENTApP Network of the ILIAS project 
RII3-CT-2004-506222.
Likewise, the work of D.~G. Cerde\~no  and
C. Mu\~noz, was also supported in part by the Spanish DGI of the
MEC under Proyecto Nacional FPA2006-01105, 
by the Comunidad de Madrid under Proyecto HEPHACOS, Ayudas de I+D
S-0505/ESP-0346, and by the EU research and training network
MRTN-CT-2006-035863.

\providecommand{\href}[2]{#2}
\begingroup\raggedright\endgroup

\end{document}